\documentclass{article}
\usepackage{bookstyleold}
\usepackage{graphicx}
\usepackage{cite}
\def\Tr{\mbox{Tr}}
\def\G{G}
\def\cth{\mbox{coth}}
\usepackage{epsfig}
\usepackage{bm}
\begin{document}
\author{Alex Kamenev}
\address{Department of Physics, University of Minnesota, \\  Minneapolis,
MN 55455, USA}
\title{Many--body theory of non--equilibrium systems}
%
\frontmatter
\maketitle
\mainmatter

\section{Introduction}
\label{sec_intro}

\subsection{Motivation and outline}

These lectures are devoted to the Keldysh formalism for the
treatment of out--of--equilibrium many--body systems. The name of
the technique takes its origin from the 1964 paper of
L.~V.~Keldysh \cite{Keldysh65}. Among the earlier  approaches that
are closely related to the Keldysh technique, one should mention
J.~Schwinger \cite{Schwinger61} and R.~P.~Feynman and F.~L.~Vernon
\cite{Feynman63}.  The classical counterparts of the Keldysh
technique are extremely useful and interesting on their own. Among
them the Wyld diagrammatic technique \cite{Wyld} and the
Martin--Siggia--Rose method \cite{MSR} for stochastic systems.

There are a number of  pedagogical presentations of the method
\cite{Landau,Mahan,Rammer}. The emphasis of these notes is on the
functional integration approach. It makes the structure of the
theory clearer and more transparent.  The notes also cover  modern
applications such as the Usadel equation and the nonlinear
$\sigma$--model.  Great attention is devoted to exposing
connections to other techniques such as the equilibrium Matsubara
method and the classical Langevin and Fokker-Planck equations, as
well as the Martin--Siggia--Rose technique.

\vskip .4cm

The Keldysh formulation of the many--body theory is useful for the
following tasks:
\begin{itemize}
\item Treatment of  systems that are not in  thermal equilibrium
(either due to the presence of external  fields, or in a transient
regime) \cite{Keldysh65,Landau,Rammer}.
\item Calculation of the full counting statistics of a
quantum mechanical observable (as opposed to an average value or
correlators) \cite{Levitov,Nazarov}.
\item As an alternative to the replica and the supersymmetry  methods in the
theory of disordered and glassy systems
\cite{Sompolinsky,Feigelman,Cugliandolo,Kamenev99,Shamon99}.
\item Treatment of equilibrium problems, in which the  Matsubara analytical continuation may
prove to be cumbersome.

\end{itemize}

The outline of these lectures is as follows. The technique is
introduced and explained for the simplest possible system, that of
a single bosonic state (harmonic oscillator), which is later
generalized  to  real (phonons), or complex (atoms) bosonic
fields. Their action and Green functions are introduced in Chapter
\ref{sec_1}. Boson interactions, the diagrammatic technique and
the quantum kinetic equation are treated in Chapter \ref{sec_2}.
Chapter \ref{sec_4} is devoted to a bosonic particle in contact
with a dissipative environment (bath). This example is used to
establish connections with the classical methods (Langevin and
Fokker--Planck) as well with the quantum equilibrium technique
(Matsubara). Fermions and fermion--boson systems are treated in
Chapter \ref{sec_5}. Covered topics include the random phase
approximation and the quantum kinetic equation. Non--interacting
Fermions in the presence of  quenched disorder are treated in
Chapter \ref{sec_6} with the help of the   Keldysh non-linear
$\sigma$-model.

\subsection{Closed time contour}
\label{sec_closed}

The standard construction of the zero temperature (or equilibrium)
many--body theory (see e.g. \cite{Mahan,AGD}) involves the
adiabatic switching  ``on'' of interactions  at a distant past,
and ``off'' at  a distant future. A typical correlation function
has the form of a time ordered product of operators in the
Heisenberg representation: $C(t,t')\equiv \langle 0| T \hat
A(t)\hat B(t')|0\rangle$, where $|0\rangle$ is the ground-state
(or thermal equilibrium state) of the {\em interacting}
Hamiltonian, $\hat H$. This state is supposed to be given by
$|0\rangle =\hat S(0,-\infty)|\rangle_0$, where $|\rangle_0$ is
the  (known) ground-state of the {\em non--interacting}
Hamiltonian, $\hat H_0$, at $t=-\infty$. The $\hat S$--matrix
operator $\hat S(t,t')= e^{i\hat H_0t} e^{-i\hat H(t-t')}
e^{-i\hat H_0t'}$ describes the evolution due to the interaction
Hamiltonian, $\hat H-\hat H_0$, and  is thus responsible for the
adiabatic switching ``on'' of the interactions. An operator in the
Heisenberg representation is given by $\hat A(t) = [\hat
S(t,0)]^\dagger \hat {\cal A}(t) \hat S(t,0)=\hat S(0,t) \hat
{\cal A}(t) \hat S(t,0)$, where $\hat {\cal A}(t)$ is the operator
in the interaction representation. As a result, the correlation
function takes the form:
\begin{eqnarray}
C(t,t')&=&\left._0\langle|\right.T\hat S(-\infty,0)\hat
S(0,t)\hat{\cal A}(t) \hat S(t,t') \hat {\cal B}(t')\hat S(t',0)
\hat
S(0,-\infty)|\rangle_0 \nonumber \\
 &=&
 \frac{\left._0\langle|\right.T\hat{\cal
A}(t) \hat {\cal B}(t') \hat
S(\infty,-\infty)|\rangle_0}{\left._0\langle|\right.\hat
S(\infty,-\infty)|\rangle_0}\,\, ,
                              \label{equilibrium}
\end{eqnarray}
where one  employed: $\left._0\langle|\right. \hat S(-\infty,0)=
e^{-iL}\left._0\langle|\right.\hat S(\infty,-\infty) \hat
S(-\infty,0)$, and interchanged the order of operators, which is
always allowed under the $T$--operation (time ordering). The idea
is that, starting at $t=-\infty$ at the ground (or equilibrium)
state, $|\rangle_0$, of the non--interacting system and then
adiabatically switching interactions ``on'' and ``off'', one
arrives at $t=+\infty$ at the state $|\infty\rangle$. The crucial
{\em assumption} is that this state is unique, independent of the
details of the switching procedure and is again the ground--state,
up to a possible phase factor: $e^{iL} =
\left._0\langle|\right.|\infty\rangle=\left._0\langle|\right.\hat
S(\infty,-\infty)|\rangle_0$.

Clearly this is {\em not} the case out of equilibrium. Starting
from some arbitrary non--equilibrium state and then switching
interactions ``on'' and ``off'',  the system evolves to some
unpredictable state. The latter depends, in general, on the
peculiarities of the switching procedure. The  entire construction
sketched above fails since we have no knowledge of the final
state.

\begin{figure}[t]
\fbox{\vtop to4cm{\vss\hsize=.975\hsize \vglue 0cm
\hspace{0.01\hsize} \hskip -.5cm \epsfxsize=1\hsize
\epsffile{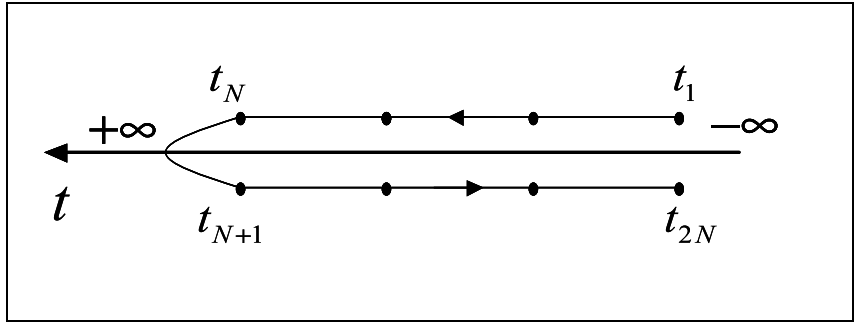} \hspace{0.1\hsize} \vss }}
\label{contour}
 \caption{
  The closed  time contour ${\cal C}$. Dots on the forward
and the backward branches of the contour denote  discrete time
points. }
\end{figure}

One would like, thus, to build a theory that avoids references to
the state at $t=+\infty$. Since  traces are calculated, one still
needs to know the final state. Schwinger's suggestion is to take
the final state to be exactly the same as the initial one. The
central idea is to let the quantum system  evolve first in the
forward direction in time and then to ``unwind'' its evolution
backwards, playing the ``movie'' in the backward direction. One
ends up, thus, with the need to construct a theory with the time
evolution along the two--branch contour, ${\cal C}$, depicted on
Fig.~\ref{contour}. Then, no matter what  the state at $t=+\infty$
is, after the backward evolution the system returns back to the
known initial state. As a result, the unitary evolution operator,
$\hat U_{t,t'}\equiv e^{-iH(t-t')}$,  along such a closed time
contour is always a unit operator:
\begin{equation}
\hat U_{\cal C}\equiv 1\, .
                               \label{evolution}
\end{equation}

In this construction there is no switching of interactions in the
future. Both switchings ``on'' and ``off'' take place in the past:
``on'' -- at the forward branch of the contour and ``off'' -- at
the backward one. This way the absence of information about the
$t=+\infty$ state is bypassed. There is a price to pay for such
luxury: a doubling of degrees of freedom. Indeed at every moment
of time one needs to specify a field residing on the forward
branch as well as on the backward branch of the contour. As a
result, the algebraic structure of the theory is more complicated.
The difficulties may be minimized, however, by a proper choice of
variables based  on the internal symmetries of the theory.

\vskip 1cm

\section{Free boson systems}
\label{sec_1}

\subsection{Partition function}
\label{sec_11}

Let us consider the simplest possible many--body system: bosonic
particles occupying a single quantum state with an energy
$\omega_0$. It is completely equivalent, of course, to a harmonic
oscillator. The secondary quantized Hamiltonian has the form:
\begin{equation}
\hat H = \omega_0\, a^{\dagger} a\, ,
                                         \label{Ham}
\end{equation}
where $a^{\dagger}$ and $a$ are bosonic creation and annihilation
operators with the commutation relation $[a,a^{\dagger}]=1$. Let
us  define  ``partition function''  as:
\begin{equation}
Z=\frac{\Tr\{ \hat U_{\cal C}\hat\rho \} }{ \Tr\{\hat \rho \} } \,
.
                                                              \label{e1}
\end{equation}
If one assumes that all  external fields are exactly the same on
the forward and backward branches of the contour, then $\hat
U_{\cal C}=1$ and therefore $Z=1$.

\begin{table}[b]
\begin{tabular*}{\columnwidth}{@{\extracolsep{\fill}}@{}lcccccccc@{}}
\hline \hline
\end{tabular*}
{\bf  Reminder:} \normalsize the bosonic coherent state
$|\phi\rangle$ ($\langle\phi|\,$), parameterized by a complex
number $\phi$,   is defined as a right (left) eigenstate of the
annihilation (creation) operator: $a|\phi\rangle
=\phi|\phi\rangle$ ($\langle\phi|a^\dagger= \langle\phi|\bar\phi\,
$). The matrix elements of a {\em normally ordered} operator, such
as the Hamiltonian, take the form $\langle\phi|\hat
H(a^\dagger,a)|\phi'\rangle = H(\bar\phi,\phi')
\langle\phi|\phi'\rangle$. The overlap between two coherent states
is \cite{Negele} $\langle\phi|\phi'\rangle=\exp\{\bar\phi\phi'\}$.
Since the coherent state basis is overcomplete, the trace of an
operator, $\hat A$, is calculated with the weight: $\Tr\{\hat
A\}=\pi^{-1} \int\!\!\!\int d(\Re \phi)\, d(\Im \phi)\,
e^{-|\phi|^2}\, \langle\phi|\hat A|\phi\rangle $.
\end{table}

The initial density matrix $\hat \rho=\hat\rho(\hat H)$ is some
operator--valued function of the Hamiltonian. To simplify the
derivations one may choose it to be the equilibrium density
matrix, $\hat\rho_0 = \exp\{-\beta (\hat H-\mu \hat
N)\}=\exp\{-\beta(\omega_0-\mu)a^\dagger a\}$. Since  arbitrary
external perturbations may be switched on (and off) at a later
time, the choice of the equilibrium  initial density matrix does
not prevent one from treating non--equilibrium dynamics. For the
equilibrium initial density matrix:
\begin{equation}
\Tr\{\hat \rho_0 \}=\sum\limits_{n=0}^\infty
e^{-\beta(\omega_0-\mu)n}= [1-\rho(\omega_0)]^{-1}\, ,
                                       \label{densitytrace}
\end{equation}
where $\rho(\omega_0)=e^{-\beta(\omega_0-\mu)}$. An important
point is that, in general, $\Tr\{\hat \rho \}$ is an interaction--
and disorder--{\em independent} constant. Indeed, both
interactions and disorder are supposed to be switched on (and off)
on the forward (backward) parts of the contour sometime after
(before) $t=-\infty$. This constant is, therefore, frequently
omitted -- it never causes a confusion.

The next step is to divide the ${\cal C}$ contour into $(2N-2)$
time steps of  length $\delta_t$, such that $t_1=t_{2N}=-\infty$
and $t_{N}=t_{N+1}=+\infty$ as shown in Fig.~\ref{contour}. One
then inserts the resolution of unity in the coherent state
overcomplete basis \cite{Negele}
\begin{equation}
1=\int\!\!\!\int \frac{d(\Re \,\phi_j)\, d(\Im\,\phi_j)}{\pi}\,\,
e^{-|\phi_j|^2}\, |\phi_j\rangle\langle\phi_j|\,
                                        \label{resunity}
\end{equation}
at each point $j=1,2,\ldots, 2N$ along the contour. For example,
for $N=3$ one obtains the following sequence in the expression
for $\Tr\{\hat U_{\cal C}\hat\rho_0\}$ (read from right to left):
\begin{equation}
\hskip -.5cm
  \langle\phi_6|\hat U_{-\delta_t}|\phi_5\rangle
  \langle\phi_5|\hat U_{-\delta_t}|\phi_4\rangle
  \langle\phi_4|\hat 1|\phi_3\rangle
  \langle\phi_3|\hat U_{+\delta_t}|\phi_2\rangle
  \langle\phi_2|\hat U_{+\delta_t}|\phi_1\rangle
  \langle\phi_1|\hat \rho_0|\phi_6\rangle\, ,
                                                   \label{succession}
\end{equation}
where $\hat U_{\pm \delta_t}$ is the evolution operator during the
time interval $\delta_t$ in the positive (negative) time
direction. Its matrix elements are given by:
\begin{equation}
\hskip -.5cm \langle\phi_{j+1}|\hat U_{\pm\delta_t}|\phi_j\rangle
\equiv \langle \phi_{j+1}| e^{\mp i\hat
H(a^\dagger,a)\delta_t}|\phi_{j}\rangle \approx \langle
\phi_{j+1}| \phi_{j}\rangle\, e^{\mp i
H(\bar\phi_{j+1},\phi_j)\delta_t}\, ,
                                    \label{matrixelement}
\end{equation}
where the last approximate equality is valid up to the linear
order in $\delta_t$. Obviously this result is not restricted to
the toy example, Eq~(\ref{Ham}), but holds for any {\em
normally--ordered} Hamiltonian. Notice that there is no evolution
operator inserted between $t_N$ and $t_{N+1}$. Indeed, these two
points are physically indistinguishable and thus the system does
not evolve during this time interval.

\begin{table}[h]
\begin{tabular*}{\columnwidth}{@{\extracolsep{\fill}}@{}lcccccccc@{}}
\hline \hline
\end{tabular*}
{\bf Exercise:} show that
$ \langle\phi| e^{\kappa a^\dagger a}|\phi'\rangle =
\exp\left\{\bar\phi\phi' e^{\kappa}\right\}\, . $
Putting $\kappa=-\beta(\omega_0-\mu)$, one finds $\langle\phi_1|
\hat\rho_0|\phi_{2N}\rangle = \exp\left\{\bar\phi_1\phi_{2N}
\rho(\omega_0) \right\}$.
\begin{tabular*}{\columnwidth}{@{\extracolsep{\fill}}@{}lcccccccc@{}}
\hline \hline
\end{tabular*}
\end{table}

Combining all such matrix elements along the contour together with
the exponential factors from the resolutions of unity,
Eq.~(\ref{resunity}), one finds for the partition function
(\ref{e1}):
\begin{equation}
Z={1\over \Tr\{\rho_0\} }  \int \!\!\!\int
\prod\limits_{j=1}^{2N}\left[\frac{d(\Re \,\phi_j)\,
d(\Im\,\phi_j)}{\pi}\right] \,\, e^{\,i \sum\limits_{j,j'=1}^{2N}
\bar\phi_j {\G}^{-1}_{jj'}\phi_{j'}}\, ,
                                              \label{partitiondiscrete}
\end{equation}
where the $2N \times 2N$ matrix $i{\G}^{-1}_{jj'}$ stands for:
\begin{equation}
  \label{Dmatrix}
  i\, {\G}^{-1}_{jj'}\equiv
\left[\begin{array}{rrr|rrr}
 -1   &      &    &   &   &   \rho(\omega_0) \\
  1\!-\!h & -1   &    &   &   &                 \\
      &  1\!-\!h & -1 &   &   &                 \\ \hline
     &     &  1 & -1 &    &                 \\
     &     &    &1\!+\!h & -1 &                 \\
     &     &    &    & 1\!+\!h&  -1
 \end{array} \right]\, ,
                                                  \label{matrix}
\end{equation}
and $h\equiv i\omega_0\delta_t$.  It is straightforward to
evaluate the determinant of such a  matrix
\begin{equation}
\hskip -.5cm \mbox{det} \big[ i{\G}^{-1} \big] = 1-
\rho(\omega_0)(1-h^2)^{N-1} \approx 1- \rho(\omega_0)\,
e^{(\omega_0\delta_t)^2(N-1)}\to 1- \rho(\omega_0) \, ,
                                                \label{determinant}
\end{equation}
where one has used that $\delta_t^2N\to 0$ if $N\to \infty$
(indeed, the assumption was $\delta_tN \to \mbox{const}$).
Employing Eqs.~(\ref{Gauss}) and (\ref{densitytrace}), one finds:
\begin{equation}
Z= \frac{ \mbox{det}^{-1}\big[ i{\G}^{-1} \big]}{\Tr\{\rho_0 \} }
= 1\, ,
                                                \label{unity}
\end{equation}
as it should be, of course. Notice, that keeping the upper--right
element of the discrete matrix, Eq.~(\ref{matrix}), is crucial to
maintain this normalization identity.

One may  now take the limit $N\to \infty$ and formally write the
partition function in the continuous notations, $\phi_j\to
\phi(t)$:
\begin{equation}
Z=  \int\!\!  {\cal D} \bar\phi \phi\,\, e^{\,iS[\bar\phi,\phi] }
= \int\!\!  {\cal D} \bar\phi \phi\, \exp\left\{i
\int\limits_{\cal C}\! \left[ \bar\phi(t)\,
{\G}^{-1}\phi(t)\right] dt\right\} \, ,
                                                              \label{e2}
\end{equation}
where according to Eqs.~(\ref{partitiondiscrete}) and
(\ref{matrix}) the action is given by
\begin{equation}
\hskip -.5cm S[\bar\phi,\phi] =
\sum\limits_{j=2}^{2N}\!\left[i\bar\phi_j\,\frac{\phi_j-\phi_{j-1}}{\delta
t_j} -\omega_0\bar\phi_j\phi_{j-1}\right]\delta
t_j\,+i\,\bar\phi_1\Big(\phi_1-\rho(\omega_0)\phi_{2N}\Big) \, ,
                                                              \label{e3}
\end{equation}
where $\delta t_j\equiv t_j-t_{j-1}=\pm \delta_t$. Thus a
continuous form of the operator ${\G}^{-1}$ is:
\begin{equation}
{\G}^{-1}= i\partial_t - \omega_0 \,.
                                             \label{Gcontinious}
\end{equation}
It is important to remember that this continuous notation is only
an abbreviation that represents  the large discrete matrix,
Eq.~(\ref{matrix}). In particular, the upper--right element of the
matrix (the last term in Eq.~(\ref{e3})), that contains the
information about the distribution function, is seemingly absent
in the continuous notations.

To avoid integration along the closed time contour, it is
convenient to split the bosonic field $\phi(t)$ into the two
components $\phi_+(t)$ and $\phi_-(t)$ that reside on the forward
and the backward parts of the time contour correspondingly. The
continuous action may  then be rewritten as
\begin{equation}
S=\int\limits_{-\infty}^{\infty}\!\!\! dt\,
\bar\phi_+(t)[i\partial_t - \omega_0]
\phi_+(t)-\int\limits_{-\infty}^{\infty}\!\!\! dt\,
\bar\phi_-(t)[i\partial_t - \omega_0] \phi_-(t)\, ,
                                         \label{plusminus}
\end{equation}
where the relative minus sign comes from the reverse  direction of
the time integration on the backward part of the contour. Once
again, the continuous notations are somewhat misleading. Indeed,
they create an undue impression that the $\phi_+(t)$ and
$\phi_-(t)$ fields are completely independent from each other. In
fact, they are connected due to the presence of the non--zero
off--diagonal blocks in the discrete matrix, Eq.~(\ref{matrix}).


\subsection{Green functions}

One would like  to define the Green functions  as:
\begin{equation}
G(t,t') = -i\!\! \int\!\! {\cal D} \bar\phi \phi\,\,
e^{iS[\bar\phi,\phi] }\,\, \phi(t)\bar\phi(t')\equiv -i\langle
\phi(t)\bar\phi(t')\rangle\, ,
                                          \label{corrfunction}
\end{equation}
where both time arguments reside somewhere on the Keldysh contour.
Notice, the absence of the factor $Z^{-1}$ in comparison with the
analogous definition in the equilibrium theory \cite{Negele}.
Indeed, in the present construction $Z=1$. This seemingly minor
difference turns out to be the major issue in the theory of
disordered systems, Chapter \ref{sec_6}.

According to the general property of  Gaussian integrals (see
\ref{app_Gaussian}),  the Green function is  the inverse of the
correlator matrix $G^{-1}$, Eq.~(\ref{matrix}), standing in the
quadratic action. Thus, one faces the unpleasant task of inverting
the large $2N\times 2N$ matrix, Eq.~(\ref{matrix}). It may seem
more attractive to invert the differential operator,
Eq.~(\ref{Gcontinious}). Such an inversion, however, is undefined
due to the presence of the zero mode ($\sim e^{-i\omega_0t}$). The
necessary regularization is  provided by the off--diagonal blocks
of the discrete matrix. The goal is to develop a formalism that
avoids dealing with the large discrete matrices and refers to the
continuous notations only.

The easiest way to proceed is to recall \cite{Negele} that the
Green functions are traces of  {\em time--ordered} products of the
field operators (in the Heisenberg representation), where the
ordering is done along the contour ${\cal C}$. Recalling also that
the time arguments on the backward branch are always {\em after}
those on the forward, one finds:
\begin{eqnarray}
                                         \label{corr}
&&\hskip -1.3cm   \langle \phi_+(t)\bar\phi_-(t')\rangle \equiv
i{\G}^<(t,t')\! =\! {\Tr\{a^\dagger(t') a(t) \hat \rho_0\}\over
\Tr\{\hat\rho_0\}}= { \Tr\{e^{i\hat H t'} a^\dagger e^{i\hat
H(t-t')} a\, e^{-i\hat Ht}
\hat \rho_0\}\over \Tr\{\hat\rho_0\}}  \,  \nonumber \\
&& ={e^{-i\omega_0(t-t')}  \over \Tr\{\hat\rho_0\}}\,
\sum\limits_{m=0}^\infty m[\rho(\omega_0)]^m
=ne^{-i\omega_0(t-t')}\, ;
\nonumber\\
&& \hskip -1.3cm\langle \phi_-(t)\bar\phi_+(t')\rangle \equiv
i{\G}^>(t,t') \! =\! {\Tr\{a(t) a^\dagger(t') \hat \rho_0\}\over
\Tr\{\hat\rho_0\}}= { \Tr\{e^{i\hat H t} a e^{i\hat H(t'-t)}
a^\dagger\, e^{-i\hat Ht'}
\hat \rho_0\}\over \Tr\{\hat\rho_0\}}  \,  \nonumber \\
&& = {e^{i\omega_0(t'-t)}  \over \Tr\{\hat\rho_0\}}\,
\!\!\sum\limits_{m=0}^\infty\! (m+1)[\rho(\omega_0)]^m
\!\!=\!\!(n+1)e^{-i\omega_0(t-t')}\, ;
\nonumber\\
&& \hskip -1.3cm\langle\phi_+(t)\bar\phi_+(t')\rangle \equiv
i{\G}^T(t,t')\!=\!
{ \Tr\{ T[ a(t) a^\dagger(t')]\hat \rho_0\} \over \Tr\{\hat\rho_0\}  } \\
&& =\theta(t-t')i{\G}^>(t,t') + \theta(t'-t)i{\G}^<(t,t') \, ;
\nonumber \\
&& \hskip -1.3cm\langle\phi_-(t)\bar\phi_-(t')\rangle \equiv
i{\G}^{\tilde T}(t,t') \! =\!
 { \Tr\{ \tilde T[ a(t) a^\dagger(t')]\hat \rho_0\} \over \Tr\{\hat\rho_0\}   }
\nonumber\\
&& =\theta(t'-t)i{\G}^>(t,t') + \theta(t-t')i{\G}^<(t,t') \, ;
\nonumber
\end{eqnarray}
where the symbols $T$ and $\tilde T$ denote time--ordering and
anti--time--ordering correspondingly. Hereafter the time arguments
reside on the open time axis $t\in ]-\infty,\infty[$. The Planck
occupation number $n$ stands for $n(\omega_0) \equiv
\rho(\omega_0)/(1-\rho(\omega_0))$.

Notice the presence of non--zero off--diagonal Green functions
$\langle \phi_-\bar\phi_+\rangle$ and $\langle
\phi_+\bar\phi_-\rangle$. This is seemingly inconsistent with the
continuous action, Eq.~(\ref{plusminus}). This is due to the
presence of the off--diagonal blocks in the discrete matrix, that
are lost in the continuous notations. The existence of the
off--diagonal Green functions does not contradict to  continuous
notations. Indeed, $[i\partial_t - \omega_0]{\G}^{>,<} = 0$, while
$[i\partial_t - \omega_0]{\G}^{T,\tilde T} = \pm \delta(t-t')$.
Therefore in the obvious $2\times 2$ matrix notations
$G^{-1}\circ{\G} = 1$,  as it should be. The point is that the
inverse of the operator $[i\partial_t - \omega_0]$  is not
well-defined (due to the presence of the eigenmode ($\sim
\exp\{-i\omega_0 t\}$) with zero eigenvalue). A regularization
must be specified and the off-diagonal blocks of the discrete
matrix  do exactly this.

The $\theta$--function in Eq.~(\ref{corr}) is the usual Heaviside
step function. There is an uncertainty, however, regarding its
value at coinciding time arguments. To resolve it, one needs to
refer  to the discrete representation one last time. Since the
fields $\bar\phi$ always appear one time step $\delta_t$ {\em
after} the fields $\phi$  on the Keldysh contour, cf.
Eq.~(\ref{matrixelement}), the proper convention is:
\begin{equation}
G^T(t,t)= G^{\tilde T}(t,t) = G^<(t,t)=n\, .
                                       \label{coincidingarguments}
\end{equation}
Obviously not all four Green functions defined above are
independent. Indeed, a direct inspection shows that for $t\neq
t'$:
\begin{equation}
{\G}^T + {\G}^{\tilde T} = {\G}^> + {\G}^< \,.
                                                \label{relation}
\end{equation}
One would like therefore to perform a linear transformation of the
fields to benefit  explicitly from this relation. This is achieved
by the  Keldysh rotation.

\subsection{Keldysh rotation}

Define  new fields as
\begin{equation}
\phi_{cl}(t) = {1\over \sqrt{2}}\big(\phi_+(t) + \phi_-(t)\big)
\,; \,\,\,\,\,\,\,\,\, \phi_{q}(t) =  {1\over
\sqrt{2}}\big(\phi_+(t) - \phi_-(t)\big)\,
                                                \label{rotation}
\end{equation}
with the analogous transformation for the conjugated fields. The
subscripts {\em ``cl''} and {\em ``q''} stand for the {\em
classical} and the {\em quantum} components of the fields
correspondingly. The rationale  for these notations will become
clear shortly. First, a simple algebraic manipulation of
Eq.~(\ref{corr}) shows that
\begin{equation}
-i\langle \phi_\alpha(t)\bar\phi_\beta(t')\rangle\equiv \hat
G^{\alpha\beta} = \left(\begin{array}{cc}
{\G}^K(t,t') & {\G}^{R}(t,t') \\
{\G}^{A}(t,t') & 0
\end{array}\right)\, ,
                                                \label{Green}
\end{equation}
where hereafter $\alpha, \beta = (cl,q)$. The cancellation of the
$(q,q)$ element of this matrix is a manifestation of
identity~(\ref{relation}). Superscripts $R,A$ and $K$ stand for
the {\em retarded, advanced} and {\em Keldysh} components of the
Green function correspondingly. These three Green functions are
the fundamental objects of the Keldysh technique. They are defined
as
\begin{eqnarray}
 \label{Greens}
{\G}^{R}(t,t') &=& {1\over 2}\left({\G}^{T} -G^{\tilde T} -
{\G}^{<}+G^>\right)=
\theta(t-t')({\G}^{>} - {\G}^{<})\, ;   \nonumber \\
{\G}^{A}(t,t') &=& {1\over 2}\left({\G}^{T} - G^{\tilde
T}+G^<-{\G}^{>}\right)=
\theta(t'-t)({\G}^{<} - {\G}^{>})\, ;    \\
{\G}^{K}(t,t') &=& {1\over 2}\left(G^T+G^{\tilde T}+{\G}^{>} +
{\G}^{<}\right)=G^<+G^>\, . \nonumber
\end{eqnarray}

In the discrete representation each of these three Green functions
is represented by an $N\times N$ matrix. Since both $G^<$ and
$G^>$ are, by definition, anti-Hermitian (cf. Eq.~(\ref{corr})),
Eq.~(\ref{Greens}) implies:
\begin{equation}
{\G}^{A} = \left[ {\G}^R \right]^{\dagger}\hskip 2cm
G^K=-[G^K]^\dagger\, ,
                                              \label{conj}
\end{equation}
where the Hermitian  conjugation includes complex conjugation, as
well as transposition of the time arguments. The retarded
(advanced) Green function a is lower (upper) triangular matrix.
 Due to the algebra
of  triangular matrices, a product of any number of upper (lower)
triangular matrices is again an upper (lower) triangular matrix.
This leads to the simple rule:
\begin{eqnarray}
{\G}_1^{R} \circ{\G}_2^{R}\circ \ldots \circ{\G}_l^{R}  &=& G^R \,
;
\nonumber \\
{\G}_1^{A}\circ {\G}_2^{A}\circ \ldots \circ{\G}_l^{A}  &=& G^A \,
,
                                         \label{traces}
\end{eqnarray}
where the  circular multiplication signs are understood as
integrations over  intermediate times (discrete matrix
multiplication). At  coinciding time arguments, one finds (cf.
Eqs.~(\ref{coincidingarguments}) and (\ref{Greens})):
\begin{equation}
G^R(t,t)+G^A(t,t) = 0\, .
                                               \label{GRplusGA}
\end{equation}
Although in the discrete representation both $G^R$ and $G^A$ {\em
do} contain non--zero (pure imaginary, due to Eqs.~(\ref{conj}),
(\ref{GRplusGA})) main diagonals (otherwise the matrix $\hat G$ is
not invertible), the proper continuous convention is:
$\theta(0)=0$. The point is that in any diagrammatic calculation,
$G^R(t,t)$ and $G^A(t,t)$ always come in symmetric combinations
and  cancel each other due to Eq.~(\ref{GRplusGA}). It is thus a
convenient and noncontradictory agreement to take $\theta(0)=0$.

\begin{figure}[t]
\fbox{\vtop to3cm{\vss\hsize=.975\hsize \vglue 0cm
\hspace{0.01\hsize} \hskip -.5cm \epsfxsize=1\hsize
 \epsffile{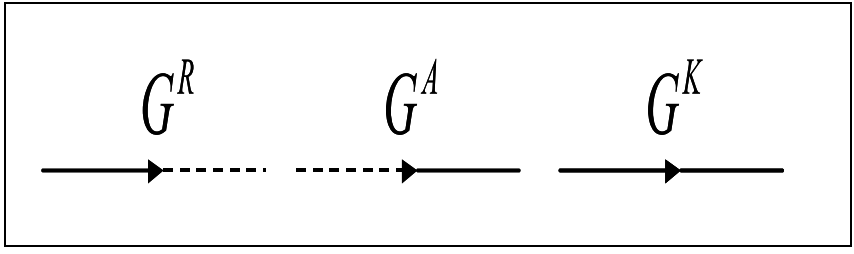}
\hspace{0.1\hsize} \vss }}
\label{fig_greens}
 \caption{Graphic representation of $G^R$, $G^A$, and $G^K$ correspondingly.
 The full line represents the classical field component, $\phi_{cl}$, while the dashed line --
 the quantum  component, $\phi_q$.}
\end{figure}

It is useful to introduce  graphic representations for the three
Green functions. To this end, let us denote the classical
component of the field by a full line and the quantum component by
a dashed line. Then the retarded Green function is represented by
a full-arrow-dashed line, the advanced by a dashed-arrow-full line
and the Keldysh by full-arrow-full line, see Fig.
\ref{fig_greens}. Notice, that the dashed-arrow-dashed line, that
would represent the $\langle \phi_q \bar\phi_q\rangle$ Green
function, is identically zero due to identity~(\ref{relation}).
The arrow shows the direction from $\phi_\alpha$  towards
$\bar\phi_\beta$.

For the single bosonic state (cf. Eq.~(\ref{corr})):  ${\G}^{>} =
-i(n+1)e^{-i\omega_0(t-t')}$ and ${\G}^{<} =
-ine^{-i\omega_0(t-t')}$, where $n=n(\omega_0)=
\rho(\omega_0)/(1-\rho(\omega_0))$ is the Planck occupation number
(since the system is non--interacting the initial distribution
function does not evolve).  Therefore:
\begin{eqnarray}
                                        \label{Greens1}
{\G}^{R}(t,t') &=&
-i\theta(t-t')\, e^{-i\omega_0(t-t')} \, ;   \nonumber \\
{\G}^{A}(t,t') &=&
i\theta(t'-t)\, e^{-i\omega_0(t-t')}\, ;    \\
{\G}^{K}(t,t') &=& -i (1+2n(\omega_0))\,
e^{-i\omega_0(t-t')}\nonumber  \, .
\end{eqnarray}
Notice that the retarded and advanced components contain
information only about the spectrum and are independent of the
occupation number, whereas the Keldysh component does depend on
it. Such a separation is common for systems that are not too far
from  thermal equilibrium. Fourier transforming with respect to
$(t-t')$  to the energy representation, one finds:
\begin{eqnarray}
                                        \label{Fourier}
\hskip -.5cm
 {\G}^{R(A)}(\epsilon)\!\!\!\!\!\! &=& (\epsilon-\omega_0\pm i0)^{-1} \, ;   \\
\hskip -.3cm{\G}^{K}(\epsilon) &=&  (1+2n(\omega_0))(-2\pi i)
\delta(\epsilon-\omega_0)=  (1+2n(\epsilon))(-2\pi i)
\delta(\epsilon-\omega_0)    \nonumber \, .
\end{eqnarray}
Therefore for the case of  thermal  equilibrium, one notices
therefore that
\begin{equation}
{\G}^{K}(\epsilon) =\mbox{coth}\,{\epsilon \over 2\,T}\, \left(
{\G}^{R}(\epsilon) - {\G}^{A}(\epsilon) \right)\, .
                                         \label{fdt}
\end{equation}
The last equation constitutes the statement of the {\em
fluctuation--dissipation theorem} (FDT). As it is shown below, the
FDT is a general property of  thermal equilibrium that is not
restricted to the toy example, considered here. It implies the
rigid relation between the response and correlation functions.

In general, it is convenient to parameterize the anti-Hermitian
(see Eq.~(\ref{conj})) Keldysh Green function by  a Hermitian
matrix $F=F^\dagger$, as:
\begin{equation}
{\G}^{K} =
 {\G}^{R}\circ F  - F \circ {\G}^{A}  \, ,
                                         \label{gfdt}
\end{equation}
where $F=F(t',t'')$ and the circular multiplication sign implies
integration over the intermediate time (matrix multiplication).
The Wigner transform (see below), $f(\tau,\epsilon)$, of the
matrix $F$ is referred to as the distribution function. In
thermal equilibrium: $f(\epsilon) = \mbox{coth}(\epsilon/2T)$.


\subsection{Keldysh action and causality}
\label{sec_24}

The Keldysh rotation from the ($\phi_+,\phi_-$) field components
to ($\phi_{cl},\phi_q$) considerably simplifies the structure of
the Green functions (cf. Eqs.~(\ref{corr}) and (\ref{Green})). It
is convenient, therefore, to write the action in terms of
$\phi_{cl},\phi_q$ as well. A simple way of doing it is to apply
the Keldysh rotation, Eq.~(\ref{rotation}), to the continuous
action, Eq.~(\ref{plusminus}), written in terms of
$\phi_+,\phi_-$. However, as was discussed above, the continuous
action, Eq.~(\ref{plusminus}), loses  the crucial information
about the off--diagonal blocks of the discrete matrix,
Eq.~(\ref{matrix}). To keep this information, one may invert the
matrix of Green functions, Eq.~(\ref{Green}), and use the result
as the correlator in the quadratic action. The inversion is
straightforward:
\begin{equation}
\hat G^{-1} = \left(\begin{array}{cc}
{\G}^K & {\G}^{R} \\
{\G}^{A} & 0
\end{array}\right)^{-1} = \left(\begin{array}{cc}
0   & [{\G}^{-1}]^{A}  \\
  \left[{\G}^{-1}\right]^{R}  & [{\G}^{-1}]^K
\end{array}\right)\, ,
                                                \label{Greeninverse}
\end{equation}
where the three components of the inverted Green function,
labelled in advance as $A,R$ and $K$, satisfy:
\begin{eqnarray}
                                            \label{inverted}
\left[G^{-1}\right]^{R(A)}\!\!\!\!\!\!& = &[G^{R(A)}]^{-1} =
i\partial_t-\omega_0 \pm i0\, ; \\
\left[G^{-1}\right]^K& = &-[G^R]^{-1}\circ G^K\circ
[G^A]^{-1}=[G^R]^{-1}\circ F -F \circ [G^A]^{-1}\, , \nonumber
\end{eqnarray}
where parameterization~(\ref{gfdt}) was employed in the last line.
It is easy to see that $[G^R]^{-1}$ and $[G^A]^{-1}$ are lower and
upper triangular matrices correspondingly, thus justifying their
superscripts. The continuous notations may create an impression
that $\left[G^{-1}\right]^K=(2i0)F$ and thus may be omitted. One
should remember, however, that this component is non--zero in the
discrete form and therefore it is important to acknowledge its
existence (even if it is not written explicitly).

Once the correlator Eqs.~(\ref{Greeninverse}), (\ref{inverted}) is
established, one may immediately write down the corresponding
action:
\begin{equation}
\hskip -.3cm S[\phi_{cl},\phi_q] =
\int\!\!\!\!\!\int\limits_{-\infty}^{\infty}\!\! dt dt'\,
(\bar\phi_{cl},\bar\phi_q)_t \left(\begin{array}{cc}
0   & [{\G}^{A}]^{-1}  \\
  \left[{\G}^{R}\right]^{-1}  & [{\G}^{-1}]^K
\end{array}\right)_{t,t'}
\left(\begin{array}{c} \phi_{cl} \\ \phi_q
\end{array}\right)_{t'} ,
                                                              \label{action}
\end{equation}
where it is acknowledged that the correlator is, in general, a
non--local function of time. The Green functions,
Eq.~(\ref{Green}), follow from the Gaussian integral with this
action, by construction. Notice that the presence of
$[{\G}^{-1}]^K=(2i0)F$ (with a positive imaginary part) is
absolutely necessary for the convergence of the corresponding
functional integral.

The structure of the Gaussian action given by Eq.~(\ref{action})
is very general and  encodes regularization of the functional
integral. Since the Keldysh component carries the information
about the density matrix, there is no further need  to recall the
discrete representation. The main features of this structure are:
\begin{itemize}
\item The $cl-cl$ component is zero.

This zero may be traced back to identity~(\ref{relation}). It has,
however,  a much simpler interpretation. It reflects the fact that
for a pure classical field configuration ($\phi_q=0$) the action
is zero. Indeed, in this case $\phi_+=\phi_-$ and the action on
the forward part of the contour is cancelled exactly by that on
the backward part. The very general  statement is, therefore, that
\begin{equation}
S[\phi_{cl},\phi_q=0] = 0\, .
                                                              \label{causality}
\end{equation}
Obviously Eq.~(\ref{causality}) is not restricted to a Gaussian
action.

\item The $cl-q$ and $q-cl$ components are mutually Hermitian conjugated
upper and lower (advanced and retarded) triangular matrices in the
time representation. This property is responsible for the
causality of the response functions as well as for protecting the
$cl-cl$ component from a perturbative renormalization (see below).

\item The $q-q$ component is an anti-Hermitian matrix (cf. Eq.~(\ref{conj}))
with a positive--definite imaginary spectrum. It is responsible
for the convergence of the functional integral. It also keeps the
information about the distribution function.

\end{itemize}

As was already mentioned, these three items are generic and
reproduce themselves in every order of  perturbation theory. For
the lack of a better terminology, we'll refer to them as the {\em
causality structure}.

\subsection{Free bosonic fields}

It is a straightforward matter to generalize the entire
construction to bosonic systems   with more than one state.
Suppose the states are labelled by an index $k$, that may be,
e.g., a momentum vector. Their energies are given by a function
$\omega_k$, for example $\omega_k=k^2/(2m)$, where $m$ is the mass
of the bosonic atoms. One introduces then a doublet of complex
fields (classical and quantum) for every state $k\,$:
$(\phi_{cl}(t;k), \phi_q(t;k))$ and writes down the action in the
form of Eq.~(\ref{action})  including a  summation over the index
$k$. Away from  equilibrium, the Keldysh component may be
non--diagonal in the index $k$: $F=F(t,t';k,k')$. The retarded
(advanced) component, on the other hand, has the  simple form
$[G^{R(A)}]^{-1}=i\partial_t -\omega_k$.

If $k$ is momentum, it is more instructive to Fourier transform to
 real space and to deal with
$(\phi_{cl}(t;r), \phi_q(t;r))$. Introducing a combined
time--space index $x=(t;r)$, one may write down for the action of
the free complex bosonic field (atoms):
\begin{equation}
\hskip -.3cm S_0 = \int\!\!\!\!\!\int\limits\!\! dx\, dx'\,
\big(\bar\phi_{cl},\bar\phi_q\big)_x \left(\begin{array}{cc}
0   & [{\G}^{A}]^{-1}  \\
  \left[{\G}^{R}\right]^{-1}  & [{\G}^{-1}]^K
\end{array}\right)_{x,x'}
\left(\begin{array}{c} \phi_{cl} \\ \phi_q
\end{array}\right)_{x'}  ,
                                                              \label{action1}
\end{equation}
where in the continuous notations
\begin{equation}
[G^{R(A)}]^{-1}(x,x')=\delta(x-x')\left( i\partial_{t'} +{1\over
2m} \nabla_{r'}^2  \right)\, ,
                                             \label{gradient}
\end{equation}
while in the discrete form it is  a lower (upper) triangular
matrix  in time (not in space). The $[{\G}^{-1}]^K$ component for
the free field is only the regularization factor, originating from
the (time) boundary terms. It is, in general, non--local in $x$
and $x'$, however, being a pure boundary term it is frequently
omitted. It is kept here as a reminder that the inversion, $\hat
G$, of the correlator matrix must posses the causality structure,
Eq.~(\ref{Green}).

In an analogous way, the action of free real bosons (phonons) is
(cf. Eq.~(\ref{Keldyshphonons1})):
\begin{equation}
\hskip -.3cm S_0 = \int\!\!\!\!\!\int\limits\!\! dx\, dx'\,
\big(\varphi_{cl},\varphi_q\big)_x \left(\begin{array}{cc}
0   & [{D}^{A}]^{-1}  \\
  \left[{D}^{R}\right]^{-1}  & [{D}^{-1}]^K
\end{array}\right)_{x,x'}
\left(\begin{array}{c} \varphi_{cl} \\ \varphi_q
\end{array}\right)_{x'} \, ,
                                                              \label{action2}
\end{equation}
where
\begin{equation}
[D^{R(A)}]^{-1}(x,x')=\delta(x-x')\big( -\partial_{t'}^{\,2} +
v_s^2\, \nabla_{r'}^{\,2} \big)\, ,
                                             \label{gradient2}
\end{equation}
in the continuous notations. In the discrete representations
$[D^{R(A)}]^{-1}$ are again the lower (upper) triangular matrices.
Here too the Keldysh component, $[{D}^{-1}]^K$, here too is just a
regularization, originating from the (time) boundary terms. It is
kept in Eq.~(\ref{action2})  to emphasize  the casuality structure
of the real boson Green function $\hat D(x,x')$, analogous to
Eq.~(\ref{Green}):
\begin{equation}
\hskip -.4cm \hat D (x,x')=\left(\begin{array}{cc}
{D}^K & {D}^{R} \\
{D}^{A} & 0
\end{array}\right)\, ;
\hskip .4cm
\begin{array}{l}
{D}^{R(A)}(\epsilon,k)=((\epsilon\pm i0)^2-v_s^2k^2)^{-1}\,; \\
{D}^{K} =D^R\circ F - F\circ D^A\,,
\end{array}
                                               \label{Greenreal}
\end{equation}
where $F=F(t,t';r,r')$ is a symmetric distribution function
matrix.

\vskip .5cm

\section{Collisions and kinetic equation}
\label{sec_2}

\subsection{Interactions}
\label{sec_31}

\begin{figure}[t]
\fbox{\vtop to4cm{\vss\hsize=.975\hsize \vglue 0cm
\hspace{0.01\hsize} \hskip -.5cm \epsfxsize=1\hsize
 \epsffile{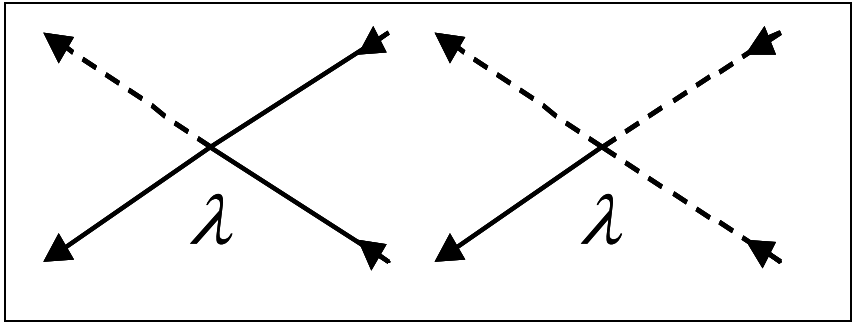}
\hspace{0.1\hsize} \vss }}
\label{fig_phi4}
 \caption{Graphic representation of the two interaction vertexes
 of the $|\phi|^4$ theory. There are also two conjugated vertexes
 with a reversed direction of all arrows.}
\end{figure}

The short range two--body collisions of bosonic atoms are
described by the local ``four--boson'' Hamiltonian:
$H_{int}=\lambda\sum_r a^{\dagger}_r a^{\dagger}_r a_r a_r$, where
index $r$ ``numerates'' spatial locations. The interaction
constant, $\lambda$, is related to a commonly used $s$--wave
scattering length, $a_s$, as $\lambda=4\pi a_s/m$
\cite{Leggett01}. The corresponding term in the continuous Keldysh
action takes the form:
\begin{equation}
S_{int} = -\lambda\!\int\!\!dr\!\!\int\limits_{\cal C}\!\! dt\,
(\bar\phi \phi)^2 =
-\lambda\!\int\!\!dr\!\!\!\int\limits_{-\infty}^\infty\!\! dt\,
\big[(\bar\phi_+ \phi_+)^2 - (\bar\phi_- \phi_-)^2\big]\, .
                                         \label{collisonaction}
\end{equation}
It is important to remember that there are no interactions in the
distant past, $t=-\infty$ (while they are present in the future,
$t=+\infty$). The interactions are supposed to be adiabatically
switched on and off on the forward and backward branches
correspondingly. That guarantees that the off--diagonal blocks of
the matrix, Eq.~(\ref{matrix}), remain intact.  Interactions
modify only those matrix elements of the evolution operator,
Eq.~(\ref{matrixelement}), that are away from $t=-\infty$. It is
also worth remembering that in the discrete time form the
$\bar\phi$ fields are taken  one time step $\delta_t$ {\em after}
the $\phi$ fields along the Keldysh contour ${\cal C}$. Therefore
the two terms on the r.h.s. of the last equation should be
understood as $\big(\bar\phi_+(t+\delta_t)\phi_+(t)\big)^2$ and
$\big(\bar\phi_-(t)\phi_-(t+\delta_t)\big)^2$ correspondingly.
Performing the Keldysh rotation, Eq.~(\ref{rotation}), one finds
\begin{equation}
S_{int}[\phi_{cl},\phi_q] = - \lambda\!
\int\limits_{-\infty}^{\infty}\!\!\! dt \left[\bar\phi_q
\bar\phi_{cl}( \phi_{cl}^2  + \phi_q^2 ) + c.c.\right]\, ,
                                                              \label{int}
\end{equation}
where $c.c.$ stands for the complex conjugate of the first term.
The collision action, Eq.~(\ref{int}), obviously satisfies the
causality condition, Eq.~(\ref{causality}). Diagrammatically the
action (\ref{int}) generates two types of vertexes depicted in
Fig.~\ref{fig_phi4} (as well as two complex conjugated vertexes,
obtained by reversing the direction of the arrows): one with three
classical fields (full lines) and one quantum field (dashed line)
and the other with one classical field and three quantum fields.

Let us demonstrate that  adding  the collision term to the action
does not violate the fundamental normalization, $Z=1$. To this end
one may expand $e^{iS_{int}}$ in powers of $\lambda$ and then
average term by term with the Gaussian action,
Eq.~(\ref{action1}). To show that the normalization, $Z=1$, is not
altered by the collisions, one needs to show that $\langle
S_{int}\rangle = \langle S_{int}^{\,2}\rangle = \ldots=0$.
Applying the Wick theorem, one finds for the terms that are linear
order in $\lambda$: $\langle\bar\phi_q \bar\phi_{cl}\phi_{cl}^2
+c.c.\rangle \sim \big[ G^R(t,t)+G^A(t,t)\big] G^K(t,t)=0$, and
$\langle \bar\phi_q \bar\phi_{cl} \phi_q^2+c.c\rangle=0$. The
first term vanishes due to identity (\ref{GRplusGA}), while the
second one vanishes because $\langle \phi_q\bar\phi_q\rangle=0$.
There are two families of terms that are second order in
$\lambda$. The first one is $\langle\bar\phi_q
\bar\phi_{cl}\phi_{cl}^2
\phi'_q\phi'_{cl}(\bar\phi'_{cl})^2\rangle \sim
G^R(t',t)G^A(t',t)[G^K(t,t')]^2$, while the  second is
$\langle\bar\phi_q \bar\phi_{cl}\phi_{cl}^2
\phi'_q\phi'_{cl}(\bar\phi'_{q})^2\rangle \sim
[G^R(t,t')]^2G^R(t',t)G^A(t',t)$, where
$\phi'_\alpha\equiv\phi_\alpha(t')$. Both of these terms are zero,
because $G^R(t',t)\sim\theta(t'-t)$, while $G^A(t',t)\sim
G^R(t,t')^*\sim\theta(t-t')$ and thus their product has no support
 \footnote{Strictly speaking, $G^R(t',t)$ and $G^A(t',t)$ are both
 simultaneously non--zero at the diagonal: $t=t'$. The
 contribution of the diagonal to the integrals is, however, $\sim
 \delta_t^2N\to 0$, when $N\to \infty$. }.
It is easy to see that, for exactly the same reasons, all higher
order terms vanish and thus the normalization is unmodified  (at
least in a perturbative expansion).

\begin{figure}[t]
\fbox{\vtop to5.5cm{\vss\hsize=.975\hsize \vglue 0cm
\hspace{0.01\hsize} \hskip -.5cm \epsfxsize=1\hsize
 \epsffile{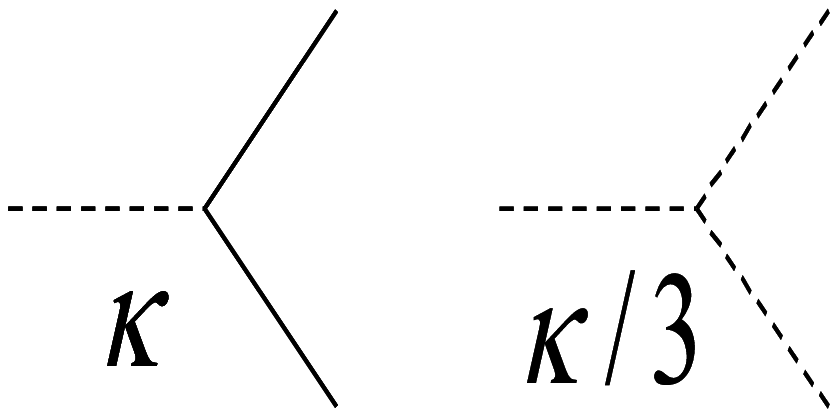}
\hspace{0.1\hsize} \vss }}
\label{fig_phi3}
 \caption{Graphic representation of the two interaction vertexes
 of the $\varphi^3$ theory. Notice the relative factor of one third between them. }
\end{figure}

As another example, consider the real boson field,
Eq.~(\ref{action2}), with the cubic nonlinearity:
\begin{equation}
\hskip -.5cm
 S_{int}\! = \!{\kappa\over
6}\!\int\!\!dr\!\!\int\limits_{\cal C}\!\! dt\, \varphi^3\! =\!
{\kappa\over 6} \!\int\!\!dr\!\!\!\int\limits_{-\infty}^\infty\!\!
dt\, \big[\varphi_+^3 - \varphi_-^3\big]\!=\!
\kappa\!\int\!\!dr\!\!\!\int\limits_{-\infty}^\infty\!\! dt\,
\big[\varphi_{cl}^2\varphi_q +{1\over 3}\, \varphi_q^3\big] \, .
                                         \label{phi3}
\end{equation}
The causality condition, Eq.~(\ref{causality}), is satisfied
again. Diagrammatically the cubic nonlinearity generates two types
of vertexes, Fig.~\ref{fig_phi3}: one with two classical fields
(full lines) and one quantum field (dashed line),  and the other
with three quantum fields. The former vortex carries the factor
$\kappa$, while the latter has a weight of $\kappa/3$. Notice that
for a real field the direction of the lines is not specified by
arrows.
\begin{table}[h]
\begin{tabular*}{\columnwidth}{@{\extracolsep{\fill}}@{}lcccccccc@{}}
\hline \hline
\end{tabular*}
{\bf Exercise:} Show that  there are no corrections of second
order in $\kappa$ to the partition function, $Z=1$. Check, that
the same is true for the higher orders, as well.
\par
\begin{tabular*}{\columnwidth}{@{\extracolsep{\fill}}@{}lcccccccc@{}}
\hline \hline
\end{tabular*}
\end{table}

\subsection{Saddle point equations}

Before developing  the perturbation theory further, one has to
discuss the saddle points of the action. According to
Eq.~(\ref{causality}), there are no terms in the action that have
zero power of both $\bar\phi_q$ and $\phi_q$. The same is
obviously true regarding $\delta S/ \delta \bar\phi_{cl}$ and
therefore one of the saddle point equations:
\begin{equation}
{\delta S\over \delta \bar\phi_{cl} } = 0\,
                                             \label{sp1}
\end{equation}
may  always be  solved by
\begin{equation}
\phi_q = 0 \, ,
                                             \label{cl1}
\end{equation}
irrespectively of what the  classical component, $\phi_{cl}$, is.
One may check that this is indeed the case for the action given by
Eqs.~(\ref{action1}) plus (\ref{int}). Under
condition~(\ref{cl1}) the second saddle point equation takes the
form:
\begin{equation}
 {\delta S\over \delta \bar\phi_{q} } =
\left(\left[G^R\right]^{-1} - \lambda\,|\phi_{cl}|^2\right)
\phi_{cl} =\left(i\partial_t + {\nabla^2_r\over 2m} - \lambda\,
|\phi_{cl}|^2\right) \phi_{cl}=0 \, ,
                                             \label{sp2}
\end{equation}
This is  the non--linear time--dependent (Gross--Pitaevskii)
equation \cite{Leggett01}, that uniquely determines the  classical
field configuration, provided some initial and boundary conditions
are specified.

The message is that among the possible solutions of the
saddle--point equations for the Keldysh action, there is always
one with a zero quantum component and with a classical component
that obeys the classical (non--linear) equations of motion. We
shall call such a saddle point -- {\em ``classical''}. Thanks to
Eqs.~(\ref{causality}) and (\ref{cl1}), the action at the
classical saddle--point field configurations is identically zero.
As was argued above, the perturbative expansion in small
fluctuations around the classical saddle point leads to a properly
normalized partition function, $Z=1$. This seemingly excludes the
possibility of having any other saddle points. Yet, this
conclusion is premature. The system may posses ``non--classical''
saddle points -- such that $\phi_q\neq 0$. Such saddle points do
not contribute to the partition function (and thus do not alter
the fundamental normalization, $Z=1$), however, they may
contribute to the correlation functions. In general, the action at
a {\em non--classical} saddle point is non--zero. Its contribution
is thus associated with exponentially small (or oscillatory)
terms. Examples include: tunnelling, thermal activation
(considered in the next chapter), Wigner-Dyson level statistics,
etc.

Let us develop now a systematic perturbative expansion in
deviations from the {\em classical} saddle point. As was discussed
above, it does not bring any new information about the partition
function. It does, however, provide  information about the Green
functions (and thus various observables). Most notably, it
generates the kinetic equation for the distribution function. To
simplify  further consideration, let us assume that $\phi_{cl}=0$
is the proper solution of the classical saddle--point equation
(\ref{sp2}) (i.e. there is no Bose condensate).

\subsection{Dyson equation}
\label{sec_33}

The goal is to calculate the {\em dressed} Green function, defined
as:
\begin{equation}
{\bf  G^{\alpha\beta}}(t,t') = -i\!\! \int\!\! {\cal D} \bar\phi
\phi\,\, e^{\, i(S_0+S_{int}) }\,\,
\phi_\alpha(t)\bar\phi_\beta(t')\, ,
                                          \label{dressed}
\end{equation}
where $\alpha,\beta=(cl,q)$ and the action is given by
Eqs.~(\ref{action1}) and (\ref{int}) (or for real bosons:
Eqs.~(\ref{action2}) and (\ref{phi3}), with $\phi\to \varphi$). To
this end one may expand the exponent in deviations from the
classical saddle point: $\phi_q\equiv 0$ and (in the simplest
case) $\phi_{cl}=0$. The functional integration with the remaining
Gaussian action is then performed using the Wick theorem. This
leads to the standard diagrammatic series. Combining all
one--particle irreducible diagrams into the self--energy matrix
$\hat\Sigma$, one obtains:
\begin{equation}
{\bf \hat G} = \hat G +  \hat G\circ\hat\Sigma\circ\hat G + \hat
G\circ\hat\Sigma\circ\hat G\circ\hat\Sigma\circ\hat G +\ldots=
\hat G\circ\left(\hat 1+\hat\Sigma\circ{\bf \hat G}\right)  \, ,
                                          \label{expansion}
\end{equation}
where $\hat G$ is given by Eq.~(\ref{Green}) and the circular
multiplication sign implies integrations over intermediate times
and coordinates as well as a $2\times 2$  matrix multiplication.
The only difference compared with the text--book \cite{Negele}
diagrammatic expansion is the presence of the $2\times 2$  Keldysh
matrix structure. The fact that the series is arranged as a
sequence of matrix products is of no surprise. Indeed, the Keldysh
index, $\alpha=(cl,q)$, is just one more index in addition to
time, space, spin, etc. Therefore, as with any other index, there
is a summation (integration) over all of its intermediate values,
hence the matrix multiplication. The concrete  form of the
self--energy matrix, $\hat \Sigma$, is of course specific to the
Keldysh technique and is discussed below in some details.

Multiplying both sides of Eq.~(\ref{expansion}) by $\hat G^{-1}$
from the left, one obtains the Dyson equation for the exact
dressed Green function, ${\bf \hat G}$:
\begin{equation}
\left(\hat G^{-1} -\hat \Sigma\right)\circ {\bf \hat G}=\hat 1\, ,
                                               \label{Dyson}
\end{equation}
where $\hat 1$ is the unit matrix. The very non--trivial feature
of the Keldysh technique is that the self energy matrix,
$\hat\Sigma$, possesses the same causality structure as $\hat
G^{-1}$, Eq.~(\ref{Greeninverse}):
\begin{equation}
\hat\Sigma=\left(\begin{array}{cc}
0   & \Sigma^{A}  \\
  \Sigma^{R}  & \Sigma^K
\end{array}\right)\, ,
                                                \label{Sigma}
\end{equation}
where $\Sigma^{R(A)}$ are lower (upper) triangular matrices in the
time domain, while $\Sigma^K$ is an anti-Hermitian matrix. This
fact will be demonstrated below. Since both $\hat G^{-1}$ and
$\hat \Sigma$ have the same structure, one concludes that the
dressed Green function, ${\bf \hat G}$, also possesses the
causality structure, like Eq.~(\ref{Green}). As a result, the
Dyson equation acquires the form:
\begin{equation}
\left(\begin{array}{cc}
0   &[G^A]^{-1}- \Sigma^{A}  \\
  \left[G^R\right]^{-1}- \Sigma^{R}  & -\Sigma^K
\end{array}\right)\circ
\left(\begin{array}{cc}
{\bf G^K}   & {\bf G^R}  \\
  {\bf G^A}  & 0
\end{array}\right) =\hat 1\, ,
                                                \label{Dysonmatrix}
\end{equation}
where one took into account that $[G^{-1}]^K$ is a pure
regularization ($\sim i0F$) and thus may be omitted in the
presence of a non--zero $\Sigma^K$. Employing the specific form of
$[G^{R(A)}]^{-1}$, Eq.~(\ref{gradient}), one  obtains for the
retarded (advanced) components:
\begin{equation}
\left( i\partial_t + {1\over 2m}\nabla_r^2 \right) {\bf G^{R(A)}}
= \Sigma^{R(A)}\circ {\bf G^{R(A)}}\, .
                                            \label{DysonRA}
\end{equation}
Provided the self--energy component $\Sigma^{R(A)}$ is known (in
some approximation), Eq.~(\ref{DysonRA}) constitutes a closed
equation for the retarded (advanced) component of the dressed
Green function. The latter carries the information about the
spectrum of the interacting system.

To write down the equation for the Keldysh component, it is
convenient to parameterize it as ${\bf G^K}={\bf G^R}\circ {\bf
F}- {\bf F}\circ {\bf G^A}$, where ${\bf F}$ is a Hermitian matrix
in the time domain. The equation for the Keldysh component then
takes the form: $([G^R]^{-1} -\Sigma^R)\circ ({\bf G^R}\circ {\bf
F}-{\bf F} \circ {\bf G^A})=\Sigma^K\circ{\bf G^A}$. Multiplying
it from the right by $([G^A]^{-1}-\Sigma^A)$ and employing
Eq.~(\ref{DysonRA}), one finally finds:
\begin{equation}
\left[{\bf F}, \left( i\partial_t + {1\over 2m}\nabla_r^2 \right)
\right]_- =\, \Sigma^K-\left(\Sigma^{R}\circ {\bf F} - {\bf F}
\circ\Sigma^A\right)\, ,
                                            \label{DysonF}
\end{equation}
where the symbol $[\,\,,\,]_-\,$ stands for the commutator. This
equation is the quantum kinetic equation for the distribution
matrix ${\bf F}$. Its l.h.s. is called the {\em kinetic} term,
while the r.h.s. is the {\em collision integral} (up to a factor).
As is shown below, $\Sigma^K$ has the meaning of an ``incoming''
term, while $\Sigma^{R}\circ {\bf F} - {\bf F} \circ\Sigma^A$ is
an ``outgoing'' term. In equilibrium  these two channels cancel
each other (the kinetic term vanishes) and the self-energy has the
same structure as the Green function: $\Sigma^K=\Sigma^{R}\circ
{\bf F} - {\bf F} \circ\Sigma^A$. This is not the case, however,
away from the equilibrium.

\subsection{Self-energy}
\label{sec_34}

Let us demonstrate in the case of one specific example, that the
self-energy matrix, $\hat \Sigma$, indeed possesses the causality
structure, Eq.~(\ref{Sigma}). To this end, we consider the real
boson action, Eq.~(\ref{action2}), with the $\kappa\varphi^3$
nonlinearity, Eq.~(\ref{phi3}), and perform  the calculations up
to the second order in the parameter, $\kappa$. Employing the two
vertexes of Fig.~\ref{fig_phi3} one finds that:

\begin{figure}[t]
\fbox{\vtop to6cm{\vss\hsize=.975\hsize \vglue 0cm
\hspace{0.01\hsize} \hskip -.5cm \epsfxsize=1\hsize
 \epsffile{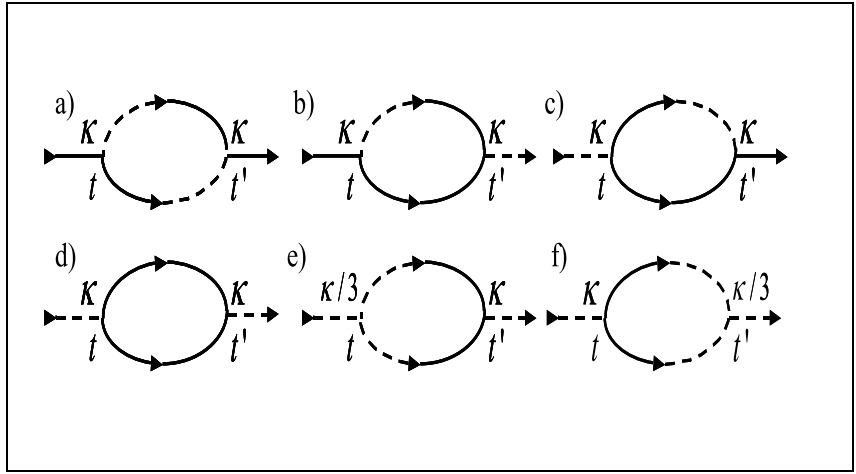}
\hspace{0.1\hsize} \vss }}
\label{fig_sigma}
 \caption{Self-energy diagrams for the $\varphi^3$ theory.  }
\end{figure}

{\em the} $cl-cl$ {\em component}  is given by the single diagram,
depicted in Fig.~\ref{fig_sigma}a. The corresponding analytic
expression is $\Sigma^{cl-cl}\!(t,t')=4i \kappa^2
D^R\!(t,t')D^A\!(t,t')\!=\!0$. Indeed, the product
$D^R(t,t')D^A(t,t')$ has no support (see, however, the footnote in
section \ref{sec_31}).

{\em the cl-q (advanced) component} is given by the single
diagram, Fig.~\ref{fig_sigma}b. The corresponding expression is:
\begin{equation}
\Sigma^A(t,t') = 4i \kappa^2 D^A(t,t')D^K(t,t')\,.
                                                  \label{SigmaA}
\end{equation}
Since $\Sigma^A(t,t')\sim D^A(t,t')\sim\theta(t'-t)$, it is,
indeed, an advanced (upper triangular) matrix. There is a
combinatoric factor of $4$, associated with the diagram (4 ways of
choosing  external legs $\times$ 2 internal permutations $\times$
1/(2!) for having two identical vertexes).

{\em the q-cl (retarded) component} is given by the  diagram of
Fig.~\ref{fig_sigma}c:
\begin{equation}
\Sigma^R(t,t') = 4i\kappa^2 D^R(t,t')D^K(t,t')\,,
                                                  \label{SigmaR}
\end{equation}
that could be obtained, of course, by the Hermitian conjugation of
Eq.~(\ref{SigmaA}) with the help of Eq.~(\ref{conj}):
$\Sigma^R=\left[\Sigma^A\right]^\dagger$. Since
$\Sigma^R(t,t')\sim D^R(t,t')\sim\theta(t-t')$, it is, indeed, a
retarded (lower triangular) matrix.

{\em the q-q (Keldysh) component} is given by the three diagrams,
Fig.~\ref{fig_sigma}d--f. The corresponding expressions are:
\begin{eqnarray}
\hskip -.4cm
 \Sigma^K(t,t')\!\!\! &=&\!\!\! 2i\kappa^2\!
\left[D^K(t,t')\right]^2\! +\! 6i\!\left({\kappa\over
3}\right)\!\kappa \left[D^A(t,t')\right]^2\! +\!
6i\kappa\left({\kappa\over 3}\right)\! \left[D^R(t,t')\right]^2
 \nonumber \\
&=&\!\!\! 2i\kappa^2\! \left( \left[D^K(t,t')\right]^2 +
\left[D^R(t,t') - D^A(t,t')\right]^2\right) \,,
                                                  \label{SigmaK}
\end{eqnarray}
where the combinatoric factors are: 2 for diagram d and 6 for e
and f. In the last equality, the fact that $G^R(t,t')G^A(t,t')=0$,
due to the absence of support in the time domain, has been used
again. Employing Eq.~(\ref{conj}), one finds
$\Sigma^K=-\left[\Sigma^K\right]^\dagger$. This completes the
proof of the statement that $\hat \Sigma$ possesses the same
structure as $\hat D^{-1}$. One may check that the statement holds
in  higher orders as well. In Eqs.~(\ref{SigmaA})--(\ref{SigmaK})
one has omitted the spatial coordinates, that may be restored in
an obvious way.

\begin{table}[h]
\begin{tabular*}{\columnwidth}{@{\extracolsep{\fill}}@{}lcccccccc@{}}
\hline \hline
\end{tabular*}
{\bf Exercise:} Calculate the self--energy matrix for the
$|\phi|^4$ theory to the second order in $\lambda$. Show that it
possesses  the causality structure.
\par
\begin{tabular*}{\columnwidth}{@{\extracolsep{\fill}}@{}lcccccccc@{}}
\hline \hline
\end{tabular*}
\end{table}

\subsection{Kinetic term}
\label{sec_35}

\begin{table}[b]
\begin{tabular*}{\columnwidth}{@{\extracolsep{\fill}}@{}lcccccccc@{}}
\hline \hline
\end{tabular*}
{\bf Reminder:} The Wigner transform of a matrix $A(r,r')$ is
defined as
$$
a(\rho,k)\equiv\int\!\!d r_1  \,  A\left(\rho+{r_1\over 2},
\rho-{r_1\over 2}\right)\,e^{ikr_1}\,.
                                             \label{Wignertransform}
$$
One may show  that the Wigner transform of the matrix $C=A\circ B$
is equal to:
$$
c(\rho,k)=\!\!\int\!\!\!\! \int\!\! dr_1 dr_2\!\! \int\!\!\!\!
\int\!\! {dk_1 dk_2\over (2\pi)^{2d}}\,\,
a\!\left(\!\rho+{r_1\over 2},k+k_1\!\right)
b\!\left(\!\rho+{r_2\over 2},k+k_2\!\right)\,
e^{i(k_1r_2-k_2r_1)}.
$$
Expanding the functions under the integrals in $k_i$ and $r_i$,
one finds:
$$
c(\rho,k)= {a(\rho,k)\,b(\rho,k)}+(2i)^{-1}\big(\nabla_\rho
a\nabla_k b - \nabla_k a\nabla_\rho b\big) +\ldots\,\, .
$$
%
\end{table}

To make  further progress in the discussion of the kinetic
equation it is  convenient to  perform the Wigner transformation
(WT). The WT of a distribution function matrix, ${\bf
F}(t,t';r,r')$, is a function: ${\bf f}(\tau,\epsilon;\rho,k)$,
where $\tau$ and $\rho$ are the ``center of mass'' time and
coordinate correspondingly. According to definition~(\ref{gfdt}),
the ${\bf F}$ matrix appears in a product with $G^R-G^A$ (or
$D^R-D^A$). Since the latter is a sharply peaked function at
$\epsilon=\omega_k$ (cf. Eq.~(\ref{Fourier}) for free particles,
while for  interacting systems this is the condition for having
well-defined quasi--particles), one frequently writes ${\bf
f}(\tau,\rho,k)$, understanding that $\epsilon=\omega_k$.

To rewrite the kinetic term (the l.h.s. of Eq.~(\ref{DysonF})) in
the Wigner representation, one notices that the WT of
$i\partial_t$ is $\epsilon$, while the WT of $\nabla^2_r$ is
$-k^2$. Then e.g. $[{\bf F},\nabla_r^2]_-\to [k^2,{\bf f}]_-
+i\nabla_k k^2 \nabla_\rho {\bf f}=2ik\nabla_\rho {\bf f}$, where
the commutator vanishes, since WT's commute.  In a similar way:
$[{\bf F},i\partial_t]_-\to -i\partial_\tau {\bf f}$. If there is
a scalar potential $V(r)a^\dagger_r a_r$ in the Hamiltonian, it
translates into the term
$-V(\bar\phi_{cl}\phi_q+\bar\phi_q\phi_{cl})$ in the action and
thus $-V(r)$ is added to $[G^{R(A)}]^{-1}$. This, in turn, brings
the term $-[{\bf F},V]_-$ to the l.h.s. of the Dyson equation
(\ref{DysonF}), or after the WT: $iE\nabla_k {\bf f}$, where
$E\equiv -\nabla_\rho V$ is the electric field. As a result, the
WT of the Dyson equation (\ref{Dyson}) takes the form:
\begin{equation}
\Big(\partial_\tau -v_k\nabla_\rho - E\nabla_k\Big){\bf
f}(\tau,\rho,k) = I_{col}[{\bf f}]\, ,
                                      \label{kinetic}
\end{equation}
where $v_k\equiv k/m$ and $I_{col}[{\bf f}]$ is the WT of the
r.h.s. of Eq.~(\ref{DysonF}) (times $i$). This is the kinetic
equation for the distribution function.

For  real bosons with the dispersion relation $\epsilon=\omega_k$,
the kinetic term is (cf. Eq.~(\ref{gradient2})): $[\epsilon^2 -
\omega_k^2,{\bf F}]_-\to 2i\big(\epsilon\,
\partial_\tau - \omega_k(\nabla_k\omega_k)\nabla_\rho\big){\bf
f}= 2i\epsilon \big( \partial_\tau - v_k\nabla_\rho\big){\bf f}$,
where  $v_k\equiv \nabla_k\omega_k$ is the group velocity. As a
result, the kinetic equation takes the form: $\big(
\partial_\tau - v_k\,\nabla_\rho\big)\,{\bf f}(\tau,\rho,k) =
I_{col}[{\bf f}]$, where the collision integral $I_{col}[{\bf f}]$
is the WT of the r.h.s. of Eq.~(\ref{DysonF}), divided by
$-2i\epsilon$.

\subsection{Collision integral}
\label{sec_36}

Let us discuss the collision integral, using the $\varphi^3$
theory calculations of section \ref{sec_34} as an example. To
shorten the algebra, let us consider  a system that is spatially
uniform and isotropic in momentum space. One, thus,  focuses on
the energy relaxation only. In this case the distribution function
is ${\bf f}(\tau,\rho,k)={\bf f}(\tau,\omega_k)={\bf
f}(\tau,\epsilon)$, where the dependence on the modulus of the
momentum is substituted by the $\omega_k=\epsilon$ argument.
Employing Eqs.~(\ref{SigmaA})--(\ref{SigmaK}), one finds for the
WT of the r.h.s. of Eq.~(\ref{DysonF})
\footnote{Only products of WT's are retained, while all the
gradient terms are neglected, in particular ${\bf D^K}\to {\bf
f}\,({\bf d^R}-{\bf d^A})$. The energy--momentum representation is
used, instead of the time--space representation  as in
Eqs.~(\ref{SigmaA})--(\ref{SigmaK}), and in the equation for
$\Sigma^{R}\circ {\bf F} -{\bf F} \circ\Sigma^A$ one performs a
symmetrization between the $\omega$ and $\epsilon-\omega$
arguments. }
:
\begin{eqnarray}
                                       \label{wignersigma}
                                       \hskip -.6cm
&& \Sigma^{R}\circ {\bf F} -{\bf F} \circ\Sigma^A \to -2i\,{\bf
f}(\tau,\epsilon)\!\!\int\!\! d\omega\,   M(\tau,\epsilon,\omega)
\Big({\bf
f}(\tau,\epsilon-\omega)+{\bf f}(\tau,\omega)\Big)\,; \nonumber\\
 \hskip -.6cm
&&\Sigma^{K}\to  -2i\!\!\int\!\! d\omega\, M(\tau,\epsilon,\omega)
\Big({\bf f}(\tau,\epsilon-\omega){\bf f}(\tau,\omega)+1\Big)\,,
\end{eqnarray}
where the square of the  transition matrix element  is given by:
\begin{equation}
M(\tau,\epsilon,\omega) = 2\pi\kappa^2\!\sum\limits_q\, {\bf
\Delta_{d} }(\tau,\epsilon-\omega;k-q)\,  {\bf
\Delta_{d}}(\tau,\omega;q)\, .
                                  \label{transitionprob}
\end{equation}
Here ${\bf \Delta_{d} }\equiv i({\bf d^R} - {\bf d^A})/(2\pi)$ and
${\bf d^{R(A)}}(\tau,\epsilon;k)$ is the WT of the retarded
(advanced) Green function. One has substituted the dressed Green
functions into Eqs.~(\ref{SigmaA})--(\ref{SigmaK}) instead of the
bare ones to perform a partial resummation of the diagrammatic
series. (This trick is sometimes called the {\em self--consistent
Born approximation}. It still neglects the vertex corrections.)
Assuming the existence of  well defined quasi--particles at all
times, one may regard ${\bf \Delta_{d}}(\tau,\epsilon,k)$ as a
sharply peaked function at $\epsilon=\omega_k$. In this case
Eq.~(\ref{transitionprob}) simply reflects the fact that an
initial particle with $\epsilon=\omega_k$ decays into two real (on
mass-shell) particles with energies $\omega=\omega_q$ and
$\epsilon-\omega=\omega_{k-q}$. As a result, one finally obtains
for the kinetic equation:
\begin{equation}
{\partial {\bf f}(\epsilon)\over \partial\tau} =  \!
\int\!\!{d\omega}\, {M(\epsilon,\omega)\over \epsilon}\,\, \Big[
{\bf f}(\epsilon-\omega){\bf f}(\omega)+1 -{\bf
f}(\epsilon)\big({\bf f}(\epsilon-\omega)+{\bf
f}(\omega)\big)\Big]\, ,
                                    \label{kineticeq}
\end{equation}
where the time arguments are suppressed for brevity.

Due to the identity: $\cth (a-b)\,\cth( b)+1 = \cth (a) \big(\cth
(a-b) +\cth (b)\big)$, the collision integral is identically
nullified by
\begin{equation}
{\bf f}(\epsilon)=\cth\, {\epsilon \over 2\,T}\, \, .
                                             \label{cth}
\end{equation}
where $T$ is a temperature. This is the thermal equilibrium
distribution function. According to the kinetic equation
(\ref{kineticeq}), it is stable for any temperature (the latter is
determined either by an external reservoir, or, for a closed
system, from the total energy conservation). Since the equilibrium
distribution obviously nullifies the kinetic term, according to
Eq.~(\ref{DysonF}) the {\em exact} self--energy satisfies
$\Sigma^K=\cth(\epsilon/(2T))(\Sigma^R-\Sigma^A)$. Since also the
bare Green functions obey the same relation, Eq.~(\ref{fdt}), one
concludes that in  thermal equilibrium the {\em exact} dressed
Green function satisfies:
\begin{equation}
{\bf D^K}= \cth\, {\epsilon \over 2\,T}\, \big({\bf D^R} - {\bf
D^A}\big)\,.
                                       \label{FDT}
\end{equation}
This is the statement of the {\em fluctuation--dissipation
theorem} (FDT). Its consequence is that in  equilibrium the
Keldysh component does not contain any additional information with
respect to the retarded one. Therefore, the Keldysh technique may
be, in principle, substituted by a more compact construction --
the Matsubara method. The latter does not work, of course, away
from  equilibrium.

Returning to the kinetic equation (\ref{kineticeq}), one may
identify ``in'' and ``out'' terms in the collision integral. Most
clearly it is done by writing the collision integral in terms of
the occupation numbers ${\bf n}_k$, defined as ${\bf f}=1+2\,{\bf
n}$. The expression in the square brackets on the r.h.s. of
Eq.~(\ref{kineticeq}) takes the form: $4\left[{\bf
n}_{\epsilon-\omega}{\bf n}_\omega- {\bf n}_\epsilon({\bf
n}_{\epsilon-\omega}+{\bf n}_\omega+1)\right]$. The first term:
${\bf n}_{\epsilon-\omega}{\bf n}_\omega$, gives a probability
that  a particle with energy $\epsilon-\omega$ absorbs a particle
with energy $\omega$ to populate a state with energy $\epsilon$ --
this is the ``in'' term of the collision integral. It may be
traced back to the $\Sigma^K$ part of the self-energy. The second
term: $ -{\bf n}_\epsilon({\bf n}_{\epsilon-\omega}+{\bf
n}_\omega+1)$, says that a state with energy $\epsilon$ may be
depopulated either by stimulated emission of particles with
energies $\epsilon-\omega$ and $\omega$, or by spontaneous
emission (unity). This is the ``out'' term, that may be traced
back to the $\Sigma^{R(A)}$ contributions.

Finally, let us discuss the approximations involved in the Wigner
transformations. Although Eq.~(\ref{DysonF}) is formally exact, it
is very difficult to extract any useful information from it.
Therefore, passing to an approximate, but much more tractable,
form like Eqs.~(\ref{kinetic}) or (\ref{kineticeq}) is highly
desirable. In doing it, one has to employ the approximate form of
the WT. Indeed, a formally infinite series in
$\nabla_k\nabla_\rho$ operators is truncated, usually by the first
non--vanishing term. This is a justified procedure as long as
$\delta k\,\delta\rho\gg 1$, where $\delta k$ is a characteristic
microscopic scale of the momentum dependence of  ${\bf f}$, while
$\delta \rho$ is a characteristic scale of its spatial variations.
One may ask if there is a similar requirement in the time domain:
$\delta\epsilon\, \delta \tau\gg 1$, with $\delta\epsilon$ and
$\delta \tau$ being the characteristic energy and the time scale
of ${\bf f}$, correspondingly? Such a requirement is very
demanding, since typically $\delta\epsilon \approx T$ and at low
temperature it would allow to treat only very slow processes: with
$\delta\tau \gg 1/T$. Fortunately, this is not the case. Because
of the peaked structure of ${\bf \Delta_d}(\epsilon,k)$, the
energy argument $\epsilon$ is locked to $\omega_k$ and does not
have its own dynamics as long as the peak is sharp. The actual
criterion is therefore that $\delta\epsilon$ is much larger than
the  width of the peak in ${\bf \Delta_d}(\epsilon,k)$. The latter
is, by definition, the quasi--particle life--time, $\tau_{qp}\,$,
and therefore the condition is $\tau_{qp}\gg 1/T$. This condition
is indeed satisfied by many systems with the interactions that are
not too strong.

\vskip .5cm

\section{Particle in contact with an environment}
\label{sec_4}

\subsection{Quantum dissipative action}
\label{sec_41}

Consider a particle with the coordinate $\Phi(t)$, living in a
potential $U(\Phi)$ and attached to a harmonic string
$\varphi(t;x)$. The particle may represent a collective degree of
freedom, such as the phase of a Josephson junction or the charge
on a quantum dot. On the other hand, the string serves to model a
dissipative environment. The advantage of the one--dimensional
string is that it is the simplest continuum system, having a
constant density of states. Due to this property it mimics, for
example, interactions with a Fermi sea. A continuous reservoir
with a constant density of states at small energies is sometimes
called an ``Ohmic''  environment (or bath). The environment is
supposed to be in  thermal equilibrium.

The Keldysh action of such a system is given by the three terms
(cf. Eqs.~(\ref{KeldyshFeynman}) and (\ref{action2})):
\begin{eqnarray}
                                          \label{particlestring}
&& S_p[\hat \Phi] = \int\limits_{-\infty}^\infty\!\!
dt\left[-2\,\Phi_q{d^{\,2} \Phi_{cl}\over dt^2} -
U\left(\Phi_{cl}+\Phi_q\right) + U(\Phi_{cl}- \Phi_q) \right]\, ;
\nonumber \\
&& S_{str}[\hat\varphi]= \int\limits_{-\infty}^\infty\!\!
dt\int\!\! dx \, \hat \varphi^T \hat D^{-1} \,\hat \varphi\, ;\\
&& S_{int}[\hat\Phi,\hat\varphi]=
2\sqrt{\gamma}\int\limits_{-\infty}^\infty\!\!dt\,\,
\left.\hat\Phi^T(t)\,\hat\sigma_1\,\nabla_x\hat\varphi(t,x)\right|_{x=0}\,
, \nonumber
\end{eqnarray}
where we have introduced vectors of classical and quantum
components, e.g. $\hat\Phi^T\equiv (\Phi_{cl},\Phi_q)$ and the
string correlator, $\hat D^{-1}$, is the same as in
Eqs.~(\ref{action2}), (\ref{gradient2}). The interaction term
between the particle and the string is taken to be the local
product of the particle coordinate and the string stress  at $x=0$
(so the force on the particle is proportional to the local stress
of the string). In the time domain the interaction is
instantaneous, $\Phi(t)\nabla_x\varphi(t,x)|_{x=0}\to
\Phi_+\nabla\varphi_+-\Phi_-\nabla\varphi_-$ on the Keldysh
contour. Transforming to the classical--quantum notations  leads
to: $2(\Phi_{cl}\nabla\varphi_q+\Phi_q\nabla\varphi_{cl})$, that
satisfies the causality condition, Eq.~(\ref{causality}). In the
matrix notations it takes the form of the last line of
Eq.~(\ref{particlestring}), where $\hat\sigma_1$ is the standard
Pauli matrix. The interaction constant is $\sqrt{\gamma}$.

One may now integrate out the degrees of freedom of the Gaussian
string to reduce the problem to the particle coordinate only.
According to the standard rules of  Gaussian integration (see.
\ref{app_Gaussian}), this leads to the so--called dissipative
action for the particle:
\begin{equation}
\hskip -.4cm
S_{diss}=-\gamma\int\!\!\!\!\!\int\limits_{-\infty}^{\infty}\!\!
dt dt' \, \hat\Phi^T(t) \underbrace{\left. \hat \sigma_1^T
\nabla_x\nabla_{x'}\hat D(t-t';x-x')\right|_{x=x'=0}\,
\hat\sigma_1}_{-\hat L^{-1}(t-t')} \hat\Phi(t') \, .
                                                \label{dissip}
\end{equation}
The straightforward matrix multiplication shows that the
dissipative correlator $\hat L^{-1}$ possesses the standard
causality structure of the inverse Green function, e.g.
Eq.~(\ref{Greeninverse}). Fourier transforming  its retarded
(advanced) components, one finds:
\begin{equation}
\left[L^{R(A)}(\epsilon)\right]^{-1} =
-\sum\limits_k\frac{k^2}{(\epsilon\pm i0)^2-k^2} = \pm\, {i\over
2}\,\epsilon + \mbox{const}\, ,
                                          \label{LRA}
\end{equation}
where we put $v_s=1$ for brevity. The $\epsilon$--independent
constant (same for $R$ and $A$ components) may be absorbed into
the redefinition of the harmonic part of the potential
$U(\Phi)=\mbox{const}\, \Phi^2+\ldots$ and, thus, may be omitted.
In equilibrium the Keldysh component of the correlator is set by
the FDT:
\begin{equation}
\left[L^{-1}\right]^{K}(\epsilon) = \cth{\epsilon\over 2\,T}\left(
\left[L^{R}\right]^{-1}-\left[L^{A}\right]^{-1} \right) =
i\epsilon\,\cth{\epsilon\over 2\,T}\, .
                                          \label{LK}
\end{equation}
It is an anti--Hermitian operator with a positive--definite
imaginary part, rendering  convergence of the functional integral
over $\Phi$.

In the time representation the retarded (advanced) component of
the correlator takes a simple local form:
$\left[L^{R(A)}\right]^{-1}= \mp {1\over
2}\,\delta(t-t')\,\partial_{t'}$. On the other hand, at low
temperatures the Keldysh component is a non--local function, that
may be  found by the inverse Fourier transform of Eq.~(\ref{LK}):
\begin{equation}
\left[L^{-1}\right]^{K}\!(t-t') =  \frac{i\pi T^2}{\sinh^2(\pi
T(t-t'))}\stackrel{T\to \infty}{\longrightarrow}\, i2T\delta(t-t')
\, .
                                          \label{LKt}
\end{equation}
Finally, for the Keldysh action of the particle connected to a
string, one obtains:
\begin{eqnarray}
                               \label{dissipaction}
                               \hskip -.3cm
S[\hat \Phi]\!\!\!\! &=&\!\!\!\!\!
\int\limits_{-\infty}^\infty\!\! dt\left[-2\,\Phi_q\left({d^{\,2}
\Phi_{cl}\over dt^2} +{\gamma\over 2}{d \Phi_{cl}\over dt}\right)
\!-\! U\left(\Phi_{cl}+\Phi_q\right)\! +\! U(\Phi_{cl}- \Phi_q)
\right]
\nonumber \\
 &+& i\gamma\int\!\!\!\!\!\int\limits_{-\infty}^\infty\!\! dt\, dt'\,
 \Phi_q(t)\, \frac{\pi T^2}{\sinh^2(\pi
T(t-t'))}\, \Phi_q(t')\, .
\end{eqnarray}
This action satisfies all the causality criterions  listed in
section \ref{sec_24}. Notice, that in the present case the Keldysh
($q-q$) component is not just a regularization factor, but rather
a quantum fluctuations damping term, originating from the coupling
to the string. The other manifestation of the string is the
presence of the friction term, $\sim \gamma\partial_t$ in the $R$
and the $A$ components. In equilibrium the friction coefficient
and fluctuations amplitude  are rigidly connected  by the FDT. The
quantum dissipative action, Eq.~(\ref{dissipaction}), is a
convenient  playground to demonstrate various approximations and
connections to other approaches.

\subsection{Saddle--point equation} \label{sec_42}

The {\em classical} saddle point equation (the one that takes
$\Phi_q(t)=0$) has the form:
\begin{equation}
\left. -{1\over 2}\frac{\delta S[\hat\Phi]}{\delta
\Phi_q}\right|_{\Phi_q=0} = {d^{\,2} \Phi_{cl}\over dt^2}
+{\gamma\over 2}{d \Phi_{cl}\over dt} + {\partial
U(\Phi_{cl})\over \partial \Phi_{cl} } =0 \, .
                                          \label{fricNewton}
\end{equation}
This is the deterministic classical equation of motion. In the
present case it happens to be the Newton equation with the viscous
force: $-(\gamma/2) \dot \Phi_{cl}$. This approximation neglects
both {\em quantum} and {\em thermal} fluctuations.

\subsection{Classical limit}
\label{sec_43}

One may keep the thermal fluctuations, while completely neglecting
the quantum ones. To this end it is convenient to restore the
Planck constant in the action (\ref{dissipaction}) and then take
the limit $\hbar\to 0$. For dimensional reasons, the factor
$\hbar^{-1}$ should stand in front of the action. To keep the part
of the action responsible  for the classical equation of motion
(\ref{fricNewton}) free from the Planck constant  it is convenient
to  rescale the  variables as: $\Phi_q \to \hbar \Phi_q$. Finally,
to have temperature in energy units, one needs to substitute $T\to
T/\hbar$ in the last term of Eq.~(\ref{dissipaction}). The limit
$\hbar\to 0$ is now straightforward: (i) one has to  expand
$U(\Phi_{cl}\pm \hbar\Phi_q)$ to the first order in $\hbar\Phi_q$
and neglect all higher order terms; (ii) in the last term of
Eq.~(\ref{dissipaction}) the $\hbar\to 0$ limit is equivalent to
the $T\to \infty$ limit, see Eq.~(\ref{LKt}). As a result,   the
classical limit of the dissipative action is:
\begin{equation}
                               \label{dissipclassical}
S[\hat \Phi] = 2\!\!\int\limits_{-\infty}^\infty\!\! dt
 \left[ -\Phi_q \left(
 {d^{\,2} \Phi_{cl}\over dt^2} +{\gamma\over 2}{d
\Phi_{cl}\over dt}+ {\partial U(\Phi_{cl})\over
\partial \Phi_{cl} } \right) + i\gamma\, T\, \Phi_q^2\, \right]\, .
\end{equation}
Physically   the limit $\hbar\to 0$ means that
$\hbar\tilde\Omega\ll T$, where $\tilde\Omega$ is a characteristic
classical frequency of the particle. This condition is necessary
for the last term of Eq.~(\ref{dissipaction}) to take the
time--local form. The condition for neglecting the higher order
derivatives of $U$ is $\hbar\ll \gamma \tilde \Phi_{cl}^2$, where
$\tilde\Phi_{cl}$ is a characteristic classical amplitude of the
particle motion.

\subsection{Langevin equations}
\label{sec_44}

One way to proceed with the classical action
(\ref{dissipclassical}) is to notice that the exponent of its last
term (times $i$) may be identically rewritten in the following
way:
\begin{equation}
e^{-2\gamma T\!\!\int\!\! dt\, \Phi_q^2(t)} =\int\!\! {\cal
D}\xi(t) \,\,
 e^{-\!\!\int\!\! dt\,\left[{1\over 2\gamma T}
 \xi^2(t) - 2i \xi(t)\Phi_q(t)\right]}\, .
\end{equation}
This identity is  called the Hubbard--Stratonovich transformation,
while $\xi(t)$ is an auxiliary Hubbard--Stratonovich field. The
identity is proved by completing the square in the exponent on the
r.h.s., performing the Gaussian integration at every instance of
time and multiplying the results. There is a constant
multiplicative factor hidden in the integration measure, ${\cal
D}\xi$.

Exchanging the order of the functional integration over $\xi$ and
$\hat \Phi$, one finds for the partition function:
\begin{equation}
\hskip -.4cm
 Z\!\!=\!\!\!\int\!\! {\cal D}\xi \,\,
 e^{-{1\over 2\gamma T}\int\!\! dt\,\xi^2}\!\!\!
 \int\!\! {\cal D}\Phi_{cl}\!\!\int\!\! {\cal D}\Phi_q\,
 e^{-2i\int\!\!dt\, \Phi_q\left( {d^{\,2} \Phi_{cl}\over dt^2} +{\gamma\over 2}{d
\Phi_{cl}\over dt}+ {\partial U(\Phi_{cl})\over
\partial \Phi_{cl} }-\xi \right)} .
\end{equation}
Since the last (imaginary) exponent depends only linearly on
$\Phi_q(t)$, the integration over ${\cal D}\Phi_q$ results in the
$\delta$--function of the expression in the round brackets. This
functional $\delta$--function enforces  its argument  to be zero
at every moment of time. Therefore, among all the possible
trajectories $\Phi_{cl}(t)$, only those that satisfy the following
equation contribute to the partition function:
\begin{equation}
 {d^{\,2} \Phi_{cl}\over dt^2} +{\gamma\over 2}{d \Phi_{cl}\over
dt} + {\partial U(\Phi_{cl})\over \partial \Phi_{cl} } = \xi(t) \,
.
                                          \label{Langevin}
\end{equation}
This is a Newton equation with a time dependent external force
$\xi(t)$. Since, the same arguments are applicable to any
correlation function of the classical fields, e.g. $\langle
\Phi_{cl}(t)\Phi_{cl}(t')\rangle$, a solution strategy is as
follows: (i) choose some realization of $\xi(t)$; (ii) solve
Eq.~(\ref{Langevin}) (e.g. numerically); (iii) having its
solution, $\Phi_{cl}(t)$, calculate the correlation function; (iv)
average the result over an ensemble of realizations of the force
$\xi(t)$. The statistics of the latter are dictated by the weight
factor in the ${\cal D}\xi$ functional integral. It states that
$\xi(t)$ is a Gaussian short--range  (white) noise with the
correlators:
\begin{equation}
\langle \xi(t) \rangle = 0\,; \hskip 2cm \langle \xi(t)
\xi(t')\rangle = \gamma T \delta(t-t')\,.
                                         \label{noise}
\end{equation}
Equation (\ref{Langevin}) with the white noise on the r.h.s. is
called the Langevin equation. It describes classical Newtonian
dynamics in presence of stochastic thermal fluctuations. The fact
that the noise amplitude is related to the friction coefficient,
$\gamma$ and to the temperature is a manifestation of the FDT. The
latter holds as long as the environment (string) is at  thermal
equilibrium.


\subsection{Martin--Siggia--Rose}
\label{sec_46}

In section \ref{sec_44} one derived the Langevin equation for a
classical coordinate, $\Phi_{cl}$, from the action written  in
terms of $\Phi_{cl}$ and another field, $\Phi_q$. An inverse
procedure of deriving the effective action from the Langevin
equation is known as the Martin--Siggia--Rose (MSR) \cite{MSR}
technique. It is sketched here in the form suggested by
De-Dominics \cite{MSR}.

Consider  a Langevin equation:
\begin{equation}
\hat {\cal O}[\Phi] = \xi(t)\, ,
                                             \label{hatO}
\end{equation}
where $\hat {\cal O}[\Phi]$ is a (non--linear) differential
operator acting on the coordinate $\Phi(t)$ and $\xi(t)$ is a
white noise force, specified by Eq.~(\ref{noise}). Define the
``partition function'' as:
\begin{equation}
Z[\xi]=\int\!\! {\cal D}\Phi\, {\cal J}[\hat {\cal O }
]\,\delta\big(\hat {\cal O}[\Phi] - \xi(t)\big) \equiv 1\, .
                                                      \label{MSRZ}
\end{equation}
It is identically equal to unity by virtue of  the integration of
the $\delta$--function, provided ${\cal J}[\hat {\cal O}]$ is the
Jacobian of the operator $\hat {\cal O}[\Phi]$. The way to
interpret Eq.~(\ref{MSRZ}) is to discretize the time axis,
introducing $N$--dimensional vectors $\Phi_j=\Phi(t_j)$ and
$\xi_j=\xi(t_j)$. The operator takes the form: ${\cal
O}_i=O_{ij}\Phi_j+\Gamma_{ijk}\Phi_j\Phi_k +\ldots$, where a
summation is taken  over repeated indexes. The Jacobian, ${\cal
J}$, is given by the absolute value of the determinant of the
following $N\times N$ matrix: $J_{ij}\equiv
\partial {\cal O}_i/ \partial\Phi_j=
O_{ij}+2\Gamma_{ijk}\Phi_k+\ldots$. It is  possible to choose a
proper (retarded) regularization  where the  $J_{ij}$ matrix is a
lower triangular matrix with a unity main diagonal (coming
entirely from the $O_{ii}=1$ term). Clearly, in this case, ${\cal
J}=1$. Indeed, consider, for example, $\hat {\cal
O}[\Phi]=\partial_t\Phi-W(\Phi)$. The retarded regularized version
of the Langevin equation is: $\Phi_{i}
=\Phi_{i-1}+\delta_t(W(\Phi_{i-1})+\xi_{i-1})$. Clearly in this
case $J_{ii}=1$ and $J_{i,i-1}=-1-W'(\Phi_{i-1})\delta_t$, while
all other components are zero; as a result ${\cal J}=1$.

Although the partition function (\ref{MSRZ}) is trivial, it is
clear that all the meaningful observables and the correlation
functions may be obtained by inserting a set of  factors:
$\Phi(t)\Phi(t')\ldots$ in the functional integral,
Eq.~(\ref{MSRZ}). Having this in mind, let us proceed with the
partition function. Employing the integral representation of the
$\delta$--function with the help of an auxiliary field $\Psi(t)$,
one obtains:
\begin{equation}
Z[\xi]=\int\!\! {\cal D}\Phi\!\! \int\!\! {\cal
D}\Psi\,e^{\,-2i\!\!\int\!\! dt \, \Psi(t) \big(\hat {\cal
O}^R[\Phi(t)] - \xi(t)\big) }\, ,
                                                      \label{MSRZ1}
\end{equation}
where $\hat {\cal O}^R$ stands for the retarded regularization of
the $\hat {\cal O}$ operator and thus one takes ${\cal J}=1$. One
may average now over the white noise, Eq.~(\ref{noise}),  by
performing the Gaussian integration over $\xi$:
\begin{equation}
\hskip -.3cm
 Z= \!\!\int\!\! {\cal D}\xi \,\, e^{-{1\over 2\gamma
T}\int\!\! dt\,\xi^2} Z[\xi]=\int\!\! {\cal D}\Phi \Psi\,
 e^{-\int\!\!
dt \, \left[ 2i\, \Psi(t) \hat {\cal O}^R[\Phi(t)] + 2\gamma
T\Psi^2(t)\right] }\, ,
                                                      \label{MSRZ1}
\end{equation}
The exponent  is exactly the classical limit of the Keldysh
action, cf. Eq.~(\ref{dissipclassical}) (including the retarded
regularization of the differential operator), where
$\Phi=\Phi_{cl}$ and $\Psi=\Phi_q$. The message is that the MSR
action is nothing, but the classical (high temperature) limit of
the Keldysh action. The MSR technique provides a simple way to
transform from a classical stochastic problem to its proper
functional representation. The latter is useful for an analytical
analysis. One example is given below.

\subsection{Thermal activation}
\label{sec_47}

\begin{figure}[t]
\fbox{\vtop to5cm{\vss\hsize=.975\hsize \vglue 0cm
\hspace{0.01\hsize} \hskip -.5cm \epsfxsize=1\hsize
 \epsffile{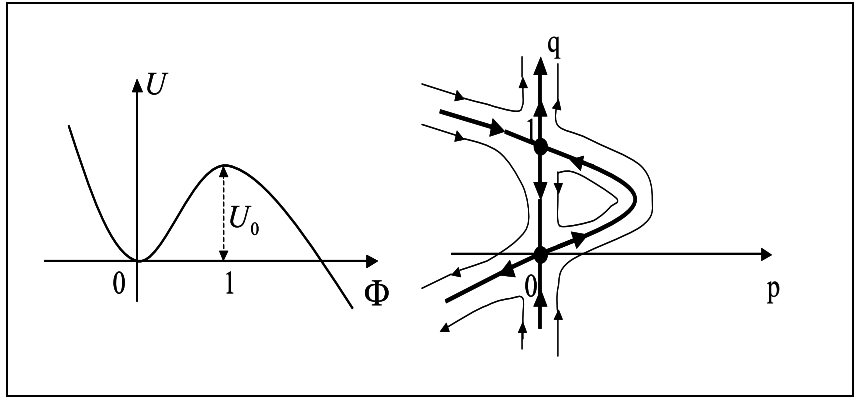}
\hspace{0.1\hsize} \vss }}
\label{fig_well}
 \caption{a) A potential with a meta-stable minimum. b) The phase portrait of the
 Hamiltonian system, Eq.~(\ref{fictHam}). Thick lines correspond
 to zero energy, arrows indicate evolution direction.
     }
\end{figure}

Consider a particle in a meta-stable potential well, plotted in
Fig.~\ref{fig_well}a. The potential has a meta-stable minimum at
$\Phi=0$ and a  maximum at $\Phi=1$ with the amplitude $U_0$. Let
us also assume that the particle's motion is over-damped, i.e.
$\gamma\gg \sqrt{U''}$. In this case one may disregard the inertia
term, leaving only viscous relaxation dynamics. The classical
dissipative action (\ref{dissipclassical}) takes the form:
\begin{equation}
                               \label{viscous}
S[\hat \Phi] = 2\!\!\int\limits_{-\infty}^\infty\!\! dt \left[
-\Phi_q \left( {\gamma\over 2}{d \Phi_{cl}\over dt}+ {\partial
U(\Phi_{cl})\over
\partial \Phi_{cl} } \right) + i\gamma\, T\, \Phi_q^2\, \right]\, .
\end{equation}
The corresponding saddle point equations are:
\begin{eqnarray}
                                      \label{saddleescape}
{\gamma\over 2}\,\dot\Phi_{cl}&=& - {\partial U(\Phi_{cl})\over
\partial \Phi_{cl} } + 2i\gamma T\, \Phi_q\, ;\\
{\gamma\over 2}\, \dot\Phi_{q}&=&  \Phi_q \,{\partial^2
U(\Phi_{cl})\over
\partial \Phi_{cl}^2 } \, \nonumber.
\end{eqnarray}
These equations possess the {\em classical} solution:
$\Phi_q(t)\equiv 0$ and $\Phi_{cl}(t)$ satisfies the classical
equation of motion: ${\gamma\over 2}\,\dot\Phi_{cl}= -\partial
U(\Phi_{cl})/\partial \Phi_{cl}$. For the  initial condition
$\Phi_{cl}(0)<1$ the latter equation predicts the viscous
relaxation towards the minimum at $\Phi_{cl}=0$. According to this
equation, there is no possibility to escape from this minimum.
Therefore the classical solution of Eqs.~(\ref{saddleescape}) does
{\em not} describe thermal activation. Thus one has to look  for
another possible solution of Eqs.~(\ref{saddleescape}), the one
with $\Phi_q\neq 0$.

To this end let us make a  simple linear change of variables:
$\Phi_{cl}(t)=q(t)$ and $\Phi_q(t)=p(t)/(i\gamma)$. Then the
dissipative action (\ref{viscous}) acquires the  form  of a
Hamiltonian action:
\begin{equation}
 iS=-\int\!\!dt\big(p\dot q - H(p,q) \big)\, ; \hskip 1cm
H(p,q)\equiv {2\over \gamma}\left[ -p\, {\partial U(q)\over
\partial q} + Tp^{\,2} \right] ,
                                         \label{fictHam}
\end{equation}
where the fictitious  Hamiltonian, $H$, is introduced
\footnote{Amazingly, this trick of rewriting viscous (or
diffusive) dynamics as a Hamiltonian one, works in a wide class of
problems. The price, one has to pay, is the doubling of the number
of degrees of freedom: $q$ and $p$ in the Hamiltonian language, or
``classical'' and ``quantum'' components in the Keldysh language.
}.
It is straightforward to see that in terms of the new variables
the equations of motion (\ref{saddleescape}) take the form of the
Hamilton equations: $\dot q=\partial H/\partial p$ and $\dot
p=-\partial H/\partial q$. One needs, thus, to investigate the
Hamiltonian system with the Hamiltonian Eq.~(\ref{fictHam}). To
visualize it, one may plot its phase portrait, consisting of lines
of constant energy $E=H(p(t),q(t))$ on the $(p,q)$ plane,
Fig.~\ref{fig_well}b. The topology is determined by the two lines
of zero energy: $p=0$ and $Tp=\partial U(q)/\partial q$, that
intersect at the two stationary points of the potential: $q=0$ and
$q=1$. The $p=0$ line corresponds to the classical (without
Langevin noise) dynamics (notice, that the action is identically
zero for  motion along this line) and thus $q=0$ is the stable
point, while $q=1$ is the unstable one. Due to Liouville theorem,
every fixed point must have one stable and one unstable direction.
Therefore, along the ``non--classical'' line: $p=T^{-1}\partial
U(q)/\partial q$, the situation is reversed: $q=0$ is unstable,
while $q=1$ is stable. It is clear now that, to escape from the
bottom of the potential well, $q=0$, the system must move along
the non--classical line of zero energy until it reaches the top of
the barrier, $q=1$, and then continue to drop according to the
classical equation of motion (moving along the classical line
$p=0$). There is a non--zero action associated with the motion
along the non--classical line:
\begin{equation}
iS= - \int \!\! dt\, p\dot q=-\int\limits_0^1\!\! p(q) dq=
-{1\over T}\int\limits_0^1\!{\partial U(q)\over \partial q}\,\, dq
= -\,{U_0\over T}\, ,
                                         \label{activation}
\end{equation}
where one has used that $H=0$ along the integration trajectory. As
a result, the thermal escape probability is proportional to
$e^{iS}=e^{-U_0/T}$, which is nothing but the thermal activation
exponent.

\subsection{Fokker-Planck equation}
\label{sec_48}

Another way to approach the action (\ref{viscous}) is to notice
that it is quadratic in $\Phi_q$ and therefore the ${\cal
D}\Phi_q$ integration may be explicitly performed. To shorten
notations and emphasize the relation to the classical coordinate,
we shall follow the previous section and denote
$\Phi_{cl}(t)\equiv q(t)$. Performing the Gaussian integration
over $\Phi_q$ of $e^{iS[\hat \Phi]}$, with $S[\Phi_{cl},\Phi_q]$
given by Eq.~(\ref{viscous}), one finds the action, depending on
$\Phi_{cl}\equiv q$ only:
\begin{equation}
                               \label{phiclonly}
iS[q] = -{1\over 2\gamma T}\!\int\limits_{-\infty}^\infty\!\!\!
dt\, \left( {\gamma\over 2}\,\dot q +  U'_q\right)^2 \, .
\end{equation}
One may now employ  the same trick, that allows to pass from the
Feynman path integral to the Schr\"odinger equation. Namely, let
us introduce the ``wave function'', ${\cal P}(q,t)$, that is a
result of the functional integration of $e^{iS[q]}$ over all
trajectories that  at time $t$ pass through the point $q_N\equiv
q$. Adding one more time step, $\delta_t$, to the trajectory, one
may write ${\cal P}(q_N,t+\delta_t)$ as an integral of ${\cal
P}(q_{N-1},t)={\cal P}(q+\delta_q,t)$ over $\delta_q\equiv
q_{N-1}-q$ :
\begin{eqnarray}
                                     \label{transferm}
{\cal P}(q,t+\delta_t)\!\!\!\!&=&\!\!\!\!C\!\int\!\! d\delta_q\,\,
e^{-{\delta_t\over 2\gamma T}\left( {\gamma\over
2}\,{-\delta_q\over\delta_t}\,\, +\,
U_q'(q+\delta_q)\right)^2}{\cal P}(q+\delta_q,t)\\
\!\!\!\!&=&\!\!\!\!C\!\int\!\! d\delta_q\,\, e^{-{\gamma\over 8 T}
{\delta_q^2\over\delta_t}}\left[\, e^{\,{\delta_q\over 2T}\,
U'_q(q+\delta_q) - {\delta_t\over 2\gamma T} \left( U'_q
\right)^2}\,{\cal P}(q+\delta_q,t)\right]\, , \nonumber
\end{eqnarray}
where the factor $C$ from the integration measure is determined by
the condition: $C\!\int\!\! d\delta_q\,
\exp\left\{-\gamma\delta_q^2/(8T\delta_t)\right\}=1$. Expanding
the  expression in the square brackets on the r.h.s. of the last
equation to the second order in $\delta_q$ and the first order in
$\delta_t$, one finds:
\begin{eqnarray}
                                     \label{transfermexp}
{\cal P}(t+\delta_t)\!\!\!\!\!&=&\!\!\!\!\! \left(\! 1\!
+\!{\langle \delta_q^2\rangle\over 2T}U''_{\!qq}\! + \! {1\over 2}
{\langle \delta_q^2\rangle\over 4T^2} \left( U'_q \right)^2 \! -
\! {\delta_t\over 2\gamma T} \left( U'_q \right)^2\!\right)\!{\cal
P}\! + \!  {\langle \delta_q^2\rangle\over 2T}U'_q{\cal P}'_q\!
\nonumber \\
 &+&\!\!\!\!\!{\langle \delta_q^2\rangle\over 2}{\cal
P}''_{\!qq}
 =\, {\cal P}(t) + \delta_t\left( {2\over
\gamma}\,U''_{\!qq}\,{\cal P} + {2\over \gamma}\, U'_q {\cal
P}'_q+ {2T\over \gamma}\,{\cal P}''_{\!qq}\right)\, ,
\end{eqnarray}
where $\langle \delta_q^2\rangle\equiv C\!\int\!\! d\delta_q
\exp\left\{-\gamma\delta_q^2/(8T\delta_t)\right\}\delta^2_q =
4T\delta_t/\gamma$. Finally, rewriting the last expression in the
differential form, one obtains:
\begin{equation}
{\partial {\cal P}\over\partial t}={2\over \gamma}\left[
 {\partial \over \partial
q}{\partial U\over \partial q}  + T{\partial^2\over \partial
q^2}\right] {\cal P}
 ={2\over \gamma}\,
 {\partial \over \partial
q}\left[ {\partial U\over \partial q}\,{\cal P}  + T\,{\partial
{\cal P}  \over
\partial q}\right].
\end{equation}
This is the Fokker--Planck (FP) equation for  the evolution of the
probability distribution function, ${\cal P}(q,t)$. The latter
describes the probability  to find the particle at the point
$q(=\Phi)$ at time $t$. If one starts from an initially sharp
(deterministic) distribution: ${\cal P}(q,0)= \delta(q-q(0))$,
then the first term on the r.h.s. of the FP equation describes the
viscous drift of the particle in the potential $U(q)$. Indeed, in
the absence of the second term ($T=0$), the equation is solved by
${\cal P}(q,t)= \delta(q-q(t))$, where $q(t)$ satisfies the
deterministic equation of motion $(\gamma/2)\dot q(t)=-\partial
U(q(t))/\partial q$
\footnote{To check this statement one may substitute ${\cal
P}(q,t)= \delta(q-q(t))$ into the $T=0$ FP equation:
$\delta'_q(q-q(t))(-\dot q(t)) =(2/\gamma)\left[U''_{\!
qq}\delta(q-q(t)) + U'_q\delta'_q(q-q(t))\right]$. Then
multiplying both parts of this equation by $q$ and integrating
over $dq$ (by performing integration by parts), one finds: $\dot
q(t) = -(2/\gamma)U'_q(q(t))$.}
. The second term describes  the diffusion spreading of the
probability distribution due to  the thermal stochastic noise
$\xi(t)$. For a confining  potential $U(q)$ (such that
$U(\pm\infty)\to \infty$) the stationary solution of the FP
equation is the equilibrium Boltzmann distribution: ${\cal
P}(q)\sim \exp\{-U(q)/T\}$.

The FP equation  may be considered as the (imaginary time)
Schr\"odinger equation: $\dot {\cal P}=\hat H {\cal P}$, where the
``Hamiltonian'', $\hat H$, is nothing but the ``quantized''
version of the classical Hamiltonian, introduced in the previous
section, Eq.~(\ref{fictHam}). The ``quantization'' rule is $p\to
\hat p\equiv -\partial/\partial q$, so the canonical commutation
relation: $[q,\hat p]_-=1$, holds. Notice that before applying
this quantization rule, the corresponding classical Hamiltonian
must be {\em normally ordered}. Namely, the momentum $\hat p$
should be to the left of the coordinate $q$, cf.
Eq.~(\ref{fictHam}). Using the commutation relation, one may
rewrite the quantized Hamiltonian as: $\hat H =T\hat p^2-\hat
pU'_q= T\left(\hat p - U'_q/(2T)\right)\left(\hat p -
U'_q/(2T)\right) - (U'_q)^2/(4T) + U''_{qq}/2$ (we took
$\gamma/2=1$) and perform the canonical transformation: $Q=q$ and
$\hat P =\hat p -U'_q/(2T)$. In terms of these new variables the
Hamiltonian takes the familiar form: $\hat H = T\hat P^2 +V(Q)$,
where $V(Q)= - (U'_Q)^2/(4T) + U''_{QQ}/2$, while the ``wave
function'' transforms as $\tilde {\cal P}(Q,t) = e^{U(Q)/(2T)}
{\cal P}$.

\subsection{From Matsubara to Keldysh}
\label{sec_49}

\begin{table}[b]
\begin{tabular*}{\columnwidth}{@{\extracolsep{\fill}}@{}lcccccccc@{}}
\hline \hline
\end{tabular*}
{\bf  Reminder:} The Matsubara technique deals with the imaginary
time $\tau$ confined to the interval $\tau\in[0,\beta[\,$, where
$\beta=1/T$. All bosonic fields must be periodic in this interval:
$\phi(\tau+\beta)=\phi(\tau)$, while the fermionic fields are
antiperiodic: $\psi(\tau+\beta)=-\psi(\tau)$. It is convenient to
introduce the discrete Fourier (Matsubara) transform, e.g.
$\phi_m=\int\limits_0^\beta d\tau \phi(\tau)\,
e^{\,i\epsilon_m\tau}$, where for bosons $\epsilon_m\equiv 2\pi m
T$, while for fermions $\epsilon_m\equiv \pi (2m+1) T$ and
$m=0,\pm 1,\ldots$.
\end{table}

In some applications it may be convenient to derive an action in
the equilibrium  {\em Matsubara} technique and change to the
Keldysh representation at a later stage to tackle
out--of--equilibrium problems. This section intends to illustrate
how such transformation  may be carried out. To this end consider
the following bosonic Matsubara action:
\begin{equation}
S[\Phi_m]=\gamma\, T\!\! \sum\limits_{m=-\infty}^\infty\!{1\over
2}\,  |\epsilon_m||\Phi_m|^2\, ,
                                                \label{matsubaraaction}
\end{equation}
where $\Phi_m=\bar\Phi_{-m}$ are the Matsubara components of a
real bosonic field, $\Phi(\tau)$. Notice, that due to the absolute
value sign: $|\epsilon_m|\neq i\partial_\tau$. In fact, in the
imaginary time representation the action (\ref{matsubaraaction})
has the non--local form:
\begin{equation}
S[\Phi]= -{\gamma\over 2} \int\!\!\!\!\int\limits_{0}^\beta
d\tau\, d\tau' \, \Phi(\tau)\, \frac{\pi T^2}{\sin^2(\pi
T(\tau-\tau'))} \,\Phi(\tau')\, .
                                                \label{matsubaratime}
\end{equation}
This action is frequently named after  Caldeira and Leggett
\cite{CL}, who used it to investigate the influence of dissipation
on quantum tunnelling.

To transforn to the Keldysh representation one needs to double the
number of degrees of freedom: $\Phi\to
\hat\Phi=(\Phi_{cl},\Phi_q)^T$. Then according to the causality
structure, section \ref{sec_24}, the general form of the time
translationally invariant Keldysh action is:
\begin{equation}
S = \gamma \int\limits\!\! {d\epsilon\over 2\pi} \,
\big(\Phi_{cl},\Phi_q\big)_{\epsilon} \left(\begin{array}{cc}
0   & [L^{A}(\epsilon)]^{-1}  \\
  \left[L^{R}(\epsilon)\right]^{-1}  & [L^{-1}]^K(\epsilon)
\end{array}\right)
\left(\begin{array}{c} \Phi_{cl} \\ \Phi_q
\end{array}\right)_{\epsilon} \, ,
                                       \label{matsubKeldysh}
\end{equation}
where $[L^{R(A)}(\epsilon)]^{-1} $ is the analytical continuation
of the Matsubara correlator $|\epsilon_m|/2$ from the {\em upper
(lower)} half--plane of the imaginary variable $\epsilon_m$  to
the real axis: $-i\epsilon_m\to \epsilon$. As a result,
$[L^{R(A)}(\epsilon)]^{-1}=\pm i\epsilon/2$. The Keldysh component
follows from the FDT: $[L^{-1}]^K(\epsilon)=i\epsilon\,\cth\,
\epsilon/(2\,T)$, cf. Eqs.~(\ref{LRA}) and (\ref{LK}).  Therefore
the Keldysh counterpart of the Matsubara action,
Eqs.~(\ref{matsubaraaction}) or (\ref{matsubaratime}) is the
already familiar dissipative action, Eq.~(\ref{dissipaction}),
(without the potential terms, of course). One may now include
external fields and allow the system to deviate from the
equilibrium.

\subsection{Dissipative chains and membranes}
\label{sec_410}

Instead of dealing with a single particle connected to a bath, let
us now consider a chain or lattice of coupled particles, with {\em
each one} connected to a bath. To this end, one (i) supplies a
spatial index, $r$, to the field: $\Phi(t)\to \Phi(t;r)$, and (ii)
adds the harmonic interaction potential between nearest neighbors
particles: $\sim(\Phi(t,r)-\Phi(t,r+1))^2\to (\nabla_r\Phi)^2$ in
the continuous limit. By changing to the classical--quantum
components and performing the spatial integration by parts (cf.
Eq.~(\ref{Keldyshphonons1})), the gradient term translates to:
$\Phi_q\nabla^2_r\Phi_{cl} + \Phi_{cl}\nabla^2_r\Phi_{q}$. Thus it
modifies  the retarded and advanced components of the correlator,
but it does {\em not} affect the $(q-q)$ Keldysh component:
\begin{equation}
[L^{R(A)}]^{-1}={1\over 2}\, \delta(t-t')\,\delta(r-r')\big( \mp
\partial_{t'} + D\nabla^2_{r'} \big)\, ,
                                               \label{diffusiveL}
\end{equation}
where $D$ is the rigidity of the chain or the membrane. In the
Fourier representation: $[L^{R(A)}(\epsilon;k)]^{-1}={1\over 2}
\big(\pm i\epsilon -Dk^2\big)$. In equilibrium the Keldysh
component is not affected by the gradient terms, and is given by
Eq.~(\ref{LK}) (in the real space representation it acquires the
factor $\delta(r-r')$). In particular, its classical limit is (cf.
Eq.~(\ref{LKt}))  $[L^{-1}]^K=i2T\delta(t-t')\delta(r-r')$. As a
result,  the action of a classical elastic chain in contact with a
bath is:
\begin{equation}
                               \label{chainclassical}
S[\hat \Phi] =
2\!\!\int\!\!dr\!\!\!\int\limits_{-\infty}^\infty\!\! dt \left[
-\Phi_q \left(\dot \Phi_{cl} -D\nabla^2_r\Phi_{cl}  + {\partial
U(\Phi_{cl})\over
\partial \Phi_{cl} } \right) + i2\, T\, \Phi_q^2\, \right]\,
,
\end{equation}
where the inertia terms have been neglected and we put
$\gamma/2=1$ for brevity.

One may introduce now an auxiliary Hubbard--Stratonovich field
$\xi(t;r)$ and write the Langevin equation according to section
\ref{sec_44}:
\begin{equation}
\dot \Phi_{cl} -D\nabla^2_r\Phi_{cl}  + {\partial
U(\Phi_{cl})\over \partial \Phi_{cl} } =\xi(t;r)\, ,
                             \label{chainLangevin}
\end{equation}
where $\xi$ is a  Gaussian noise: $\langle
\xi(t;r)\xi(t';r')\rangle=2T\delta(t-t')\delta(r-r')$ with
short--range correlations.

Let us consider an elastic chain  sitting in the bottom of  the
($r$--independent) meta-stable potential well, depicted in
Fig.~\ref{fig_well}a. If a sufficiently large piece of the chain
thermally escapes from the well, it may find it favorable to slide
down the potential, pulling the entire chain out of the well. To
find the shape of such an optimally large critical domain and its
action, let us change to the Hamiltonian variables of section
\ref{sec_47}: $q(t;r)\equiv \Phi_{cl}(t;r)$ and $p(t;r)\equiv
2i\Phi_q(t;r)$. The action (\ref{chainclassical}) takes the
Hamiltonian form:
\begin{equation}
 \hskip -.2cm
 iS=-\!\!\int\!\!\!\!\!\int\!\!drdt\,\big(p\dot q - H(p,q) \big)\, ; \hskip .5cm
H\equiv  -p\, {\partial U(q)\over
\partial q} +   p\, D\nabla_r^2 q +  Tp^{\,2}  ,
                                         \label{chainHam}
\end{equation}
and the corresponding equations of motion are:
\begin{eqnarray}
                                              \label{chaineqmotion}
\dot q &=& \frac{\delta H}{\delta p}\, =\, D\nabla^2_r q - U'_q(q)
+2Tp\, ; \\
\dot p &=&\!\!\!\!\! -\frac{\delta H}{\delta q} = -D\nabla^2_r p +
p\,U''_{\!qq}(q)\, . \nonumber
\end{eqnarray}
These are complicated partial differential equations, that cannot
be solved in general. Fortunately, the shape of the optimal
critical domain can be found. As was discussed in section
\ref{sec_47}, the minimal action trajectory corresponds to a
motion with zero energy, $H=0$. According to Eq.~(\ref{chainHam}),
this is the case  if either $p=0$ (classical zero--action
trajectory), or $Tp= U'_q(q)-D \nabla_r^2 q $ (finite--action
escape trajectory). In the latter case the equation of motion for
$q(t;r)$ takes the form of the classical equation in the reversed
time: $\dot q =- D\nabla^2_r q + U'_q(q)=Tp\,$. Thanks to the last
equality the equation of motion for $p(t;r)$ is automatically
satisfied
\footnote{Indeed, $T\dot p =\partial_t \dot q=-D\nabla^2_r \dot
q+\dot q U''_{\!qq}=T(-D\nabla^2_r p + pU''_{\!qq} )$. This
non--trivial fact reflects the existence of an  accidental
conservation law: $H\big(p(t;r),q(t;r)\big)=0$ -- {\em locally}!
While from the general principles only the total global energy has
to be conserved.}.
In the reversed time dynamics the $q(t;r)=0$ configuration is
unstable and therefore the chain develops a ``tongue'' that grows
until it reaches the stationary shape:
\begin{equation}
- D\nabla^2_r q + U'_q(q)=0\, .
                              \label{domain}
\end{equation}
The solution of this equation gives the shape of the critical
domain. Once it is formed, it  may grow further according to the
classical equation $\dot q = D\nabla^2_r q - U'_q(q)$ and $p=0$
with zero action. The action along the non--classical escape
trajectory,  paid to form the ``tongue'' is ($H(p,q)=0$):
\begin{equation}
 \hskip -.5cm
-iS\!=\!\!\!\int\!\!\!\!\!\int\!\!\!drdt\,p\dot q\!=\! {1\over
T}\!\!\int\!\!\!\!\!\int\!\!drdt\big( U'_q(q)-D \nabla_r^2 q \big)
\dot q\! = \!{1\over T}\!\!\int\!\!\! dr\Big( U(q)+ {D\over 2}
(\nabla_r q)^2 \!\Big) ,
                                  \label{chainstatic}
\end{equation}
where in the last equality an explicit integration over time is
performed. The escape action is given therefore by the static
activation expression that includes both the potential and the
elastic energies. The optimal domain, Eq.~(\ref{domain}), is found
by the minimization of this static action (\ref{chainstatic}). One
arrives, thus, at a thermodynamic Landau-type description of the
first--order phase transitions. Notice, that the effective
thermodynamic description appears due to the assumption that
$H(p,q)=0$ and, therefore, that all the processes take an
infinitely long time.

\vskip .5cm

\section{Fermions}
\label{sec_5}

\subsection{Free fermion Keldysh action}

Consider a single  quantum state, with the energy $\epsilon_0$.
This state is populated by  spin-less fermions (particles obeying
the Pauli exclusion principle). In fact, one may have either zero,
or one particle in this state. The secondary quantized Hamiltonian
of such a system has the form:
\begin{equation}
\hat H= \epsilon_0\, c^\dagger c\, ,
                                               \label{fermH}
\end{equation}
where $c^\dagger$ and $c$ are fermion creation and annihilation
operators of the state $\epsilon_0$. They obey  standard {\em
anti}commutation relations: $\{c\,,c^\dagger\}_+=1$ and
$\{c\,,c\}_+=\{c^\dagger\,,c^\dagger\}_+=0$, where $\{\,,\,\}_+$
stands for the anti-commutator.

\begin{table}[b]
\begin{tabular*}{\columnwidth}{@{\extracolsep{\fill}}@{}lcccccccc@{}}
\hline \hline
\end{tabular*}
{\bf  Reminder:} \normalsize the fermionic coherent state
$|\psi\rangle\equiv (1-\psi c^\dagger)|0\rangle$, parameterized by
a Grassmann number $\psi$ (such that
$\{\psi,\psi'\}_+=\{\psi,c\}_+=0$), is an eigenstate of the
annihilation operator: $c|\psi\rangle =\psi|\psi\rangle$.
Similarly: $ \langle\psi|c^\dagger=\langle\psi|\bar\psi$, where
$\bar\psi$ is another Grassmann number, {\em unrelated} to $\psi$.
The matrix elements of a {\em normally ordered} operator, such as
e.g. the Hamiltonian, take the form $\langle\psi|\hat
H(c^\dagger,c)|\psi'\rangle = H(\bar\psi,\psi')
\langle\psi|\psi'\rangle$. The overlap between any two coherent
states is
$\langle\psi|\psi'\rangle=1+\bar\psi\psi'=\exp\{\bar\psi\psi'\}$.
The trace of an operator, $\hat A$, is calculated as: $\Tr\{\hat
A\}= \int\!\!\!\int  d\bar\psi\, d\psi\, e^{-\bar\psi\psi}
\langle-\psi|\hat A|\psi\rangle $, where the Grassmann integrals
are {\em defined} as: $\int\! d\psi\, 1=0$ and $\int\! d\psi\,
\psi =1$.
\end{table}

One can now consider  the evolution operator along the Keldysh
contour, ${\cal C}$ and the corresponding ``partition function'',
$Z=1$, defined in exactly the same manner as for bosonic  systems:
Eq.~(\ref{e1}). The trace of the equilibrium density matrix is
$\Tr\{\rho_0\}=1+\rho(\epsilon_0)$, where the two terms stand for
the empty and the singly  occupied state. One  divides the Keldysh
contour into $(2N-2)$ time intervals of  length $\delta_t\sim
1/N\to 0$ and introduces resolutions of unity in $2N$ points along
${\cal C}$, Fig.~(\ref{contour}). The only difference from the
bosonic case in  section~\ref{sec_11} is that now one uses a
resolution of unity in the {\em fermionic} coherent state basis
\cite{Negele}:
\begin{equation}
1=\int\!\!\!\int d\bar\psi_j\, d\psi_j\,\, e^{-\bar\psi_j\psi_j}\,
|\psi_j\rangle\langle\psi_j|\, ,
                                        \label{resunityfermions}
\end{equation}
where $\bar\psi_j$ and $\psi_j$ are {\em mutually independent}
Grassmann variables. The rest of the algebra goes through exactly
as in the bosonic case, section~\ref{sec_11}. As a result, one
arrives at:
\begin{equation}
Z={1\over \Tr\{\rho_0\} }  \int \!\!\!\int
\prod\limits_{j=1}^{2N}\left[d\bar\psi_j\, d\psi_j \right] \,\,
e^{\,i \sum\limits_{j,j'=1}^{2N} \bar\psi_j {\cal
G}^{-1}_{jj'}\psi_{j'}}\, ,
                                              \label{partitiondiscreteferm}
\end{equation}
where the $2N \times 2N$ matrix ${\cal G}^{-1}_{jj'}$ stands for:
\begin{equation}
  \label{Dmatrix}
  i\, {\cal G}^{-1}_{jj'}\equiv
\left[\begin{array}{rrr|rrr}
 -1   &      &    &   &   &   -\rho(\epsilon_0) \\
  1\!-\!h & -1   &    &   &   &                 \\
      &  1\!-\!h & -1 &   &   &                 \\ \hline
     &     &  1 & -1 &    &                 \\
     &     &    &1\!+\!h & -1 &                 \\
     &     &    &    & 1\!+\!h&  -1
 \end{array} \right]\, ,
                                                  \label{matrixferm}
\end{equation}
and $h\equiv i\epsilon_0\delta_t$. The only difference from the
bosonic case is the negative sign before the $\rho(\epsilon_0)$
matrix element, originating from the minus sign in the
$\langle-\psi_{2N}|$ coherent state in the expression for the
fermionic trace. To check the normalization, let us evaluate the
determinant of such a matrix:
\begin{equation}
 \hskip -.35cm
\mbox{det} \big[ i{\cal G}^{-1} \big] = 1+
\rho(\epsilon_0)(1-h^2)^{N-1} \approx 1+ \rho(\epsilon_0)\,
e^{(\epsilon_0\delta_t)^2(N-1)}\to 1+ \rho(\epsilon_0)  .
                                                \label{determinantferm}
\end{equation}
Employing the fact that the fermionic Gaussian integral is given
by the determinant (unlike the {\em inverse} determinant for
bosons) of the correlation matrix, ~\ref{app_Gaussian}, one finds:
\begin{equation}
Z= \frac{ \mbox{det}\big[ i{\cal G}^{-1} \big]}{\Tr\{\rho_0 \} } =
1\, ,
                                                \label{unityferm}
\end{equation}
as it should be. Once again,  the upper--right element of the
discrete matrix, Eq.~(\ref{matrixferm}), is crucial to maintain
the correct normalization.

Taking the limit $N\to \infty$ and introducing the continuous
notations, $\psi_j\to \psi(t)$, one obtains:
\begin{equation}
Z=  \int\!\!  {\cal D} \bar\psi \psi\,\, e^{\,iS[\bar\psi,\psi] }
= \int\!\!  {\cal D} \bar\psi \psi\, \exp\left\{i
\int\limits_{\cal C}\! \left[ \bar\psi(t)\, {\cal
G}^{-1}\psi(t)\right] dt\right\} \, ,
                                                              \label{e2ferm}
\end{equation}
where according to Eqs.~(\ref{partitiondiscreteferm}) and
(\ref{matrixferm}) the action is given by
\begin{equation}
 \hskip -.35cm
S[\bar\psi,\psi]\! =\!
\sum\limits_{j=2}^{2N}\!\left[i\bar\psi_j\,\frac{\psi_j-\psi_{j-1}}{\delta
t_j} -\epsilon_0\bar\psi_j\psi_{j-1}\right]\delta
t_j\,+i\,\bar\psi_1\Big(\psi_1+\rho(\epsilon_0)\psi_{2N}\Big)  ,
                                                              \label{e3ferm}
\end{equation}
where $\delta t_j\equiv t_j-t_{j-1}=\pm \delta_t$. Thus the
continuous form of the operator ${\cal G}^{-1}$ is the same as for
bosons, Eq.~(\ref{Gcontinious}): ${\cal G}^{-1}= i\partial_t -
\epsilon_0$. Again the upper--right element of the discrete matrix
(the last term in Eq.~(\ref{e3ferm})), that contains  information
about the distribution function, is seemingly absent in the
continuous notations.

Splitting the Grassmann field $\psi(t)$ into the two components
$\psi_+(t)$ and $\psi_-(t)$ that reside on the forward and the
backward parts of the time contour correspondingly, one may
rewrite the action as:
\begin{equation}
S=\int\limits_{-\infty}^{\infty}\!\!\! dt\,
\bar\psi_+(t)[i\partial_t - \epsilon_0]
\psi_+(t)-\int\limits_{-\infty}^{\infty}\!\!\! dt\,
\bar\psi_-(t)[i\partial_t - \epsilon_0] \psi_-(t)\, ,
                                         \label{plusminusferm}
\end{equation}
where the dynamics of $\psi_+$ and $\psi_-$ are {\em not}
independent from each other, due to the presence of  non--zero
off--diagonal blocks in the discrete matrix,
Eq.~(\ref{matrixferm}).

The four fermionic Greens functions: ${\cal G}^{T(\tilde T)}$ and
${\cal G}^{<(>)}$ are defined in the same way as their bosonic
counterparts, Eq.~(\ref{corr}):
\begin{eqnarray}
                                         \label{corrferm}
 \hskip -.35cm
 \langle \psi_+(t)\bar\psi_-(t')\rangle \!\!\!\! &\equiv &\!\!\!\!
i{\cal G}^<(t,t') =  -{\Tr\{c^\dagger(t') c(t) \hat \rho_0\}\over
\Tr\{\hat\rho_0\}}= -n_F\,e^{-i\epsilon_0(t-t')}\, ;
\nonumber\\
\hskip -.5cm
 \langle \psi_-(t)\bar\psi_+(t')\rangle
\!\!\!\!&\equiv & \!\!\!\! i{\cal G}^>(t,t')
 = {\Tr\{c(t) c^\dagger(t') \hat \rho_0\}\over \Tr\{\hat\rho_0\}}=
(1-n_F)\,e^{-i\epsilon_0(t-t')} ;
\\
 \hskip -.35cm
\langle\psi_+(t)\bar\psi_+(t')\rangle \!\!\!\!&\equiv & \!\!\!\!
i{\cal G}^T(t,t') = \theta(t-t')i{\cal G}^>(t,t') +
\theta(t'-t)i{\cal G}^<(t,t') \, ;
\nonumber \\
 \hskip -.4cm
\langle\psi_-(t)\bar\psi_-(t')\rangle \!\!\!\! &\equiv & \!\!\!\!
i{\cal G}^{\tilde T}(t,t') =
 \theta(t'-t)i{\cal G}^>(t,t') + \theta(t-t')i{\cal G}^<(t,t')
\, ; \nonumber
\end{eqnarray}
The difference is  in the minus sign in  the expression for ${\cal
G}^<$, due to the anti--commutation relations, and the Planck
occupation number is exchanged for the Fermi one: $n\to n_F\equiv
\rho(\epsilon_0)/(1+\rho(\epsilon_0))$. Equations
(\ref{coincidingarguments}) and (\ref{relation}) hold for the
fermionic Green functions as well.


\subsection{Keldysh rotation}
\label{sec_52}

It is customary to perform the  Keldysh rotation in the fermionic
case in a different manner  from the bosonic one. Define the new
fields as:
\begin{equation}
\psi_{1}(t) = {1\over \sqrt{2}}\big(\psi_+(t) + \psi_-(t)\big) \,;
\,\,\,\,\,\,\,\,\, \psi_{2}(t) =  {1\over \sqrt{2}}\big(\psi_+(t)
- \psi_-(t)\big)\,.
                                                \label{rotationferm}
\end{equation}
This line is exactly parallel to the bosonic one,
Eq.~(\ref{rotation}). However, following Larkin and Ovchinnikov
\cite{LO}, it is agreed that the ``bar'' fields  transform in a
different way:
\begin{equation}
\bar\psi_{1}(t) = {1\over \sqrt{2}}\big(\bar\psi_+(t) -
\bar\psi_-(t)\big) \,; \,\,\,\,\,\,\,\,\, \bar\psi_{2}(t) =
{1\over \sqrt{2}}\big(\bar\psi_+(t) + \bar\psi_-(t)\big)\,.
                                                \label{rotationferm1}
\end{equation}
The point is that the Grassmann fields $\bar\psi$ are {\em not}
conjugated to $\psi$, but rather are completely independent
fields, that may be chosen to transform in an arbitrary manner (as
long as the transformation matrix has a non-zero determinant).
Notice, that there is no issue regarding the convergence of the
integrals, since the Grassmann integrals are {\em always}
convergent. We also avoid the subscripts $cl$ and $q$, because the
Grassmann variables {\em never} have a classical meaning. Indeed,
one can never write a saddle--point or any other equation in terms
of $\bar\psi,\psi$ rather they must  always be integrated out in
some stage of the calculations.

Employing Eqs.~(\ref{rotationferm}),  (\ref{rotationferm1}) along
with Eq.~(\ref{corrferm}), one finds:
\begin{equation}
-i\langle \psi_a(t)\bar\psi_b(t')\rangle \equiv \hat {\cal G}^{ab}
= \left(\begin{array}{cc}
{\cal G}^R(t,t') & {\cal G}^{K}(t,t') \\
0                & {\cal G}^{A}(t,t')
\end{array}\right)\, ,
                                                \label{Greenferm}
\end{equation}
where hereafter $a,b  = (1,2)$. The presence of zero in the
$(2,1)$ element of this matrix is a manifestation of
identity~(\ref{relation}). The {\em retarded, advanced} and {\em
Keldysh} components of the Green function are expressed in terms
of ${\cal G}^{T(\tilde T)}$ and ${\cal G}^{<(>)}$ in exactly the
same way as their bosonic analogs, Eq.~(\ref{Greens}), and
therefore posses the same symmetry properties:
Eqs.~(\ref{conj})--(\ref{GRplusGA}). An important consequence of
Eqs.~(\ref{traces}), (\ref{GRplusGA}) is:
\begin{equation}
\Tr\left\{ \hat {\cal G}_1\circ \hat {\cal G}_2\circ\ldots\circ
\hat {\cal G}_k \right\}(t,t) =0\, ,
                                                \label{tracesferm}
\end{equation}
where the circular multiplication sign involves integration over
the intermediate times along with the $2\times 2$ matrix
multiplication. The argument $(t,t)$ states that the first time
argument of $\hat {\cal G}_1$ and the last argument of $\hat {\cal
G}_k$ are the same.

Notice that the fermionic Green function has a different structure
from  its bosonic counterpart, Eq.~(\ref{Green}): the positions of
the $R,A$ and $K$ components in the matrix are exchanged. The
reason, of course, is  the different convention for transformation
of the ``bar'' fields. One could choose the fermionic convention
to be the same as the bosonic (but {\em not} the other way
around!), thus having the same structure, Eq.~(\ref{Green}), for
fermions as for bosons.  The rationale for the Larkin--Ovchinnikov
choice, Eq.~(\ref{Greenferm}),  is that the inverse Green
function, $\hat {\cal G}^{-1}$ and fermionic self energy $\hat
\Sigma_F$ have the same appearance as $\hat {\cal G}$:
\begin{equation}
\hat {\cal G}^{-1}= \left(\begin{array}{cc}
\left[{\cal G}^R\right]^{-1} & \left[{\cal G}^{-1}\right]^{K} \\
0                & \left[{\cal G}^{A}\right]^{-1}
\end{array}\right)\, ; \hskip 1cm
\hat \Sigma_F = \left(\begin{array}{cc}
\Sigma_F^R       &  \Sigma_F^{K} \\
0                &  \Sigma_F^{A}
\end{array}\right)\, ,
                                                \label{Greeninverseferm}
\end{equation}
whereas in the case of bosons $\hat G^{-1}$,
Eq.~(\ref{Greeninverse}), and $\hat \Sigma$, Eq.~(\ref{Sigma}),
look differently from $\hat G$, Eq.~(\ref{Green}). This fact gives
the form Eq.~(\ref{Greenferm}), (\ref{Greeninverseferm}) a certain
technical advantage.

For the single fermionic state (see. Eq.~(\ref{corrferm})):
\begin{eqnarray}
                                        \label{Greens1ferm}
                                        \hskip -.3cm
{\cal G}^{R}(t,t')\!\!\! &=&
- i\theta(t-t')\, e^{-i\epsilon_0(t-t')} \to (\epsilon-\epsilon_0+ i0)^{-1}\, ;  \nonumber \\
{\cal G}^{A}(t,t')\!\!\! &=&
 i\theta(t'-t)\, e^{-i\epsilon_0(t-t')} \to (\epsilon-\epsilon_0 - i0)^{-1}\, ;  \\
 \hskip -.3cm
{\cal G}^{K}(t,t') &=& -i (1-2n_F)\, e^{-i\epsilon_0(t-t')}\to
(1-2n_F(\epsilon))(-2\pi i) \delta(\epsilon-\epsilon_0) \nonumber
\, .
\end{eqnarray}
where the r.h.s. provides also the Fourier transforms. In thermal
equilibrium, one obtains:
\begin{equation}
{\cal G}^{K}(\epsilon) =\tanh{\epsilon \over 2\,T} \left( {\cal
G}^{R}(\epsilon) - {\cal G}^{A}(\epsilon) \right)\, .
                                         \label{fdtferm}
\end{equation}
This is the FDT for fermions. As in the case of bosons, the FDT
statement is a generic feature of an equilibrium system, not
restricted to the toy model. In general, it is convenient to
parameterize the anti-Hermitian  Keldysh Green function by  a
Hermitian matrix ${\cal F}={\cal F}^\dagger$ as:
\begin{equation}
{\cal G}^{K} =
 {\cal G}^{R}\circ {\cal F}  - {\cal F} \circ {\cal G}^{A}  \, ,
                                         \label{gfdtferm}
\end{equation}
The Wigner transform of ${\cal F}(t,t')$ plays the role of the
fermionic distribution function.

One may  continue  now to a system with many degrees of freedom,
counted by an index $k$. To this end, one simply changes:
$\epsilon_0\to \epsilon_k$ and perform summations over $k$. If $k$
is a momentum and $\epsilon_k=k^2/(2m)$, it is instructive to
transform to the real space representation: $\psi(t;k)\to
\psi(t;r)$ and $\epsilon_k=k^2/(2m)=-(2m)^{-1}\nabla^2_r$.
Finally, the Keldysh action for a non--interacting gas of fermions
takes the form:
\begin{equation}
S_0[\bar\psi,\psi]=\!\!\int\!\!\!\!\!\int\!\! dx\, dx'\!
\sum\limits_{a,b=1}^2 \bar\psi_a(x) \big[\hat {\cal
G}^{-1}(x,x')\big]_{ab} \,\psi_b(x')\, ,
                                                  \label{keldactionferm}
\end{equation}
where $x=(t;r)$ and the matrix correlator $[\hat {\cal
G}^{-1}]_{ab}$ has the structure of Eq.~(\ref{Greeninverseferm})
with
\begin{equation}
[{\cal G}^{R(A)}(x,x')]^{-1} = \delta(x-x')\left( i\partial_{t'} +
{1\over 2m} \nabla_{r'}^2  \right)\, .
                                             \label{gradientferm}
\end{equation}
Although in continuous notations the $R$ and the $A$ components
look seemingly the same, one has to remember that in the discrete
time representation, they are matrices with the structure below
and above the main diagonal correspondingly. The Keldysh component
is a pure regularization, in the sense that it does not have a
continuum limit  (the self-energy Keldysh component does have a
non--zero continuum representation). All this information is
already properly taken into account, however, in the structure of
the Green function, Eq.~(\ref{Greenferm}).

\subsection{External fields and sources}
\label{sec_53}

Let us introduce an external time--dependent scalar potential
$-V(t)$ defined along the contour. It interacts with the fermions
as: $S_V=\int_{\cal C}\!dt\, V(t)\bar\psi(t)\psi(t)$. Expressing
it via the field components, one finds:
\begin{eqnarray}
 \hskip -.5cm
S_V\!\!\!\!\!\!&
=&\!\!\!\!\!\!\!\!\!\int\limits_{-\infty}^\infty\!\!\!\!dt\!
\left[ V_+\bar\psi_+\psi_+ \!\! - \!\! V_-\bar\psi_-\psi_- \right]
\!\! = \!\!\!\! \int\limits_{-\infty}^\infty\!\!\!\! dt\! \left[
V_{cl}\big( \bar\psi_+\psi_+  \!\!-\!\! \bar\psi_-\psi_-\big)\!\!
+ \!\! V_{q}\big(
\bar\psi_+\psi_+ \!+\! \bar\psi_-\psi_- \big) \right] \nonumber \\
 \hskip -.5cm
& =&\!\!\!\! \int\limits_{-\infty}^\infty\!\!\! dt \left[
V_{cl}\big( \bar\psi_1\psi_1  + \bar\psi_2\psi_2\big)\!  + \!
V_{q}\big( \bar\psi_1\psi_2 + \bar\psi_2\psi_1 \big) \right]\, ,
                                            \label{sourceferm}
\end{eqnarray}
where the $V_{cl}$ and the $V_q$ components are defined in the
standard way for real bosonic fields: $V_{cl(q)} = (V_+ \pm
V_-)/2$. Notice that the physical fermionic density (symmetrized
over the two branches of the contour): $\varrho = {1\over 2} \big(
\bar\psi_+\psi_+ \!+\! \bar\psi_-\psi_- \big)$ is coupled to the
{\em quantum} component of the source field, $V_q$. On the other
hand,  the classical source component, $V_{cl}$, is nothing but an
external physical scalar potential, the same at the two branches.

Notations may be substantially compactified by introducing vertex
gamma--matrices:
\begin{equation}
\hat\gamma^{cl} \equiv \left( \begin{array}{cc}
 1 & 0   \\
 0 & 1
 \end{array} \right)\, ; \hskip 1cm
\hat\gamma^{q} \equiv \left( \begin{array}{cc}
 0 & 1   \\
 1 & 0
 \end{array} \right)\, .
                              \label{gammas}
\end{equation}
With the help of these definitions, the source action
(\ref{sourceferm}) may be written as:
\begin{equation}
S_V\! =\!\!\int\limits_{-\infty}^\infty\!\!\!dt
\!\sum\limits_{a,b=1}^2 \!\left[ V_{cl}\,
\bar\psi_a\gamma^{cl}_{ab} \psi_b + V_{q}\,
\bar\psi_a\gamma^{q}_{ab} \psi_b   \right] =
\!\!\int\limits_{-\infty}^\infty\!\!\!dt\,\, \hat{\bar\psi}
V_{\alpha}\hat\gamma^\alpha \hat\psi \, ,
                                            \label{sourcematrixferm}
\end{equation}
where the summation index is $\alpha=(cl,q)$.

Let us define now the ``generating'' function as:
\begin{equation}
Z[V_{cl},V_q] \equiv  \left\langle e^{iS_V} \right\rangle \, ,
                                                       \label{generating}
\end{equation}
where the angular brackets denote the  functional integration over
the Grassmann fields $\bar\psi$ and $\psi$ with the weight of
$e^{iS_0}$ specified by the fermionic action
(\ref{keldactionferm}). In the absence of  the quantum component,
$V_q=0$, the source field is the same at  both branches of the
contour.  Therefore, the evolution along the contour still brings
the system back to its exact initial state. Thus one expects that
the classical component alone does not change the fundamental
normalization, $Z=1$. As a result:
\begin{equation}
Z[V_{cl},V_q=0] = 1\, .
                                                       \label{knorm}
\end{equation}
One may verify this statement explicitly by  expanding the action
in powers of $V_{cl}$ and employing the Wick theorem. For example,
in the first order one finds: $Z[V_{cl},0]=1 + \int\!dt\,
V_{cl}(t)\, \Tr\{\hat\gamma^{cl} \hat {\cal G}(t,t)\}=1$, where
one uses that $\hat\gamma^{cl}=\hat 1$ along with
Eq.~(\ref{tracesferm}). It is straightforward to see that for
exactly the same reasons all higher order terms in $V_{cl}$ vanish
as well.

A lesson from Eq.~(\ref{knorm}) is that  one necessarily has  to
introduce {\em quantum} sources (that change  sign between the
forward and the backward branches of the contour). The presence of
such source fields explicitly violates causality, and thus changes
the generating function. On the other hand, these fields usually
do not have a physical meaning and play only an auxiliary role. In
most cases one uses them only to generate observables by an
appropriate differentiation. Indeed, as was mentioned above, the
physical density is coupled to the quantum component of the
source. In the end, one takes the quantum sources to be zero,
restoring the causality of the action. Notice that the classical
component, $V_{cl}$, does {\em not} have to be taken to zero.

Let us see how it works. Suppose we are interested in the average
fermion density $\varrho$ at time $t$ in the presence of a certain
physical scalar potential $V_{cl}(t)$. According to
Eqs.~(\ref{sourceferm}) and (\ref{generating}) it is given by:
\begin{equation}
\varrho(t;V_{cl}) = -{i\over 2}\,{\delta \over \delta V_q(t)}\,
Z[V_{cl},V_q] \Big|_{V_q=0}\, .
                                                       \label{density}
\end{equation}
The problem is simplified if the external field, $V_{cl}$, is weak
in some sense. Then one may restrict oneself to the linear
response, by defining the  susceptibility:
\begin{equation}
\Pi^R(t,t') \equiv {\delta \over \delta V_{cl}(t')}\,
\varrho(t;V_{cl})\Big|_{V_{cl}=0} = -{i\over 2}\,\left. {\delta^2
\, Z[V_{cl},V_q] \over \delta V_{cl}(t') \delta V_q(t)}\,
\right|_{ V_q=V_{cl}=0}\, .
                                                       \label{linear}
\end{equation}
We add the subscript $R$ anticipating on  physical grounds that
the response function must be {\em retarded} (causality). We shall
demonstrate it momentarily. First, let us introduce  the {\em
polarization} matrix as:
\begin{equation}
\hat \Pi^{\alpha\beta}(t,t') \equiv -{i\over 2}\, \left. {\delta^2
\ln Z[\hat V] \over \delta V_{\beta}(t') \delta V_{\alpha}(t)}
\right|_{\hat V=0} = \left(\begin{array}{cc}
0   & \Pi^A(t,t')  \\
\!\!\! \Pi^R(t,t')  & \Pi^K(t,t')
\end{array}\!\!\right)\, .
                                              \label{respmatrix}
\end{equation}
Due  to the fundamental normalization, Eq.~(\ref{knorm}), the
logarithm is redundant for the $R$ and the $A$ components and
therefore the two definitions (\ref{linear}) and
(\ref{respmatrix}) are not in contradiction. The fact that
$\Pi^{cl,cl} =0$ is obvious from Eq.~(\ref{knorm}). To evaluate
the polarization matrix, $\hat\Pi$, consider  the Gaussian action,
Eq.~(\ref{keldactionferm}). Adding the source term,
Eq.~(\ref{sourcematrixferm}), one finds: $\,S_0+S_V = \int\!dt\,\,
\hat{\bar\psi}[\hat {\cal G}^{-1} + V_\alpha\hat \gamma^\alpha]
\psi$. Integrating out the fermion fields $\hat{\bar\psi}$, $\hat
\psi$ according to the rules of fermionic Gaussian integration,
~\ref{app_Gaussian},  one obtains:
\begin{equation}
\hskip -.4cm
 Z[\hat V]\! =\!{1\over \Tr{\hat\rho_0}}\! \det\! \left\{\! i\hat
{\cal G}^{-1}\! + \!iV_\alpha\hat \gamma^\alpha\!\right\}\!\! =\!
\det\! \left\{\! 1\!+\! \hat {\cal G}\, V_\alpha\hat \gamma^\alpha
\! \right\}\! =\! e^{\,\Tr \ln (1 + \hat{\cal G}\, V_\alpha\hat
\gamma^\alpha)} ,
                                                              \label{trlog}
\end{equation}
where one used normalization, Eq.~(\ref{unityferm}).  Notice, that
the normalization is exactly right, since $Z[0] = 1$. One may now
expand $\ln (1 + \hat {\cal G}\, V_\alpha\hat \gamma^\alpha)$ to
the second order in $V$ and then differentiate twice. As a result,
one finds for the polarization matrix:
\begin{equation}
 \Pi^{\alpha\beta}(t,t') = -{i\over 2}\,\, \Tr\left\{\hat \gamma^{\alpha}
 \hat {\cal G}(t,t')\hat \gamma^{\beta} \hat {\cal G}(t',t) \right\}\, .
                                    \label{Pi}
\end{equation}
Substituting the explicit form of the gamma-matrices,
Eq.~(\ref{gammas}), and the Green functions,
Eq.~(\ref{Greenferm}), one obtains for the {\em response} and the
{\em correlation} components:
\begin{eqnarray}
                                       \label{Kubo}
&&                                       \hskip -1cm
\Pi^{R(A)}(t,t')\! = \! -{i\over 2}\left[ {\cal G}^{R(A)}(t,t')
{\cal G}^K(t',t) +  {\cal G}^{K}(t,t') {\cal G}^{A(R)}(t',t)
\right] ;
\\
&&\hskip -1cm \Pi^{K}(t,t')\! =\! -{i\over 2}\left[ {\cal
G}^{K}(t,t') {\cal G}^K(t',t)\! + \! {\cal G}^{R}(t,t') {\cal
G}^{A}(t',t)\! + \! {\cal G}^{A}(t,t') {\cal G}^{R}(t',t)\right] .
\nonumber
\end{eqnarray}
From the first line it is obvious that $\Pi^{R(A)}(t,t')$ is
indeed a lower (upper) triangular matrix in the time domain,
justifying their superscripts. Moreover, from the symmetry
properties of the fermionic Green functions (same as
Eq.~(\ref{conj})) one finds: $\Pi^R=[\Pi^A]^\dagger$ and
$\Pi^K=-[\Pi^K]^\dagger$. As a result, the polarization matrix,
$\hat\Pi$, possesses all the symmetry properties of the bosonic
self-energy $\hat\Sigma$, Eq.~(\ref{Sigma}).

\begin{table}[h]
\begin{tabular*}{\columnwidth}{@{\extracolsep{\fill}}@{}lcccccccc@{}}
\hline \hline
\end{tabular*}
{\bf Exercise:} In the stationary case: $\hat {\cal G}(t,t')=\hat
{\cal G}(t-t')$. Fourier transform to the energy domain and write
down expressions for $\hat \Pi(\omega)$. Assume  thermal
equilibrium and, using Eq.~(\ref{fdtferm}), rewrite your results
in terms of ${\cal G}^{R(A)}$ and the equilibrium distribution
function. Show that in equilibrium, the response,
$\Pi^{R(A)}(\omega)$, and the correlation, $\Pi^K(\omega)$,
functions are related by the bosonic FDT:
\begin{equation}
\Pi^K(\omega) =\cth\, {\omega\over 2\,T}\left(
\Pi^R(\omega)-\Pi^A(\omega)\right)\, .
                              \label{responsecorrelation}
\end{equation}
\begin{tabular*}{\columnwidth}{@{\extracolsep{\fill}}@{}lcccccccc@{}}
\hline \hline
\end{tabular*}
\end{table}

Equation (\ref{Kubo}) for $\Pi^R$ constitutes the Kubo formula for
the density--density response function. In equilibrium it may be
derived using the Matsubara technique. The Matsubara routine
involves, however, the analytical continuation from discrete
imaginary frequency $\omega_m$ to real frequency $\omega$. This
procedure may prove to be cumbersome in specific applications. The
purpose of the above discussion is to demonstrate how the linear
response problems may be compactly formulated in the Keldysh
language. The latter allows to circumvent the analytical
continuation and yields results directly in the real frequency
domain.

\subsection{Tunnelling current}
\label{sec_54}

As a simple application of the technique, let us derive the
expression for the tunnelling conductance. Our starting point is
the tunnelling Hamiltonian:
\begin{equation}
\hat H=\sum\limits_k \left[ \epsilon_k^{(c)}\, c^\dagger_k c_k +
\epsilon_k^{(d)}\, d^\dagger_k d_k \right]
+\sum\limits_{kk'}\left[ T_{kk'}\,c^\dagger_k d_{k'} + T_{kk'}^*\,
d^\dagger_{k'} c_{k}\right]  ,
                                  \label{tunhamiltonian}
\end{equation}
where the operators $c_k$ and $d_{k'}$  describe fermions in the
left and right leads, while  $T_{kk'}$ are tunnelling matrix
elements between the two. The current operator is: $\hat J =
{d\over dt} \sum_k c^\dagger_k c_k =i[\hat H, \sum_k c^\dagger_k
c_k]_- = -i\sum_{kk'}\left[ T_{kk'}\,c^\dagger_k d_{k'} -
T_{kk'}^*\, d^\dagger_{k'} c_k\right]$.

To describe the system in the Keldysh formalism, one introduces
the the four--component spinor: $\hat{\bar \psi}_k=\big( {\bar
\psi}^{(c)}_{1k}, {\bar \psi}^{(c)}_{2k}, {\bar \psi}^{(d)}_{1k},
{\bar \psi}^{(d)}_{2k}\big) $, a similarly one for the fields
without the bar, and the $4\times 4$ matrices:
\begin{equation}
                           \label{tunnelingmatrix}
                           \hskip -.5cm
\hat {\cal G}_k\! = \!\! \left(\!\!\!
 \begin{array}{cc}
   \hat{\cal G}^{(c)}_k  & 0\\
   0  &  \!\!\!\!\hat{\cal G}^{(d)}_k
   \end{array} \!\!\!\!\right)\! ; \hskip .1cm
   \hat T_{k,k'}\!=\!  \left(\!\!
 \begin{array}{cc}
   0 & \!\!\!\!\!\!\!T_{kk'} \hat \gamma^{cl}\\
   \!\!T_{kk'}^*\hat\gamma^{cl}  &  0
   \end{array} \!\!\!\right)\! ; \hskip .1cm
   \hat J_{kk'}\!=\!\! \left(\!\!
 \begin{array}{cc}
   0 & \!\!\!\!\!\!\! iT_{kk'} \hat \gamma^{q}\\
   \!\! -iT_{kk'}^*\hat\gamma^{q}  &  0
   \end{array} \!\!\!\right)\! .
\end{equation}
In addition to the  already familiar Keldysh structure the spinors
and matrices above possess the structure of the left--right space.
In terms of these objects the action and the current operator take
the form:
\begin{equation}
 \hskip -.3cm
S = \!\!\int\limits_{-\infty}^\infty \!\!\! dt \sum\limits_{kk'}
\hat{\bar\psi}_k
 \left[ \delta_{kk'}\hat{\cal G}_k^{-1}
    - \hat T_{k,k'}\right] \hat\psi_{k'} \,; \hskip .5cm
                                  \label{actioncurrent}
\hat J(t) = -\sum\limits_{kk'} \hat{\bar\psi}_k
 \hat J_{k,k'} \hat\psi_{k'}\, .
\end{equation}
The current is expressed through the $\gamma^{q}$ vertex matrix in
the Keldysh space because any observable is generated by
differentiation over the {\em quantum} component of the source
field (the classical component of the source does not change the
normalization, Eq.~(\ref{knorm})).

One is now in a position to calculate the average tunnelling
current up to the second order in the matrix elements $T_{k,k'}$.
To this end one expands the action up to the first order in
$T_{k,k'}$, and applies the Wick theorem:
\begin{eqnarray}
                                \label{tuncurrent}
&& \hskip -.5cm  J(t) = i\!\!\int\limits_{-\infty}^\infty \!\!\!
dt' \sum\limits_{kk'} \Tr\left\{
 \hat J_{kk'} \hat {\cal G}_{k'}(t,t') \hat T_{k'k}
 \hat {\cal G}_{k}(t',t) \right\} \\
 &&\hskip -1cm
 \!=\!\int\limits_{-\infty}^\infty \!\!\! dt'
\sum\limits_{kk'} |T_{kk'}|^2 \Tr\left\{\hat \gamma^q \hat {\cal
G}_{k}^{(c)}(t,t')\hat\gamma^{cl}
 \hat {\cal G}_{k'}^{(d)}(t',t)   - \hat\gamma^q\hat {\cal G}_{k'}^{(d)}(t,t')
 \hat\gamma^{cl} \hat {\cal G}_{k}^{(c)}(t',t)  \right\}=
 \nonumber\\
 &&\hskip -1.1cm
 \int\limits_{-\infty}^\infty \!\!\! dt'\!
 \sum\limits_{kk'} \!|T_{kk'}|^2 \!\!\left[\!
 {\cal G}^{(c)R}_{k(t,t')}  {\cal G}^{(d)K}_{k'(t',t)}\! +\!
 {\cal G}^{(c)K}_{k(t,t')}  {\cal G}^{(d)A}_{k'(t',t)}\! -\!
 {\cal G}^{(d)R}_{k'(t,t')}  {\cal G}^{(c)K}_{k(t',t)}\! -\!
 {\cal G}^{(d)K}_{k'(t,t')}  {\cal G}^{(c)A}_{k(t',t)}\!
 \right]. \nonumber
\end{eqnarray}
Now let us assume a stationary situation so that all the Green
functions depend on the time difference only. We shall also assume
that each lead is in  local thermal equilibrium and thus its Green
functions are related to each other via the FDT: $
 {\cal G}^{(c)K}_{k}(\epsilon)  =
(1-2n_F^{(c)}(\epsilon)) [{\cal G}^{(c)R}_{k}(\epsilon) -{\cal
G}^{(c)A}_{k}(\epsilon) ]$. Similarly for the ``$d$''--lead with a
different  occupation function $n_F^{(d)}(\epsilon)$. As a result,
one finds for the tunnelling current:
\begin{equation}
 \hskip -.3cm
J\!\!=\!\!\int\limits_{-\infty}^\infty \!\!\! {d\epsilon\over \pi}
\big[ n_F^{(d)}(\epsilon)-n_F^{(c)}(\epsilon) \big] \!
\sum\limits_{kk'} \!|T_{kk'}|^2
 \big[{\cal G}^{(c)R}_{k} -{\cal G}^{(c)A}_{k} \big]
 \big[{\cal G}^{(d)R}_{k'} -{\cal G}^{(d)A}_{k'}
 \big].
                                             \label{Landauer}
\end{equation}
If the current matrix elements may be considered as approximately
momentum independent: $|T_{kk'}|^2\approx |T|^2$, the last
expression is reduced to:
\begin{equation}
J= 4\pi |T|^2  \!\!\int\limits_{-\infty}^\infty \!\!\!
d\epsilon\,\big[ n_F^{(c)}(\epsilon)-n_F^{(d)}(\epsilon)\big]\,
\nu^{(c)}(\epsilon)\, \nu^{(d)}(\epsilon)\, ,
                                             \label{Landauerfinal}
\end{equation}
where the density of states (DOS) is defined as:
\begin{equation}
\nu(\epsilon) \equiv {i\over 2\pi} \sum\limits_{k} \big[{\cal
G}^{R}_{k}(\epsilon) -{\cal G}^{A}_{k}(\epsilon) \big]\, .
                                             \label{DOS}
\end{equation}

\subsection{Interactions}
\label{sec_55}

Consider a liquid of  fermions that interact through instantaneous
density--density interactions: $\hat H_{int}=-{1\over 2}
\!\int\!\!\int\!drdr' :\!\hat \varrho(r) U(r-r') \hat
\varrho(r')\!:\, $, where $\hat \varrho(r) =c^\dagger_r c_r$ is
the  local density operator and $:\ldots:$ stands for  normal
ordering. The corresponding Keldysh contour action has the form:
$S_{int} = {1\over 2}\int_{\cal C}\! dt \!\int\!\!\int\!drdr'
U(r-r')\bar\psi_{r}\bar\psi_{r'}\psi_r\psi_{r'}$. One may now
perform the Hubbard--Stratonovich transformation with the help of
a real boson field $\varphi(t;r)$, defined along the contour:
\begin{equation}
 \hskip -.5cm
e^{\,{i \over 2}\! \int\limits_{\cal C}\! dt
\!\!\int\!\!\!\int\!drdr'
U(r-r')\bar\psi_{r}\bar\psi_{r'}\psi_r\psi_{r'} }\!\!\!\!
=\!\!\!\int \!\!{\cal D} \varphi\, e^{\,\, i \! \int\limits_{\cal
C}\! dt\! \left[{1\over 2}\! \!\int\!\!\!\int\!drdr' \varphi_r
U^{-1}_{rr'} \varphi_{r'} +\int\!\! dr \varphi_r\bar\psi_r\psi_r
\right]}\!\!,
                                               \label{HSferm}
\end{equation}
where $U^{-1}$ is a kernel, that is inverse  to the interaction
potential: $U^{-1}\circ U=1$. One notices that the auxiliary
bosonic field, $\varphi$, enters the fermionic action in exactly
the same manner as a scalar source field. Following
Eq.~(\ref{sourceferm}), one introduces  $\varphi_{cl(q)}\equiv
(\varphi_+\pm \varphi_-)/2$ and rewrites the fermion--boson
interaction term as
$\bar\psi_a\varphi_\alpha\gamma^\alpha_{ab}\psi_b\,$, where
summations are assumed over $a,b=(1,2)$ and $\alpha=(cl,q)$. The
free bosonic term takes the form of: ${1\over 2} \varphi U^{-1}
\varphi\to \varphi_\alpha U^{-1}\hat\sigma_1^{\alpha\beta}
\varphi_\beta$.

At this stage the fermionic action is Gaussian  and one may
integrate out the Grassmann variables in the same way it was done
in Eq.~(\ref{trlog}). As a result, one finds for the generating
function, Eq.~(\ref{generating}), of the interacting fermionic
liquid:
\begin{equation}
Z[\hat V] = \!\!\int \!\!{\cal D} \varphi\,\,\,  e^{\,\,\,\, i \!
\int\limits_{-\infty}^\infty\! dt \!\int\!\!\!\int\!drdr' \hat
\varphi\, U^{-1}\hat\sigma_1 \hat\varphi\,\,  +\, \Tr \ln \big[1 +
\hat{\cal G}\, (V_\alpha+\varphi_\alpha)\hat \gamma^\alpha
\big]}\, .
                                      \label{effectiveboson}
\end{equation}
Quite generally, thus, one may reduce an interacting  fermionic
problem to a theory of an effective non--linear bosonic field
(longitudinal photons). Let us demonstrate that this bosonic
theory possesses the causality structure. To this end, one
formally expands the logarithm on the r.h.s. of
Eq.~(\ref{effectiveboson}). Employing Eq.~(\ref{tracesferm}) and
recalling that $\hat\gamma^{cl}=\hat 1$, one notices that for
$\varphi_q=V_q=0$ the bosonic action is zero. As a result,
Eq.~(\ref{causality}) holds.

To proceed we shall restrict ourselves to the, so called, {\em
random phase approximation} (RPA). It neglects all terms in the
expansion of the logarithm beyond the second order. The second
order term in the expansion is conveniently expressed through the
(bare) polarization matrix $\Pi^{\alpha,\beta}$ (see
Eq.~(\ref{Pi})) of the {\em non--interacting} fermions. The
resulting effective bosonic theory is Gaussian with the action:
\begin{equation}
 \hskip -.25cm
S_{RPA}[\hat \varphi, \hat V]\! =\!
\!\!\int\!\!\!\!\!\!\int\limits_{-\infty}^\infty\!\!\! dt dt'
\!\!\int\!\!\!\!\int\!\!drdr'\!\left[  \hat \varphi \left(
U^{-1}\hat\sigma_1\! -\! \hat \Pi\right)  \hat\varphi\,\, \! -\!
2\, \hat\varphi \hat \Pi \hat V\!  -\! \hat V \hat \Pi \hat V
\right]\, .
                                      \label{RPAaction}
\end{equation}
One notices that the bare polarization matrix plays exactly the
same role as of the self-energy, $\hat \Sigma$, cf.
Eqs.~(\ref{Dyson}), (\ref{Sigma}), in the effective bosonic
theory. As a result, the full bosonic correlator
$(U^{-1}\hat\sigma_1 - \hat \Pi)$ possesses all the causality
properties, listed in section \ref{sec_24}.

Finally, let us evaluate the {\em dressed} polarization matrix of
the interacting fermi--liquid in the RPA. To this end one may
perform the bosonic Gaussian integration in the RPA action
(\ref{RPAaction}) to find  the logarithm of the generating
function: $i\ln Z_{RPA}[\hat V]=\hat V \left(\hat \Pi
+\hat\Pi(U^{-1} \hat\sigma_1 - \hat \Pi)^{-1} \hat\Pi\right) \hat
V $. Finally, employing the definition of the polarization matrix,
Eq.~(\ref{respmatrix}), and performing  simple matrix algebra, one
finds:
\begin{equation}
{\bf \hat \Pi}_{RPA}= \hat \Pi \circ \left(1-\hat\sigma_1 U\circ
\hat \Pi\right)^{-1}\, .
                                               \label{RPApolarization}
\end{equation}
It is straightforward to demonstrate that the dressed polarization
matrix possesses the same causality structure as the bare one,
Eq.~(\ref{respmatrix}). For the response component of the dressed
polarization, ${\bf \hat \Pi}_{RPA}^R$, the second factor on the
r.h.s. of Eq.~(\ref{RPApolarization}) may be considered as a
modification of the applied field, $V_{cl}$. Indeed, cf.
Eq.~(\ref{linear}), $\varrho ={\bf \hat \Pi}_{RPA}^R\circ V_{cl} =
\hat \Pi^R \circ V^{scr}_{cl}$, where the screened external
potential $V_{cl}^{scr}$ is given by:
$V_{cl}^{scr}=\left(1-\hat\sigma_1 U\circ \hat
\Pi^R\right)^{-1}\!\!\!\!\circ\! V_{cl} $. This is the RPA result
for the screening of  an external scalar potential.

\subsection{Kinetic equation}
\label{sec_56}

According to Eq.~(\ref{Pi}) to evaluate the bare (and thus RPA
dressed, Eq.~(\ref{RPApolarization})) polarization matrix, one
needs to know the fermionic Green function, $\hat {\cal G}$. While
it is known in equilibrium, it has to be determined
self--consistently in an out--of--equilibrium situation. To this
end one employs the same idea that was used in the bosonic theory
of chapter \ref{sec_2}. Namely, one writes down the Dyson equation
for the dressed fermionic Green function:
\begin{equation}
\left(\hat {\cal G}^{-1}_0 -\hat \Sigma_F\right)\circ {\bf \hat
{\cal G} }=\hat 1\, ,
                                               \label{Dysonferm}
\end{equation}
where the subscript ``$0$'' indicates the bare Green function. The
fermionic self-energy, $\Sigma_F$ turns out to have the same
structure as $\hat{\cal G}^{-1}$, Eq.~(\ref{Greeninverseferm}).
Thus the $R$ and $A$ components of the Dyson equation take a
simple form:
\begin{equation}
\left( i\partial_t + {1\over 2m}\nabla_r^2 \right) {\bf {\cal
G}^{R(A)}} = \Sigma_F^{R(A)}\circ {\bf {\cal G}^{R(A)}}\, .
                                            \label{DysonRAferm}
\end{equation}
Employing the parameterization   ${\bf {\cal G}^K} ={\bf {\cal
G}^R}\circ {\bf {\cal F}}- {\bf {\cal F}}\circ {\bf {\cal G}^A}$,
where ${\cal {\bf {\cal F}}}$ is a Hermitian matrix,  along with
Eq.~(\ref{DysonRAferm}), one rewrites the Keldysh component of the
Dyson equation as:
\begin{equation}
 \hskip -.25cm
\left[{\bf {\cal F}}, \left( i\partial_t + {1\over 2m}\nabla_r^2
\right) \right]_- =\, \Sigma_F^K-\left(\Sigma_F^{R}\circ {\bf
{\cal F} } - {\bf{\cal F}}
\circ\Sigma_F^A\right)=-i\,I_{col}[{\cal F}] \, .
                                            \label{DysonFferm}
\end{equation}
This equation is the quantum kinetic equation for the distribution
matrix ${\bf {\cal F}}$. Its l.h.s. is  the {\em kinetic} term,
while the r.h.s. is the {\em collision integral}  with
$\Sigma_F^K$ having the meaning of an ``incoming'' term and
$\Sigma_F^{R}\circ {\bf{\cal F}} - {\bf {\cal F}} \circ\Sigma_F^A$
that of an ``outgoing'' term.

The simplest diagram for the fermionic self-energy matrix, $\hat
\Sigma_{F}^{ab}$, is obtained by expanding the
Hubbard--Stratonovich transformed action, Eq.~(\ref{HSferm}), to
the second order in the fermion--boson interaction vertex,
$\hat{\bar\psi}_a \varphi_{\alpha}\hat \gamma^{\alpha}_{ab}
\hat\psi_b$, and applying the Wick theorem for both fermion and
boson fields. As a result, one finds:
\begin{eqnarray}
&& \hskip -1cm
 \hat \Sigma_{F}^{ab}(t,t') = \left(\hat
\gamma^{\alpha}_{ac}\, \hat{\bf{\cal G}}^{cd}(t,t')\, \hat
\gamma^{\beta}_{db}\right)  \, \left\langle
\varphi_\alpha(t)\varphi_\beta(t') \right\rangle =\! \left(\!\hat
\gamma^{cl}\, \hat{\bf{\cal G}}(t,t')\, \hat
\gamma^{cl}\!\right)^{\!\!ab}\!\!\! iD^K(t,t');\! \nonumber \\
&&\hskip -1cm  +\! \left(\!\hat \gamma^{cl}\, \hat{\bf{\cal
G}}(t,t')\, \hat \gamma^{q}\!\right)^{\!\!ab} \!\!\!iD^R(t,t')\! +
\!\left(\!\hat \gamma^{q}\, \hat{\bf{\cal G}}(t,t')\, \hat
\gamma^{cl}\!\right)^{\!\!ab}\!\!\! iD^A(t,t') ,
                                           \label{sigmaferm}
\end{eqnarray}
where summations over all repeated indexes are understood and the
spatial arguments have the same general structure as the time
ones. The boson Green function is denoted as $\left\langle
\varphi_\alpha(t) \varphi_\beta(t') \right\rangle
=iD^{\alpha\beta}(t,t')$.   Finally one finds for the $R$, $A$
(i.e. $(1,1)$ and $(2,2)$) and $K$ (i.e $(1,2)$) components of the
fermionic self-energy:
\begin{eqnarray}
                              \label{SigmafermRAK}
&& \hskip -1cm\Sigma_{F}^{R(A)}(t,t') = i\left({\bf{\cal
G}}^{R(A)}(t,t')D^K(t,t') + {\bf{\cal
G}}^{K}(t,t')D^{R(A)}(t,t')\right) \,; \\
&&\hskip -1cm \Sigma_{F}^{K}(t,t') = i\Big({\bf{\cal
G}}^{K}(t,t')D^K(t,t') + {\bf{\cal G}}^{R}(t,t')D^{R}(t,t') +
{\bf{\cal
G}}^{A}(t,t')D^{A}(t,t')\Big)  \nonumber\\
&&\hskip -1cm\,\,\,\,\,\,=i\Big({\bf{\cal G}}^{K}(t,t')D^K(t',t) +
\big({\bf{\cal G}}^{R}(t,t') - {\bf{\cal G}}^{A}(t,t')\big)\big(
D^{R}(t,t')- D^{A}(t,t')\big)\Big) \, , \nonumber
\end{eqnarray}
where in the last equality one had used that ${\bf{\cal
G}}^{R(A)}(t,t')D^{A(R)}(t,t')=0$,  since these expressions have
no support in the time domain (see, however, the footnote in
section \ref{sec_31}). For the same reason: $\Sigma_{F}^{21}(t,t')
= i\big({\bf{\cal G}}^{A}(t,t')D^R(t,t') + {\bf{\cal
G}}^{R}(t,t')D^{A}(t,t')\big) =0$. As expected, the retarded and
advanced components are lower and upper triangular matrices
correspondingly, with $\Sigma^R=[\Sigma^A]^\dagger$, while
$\Sigma^K=-[\Sigma^K]^\dagger$.  Notice the close resemblance of
expressions (\ref{SigmafermRAK}) to their bosonic counterparts,
Eqs.~(\ref{SigmaA})--(\ref{SigmaK}).

If one understands the bosonic Green function, $\hat D$, as the
bare {\em instantaneous} interaction potential (i.e.
$D^R=D^A=U(r-r')\delta(t-t')$ and $D^K=0$), one finds:
$\Sigma_F^R=\Sigma_F^A = iU{\cal G}^K(t,t)\delta(t-t')$ and
$\Sigma_F^K=0$. In this approximation the r.h.s. of the kinetic
equation (\ref{DysonFferm}) vanishes (since ${\cal F}$ is a
symmetric matrix) and so there is no collisional relaxation. Thus
one has to employ an approximation for $\hat D$ that contains some
retardation. The simplest and most convenient one is the RPA,
where $\hat D=(U^{-1}\hat\sigma_1 - \hat \Pi)^{-1} $, cf.
Eq.~(\ref{RPAaction}), with a matrix $\hat\Pi$ that is non--local
in time. This relation may be rewritten as the Dyson equation for
$\hat D$, namely $(U^{-1}\hat\sigma_1 - \hat \Pi)\circ\hat D=\hat
1 $. One may easily solve it for the three components of $\hat D$
and write them in the following way:
\begin{equation}
D^{R(A)} = D^R\circ\left( U^{-1} - \Pi^{A(R)}\right)\circ D^A\, ;
 \hskip .5cm
D^{K} = D^R\circ \Pi^{K}\circ D^A .
\end{equation}

Performing the Wigner transform following sections \ref{sec_35},
\ref{sec_36}, the kinetic term (the l.h.s. of
Eq.~(\ref{DysonFferm})) is exactly the same as for the complex
boson case (one has to take into account the gradient terms to
obtain a non--zero result for the WT of the commutator). The
result is (cf. Eq.~(\ref{kinetic})):
\begin{equation}
\Big(\partial_\tau -v_k\nabla_\rho - E\nabla_k\Big){\bf
f_F}(\tau,\rho,k) = I_{col}[{\bf f_F}]\, ,
                                      \label{kineticferm}
\end{equation}
where $v_k =\partial_k\epsilon_k$, $E$ is an external electric
field and the collision integral, $I_{col}$, is $i$ times the WT
of the r.h.s. of Eq.~(\ref{DysonFferm}). On the r.h.s. one may
keep only the leading terms (without the gradients). One also
employs a parameterization of the Keldysh component of the
fermionic Green function through the corresponding distribution
function: ${\cal G}^K\to {\sf g}^K={\bf f_F}( {\mathsf g}^R -{\sf
g}^A)$, where ${\bf f_{F}}(\tau,\rho,k)$ is the WT of ${\cal F}$.
Assuming, for brevity, a spatially uniform and momentum isotropic
case, one may restrict oneself to ${\bf f_{F}}(\tau,\epsilon_k)=
{\bf f_{F}}(\tau,\epsilon)$. As a result, one finds for the
collision integral:
\begin{eqnarray}
                                       \label{collisionferm}
&&\hskip -.65cm I_{col}[{\bf f_{F}}(\epsilon)]\!=
i\!\!\int\!\!{d\omega} \sum\limits_q\, D^R(\omega,q)D^A(\omega,q)
\,{\bf \Delta_{\sf g}}(\epsilon-\omega,k-q) \\
&&\hskip -.65cm \times \!\!  \left[ \left( \Pi^R-\Pi^A\right)
\big(1- {\bf f_F}(\epsilon-\omega){\bf f_F}(\epsilon)\big) - \Pi^K
\big( {\bf f_F}(\epsilon)- {\bf f_F}(\epsilon-\omega) \big)
\right]
 \nonumber ,
\end{eqnarray}
where $\Pi^{\alpha\beta}=\Pi^{\alpha\beta}(\omega,q)$, while the
time index, $\tau$, is suppressed for brevity and the notation
\begin{equation}
{\bf \Delta_{\sf g}}(\epsilon,k)\equiv\, {i\over 2\pi}
\,\big({\mathsf g}^R(\epsilon,k) -{\sf g}^A(\epsilon,k)\big)\,
                                       \label{Deltag}
\end{equation}
is introduced. For free fermions ${\bf \Delta_{\sf
g}}(\epsilon,k)=\delta(\epsilon- \epsilon_k)$. At this stage one
may observe that if the bosonic system is at equilibrium: $\Pi^K
=\cth(\omega/2T)\left[ \Pi^R-\Pi^A\right]$, then the fermionic
collision integral is nullified by:
\begin{equation}
{\bf f_F}(\epsilon)=\tanh\, {\epsilon \over 2\,T}\, \,.
                                             \label{tanh}
\end{equation}
Indeed, $1-\tanh (b-a)\,\tanh( b) = \cth (a) \big(\tanh (b) -\tanh
(b-a)\big)$. One should take into account, however, that the
bosonic degrees of freedom are {\em not} independent from the
fermionic ones. Namely, components of the polarization matrix
$\hat \Pi$ are expressed through the fermionic Green functions
according to Eq.~(\ref{Kubo}). In the WT representation these
relations take the form:
\begin{eqnarray}
                                       \label{KuboWT}
&&\hskip -1.2cm \Pi^{R} - \Pi^{A} \! = i\pi \!\!\int\!\!
{d\epsilon'}\! \sum\limits_{k'}\! {\bf \Delta_{\sf
g}}(\epsilon',k'){\bf \Delta_{\sf g}}(\epsilon'-\omega,k'-q)
 \left[ {\bf f_F}(\epsilon'-\omega)- {\bf f_F}(\epsilon')   \right] ;
 \nonumber \\
&&\hskip -1.2cm \Pi^{K}(\omega,q)   = i\pi\! \!\int\!\!
{d\epsilon'}\! \sum\limits_{k'}\! {\bf \Delta_{\sf
g}}(\epsilon',k'){\bf \Delta_{\sf
g}}(\epsilon'\!-\!\omega,k'\!-\!q)
 \left[ {\bf f_F}(\epsilon'\!-\!\omega){\bf f_F}(\epsilon') - 1
 \right].
\end{eqnarray}
Due to the same trigonometric identity the equilibrium argument
can be made self-consistent: if the fermionic system is in
equilibrium, Eq.~(\ref{tanh}), then  components of $\hat \Pi$
satisfy the bosonic FDT, Eq.~(\ref{FDT}).

\begin{figure}[t]
\fbox{\vtop to6.5cm{\vss\hsize=.975\hsize \vglue 0cm
\hspace{0.01\hsize} \hskip -.5cm \epsfxsize=1\hsize
 \epsffile{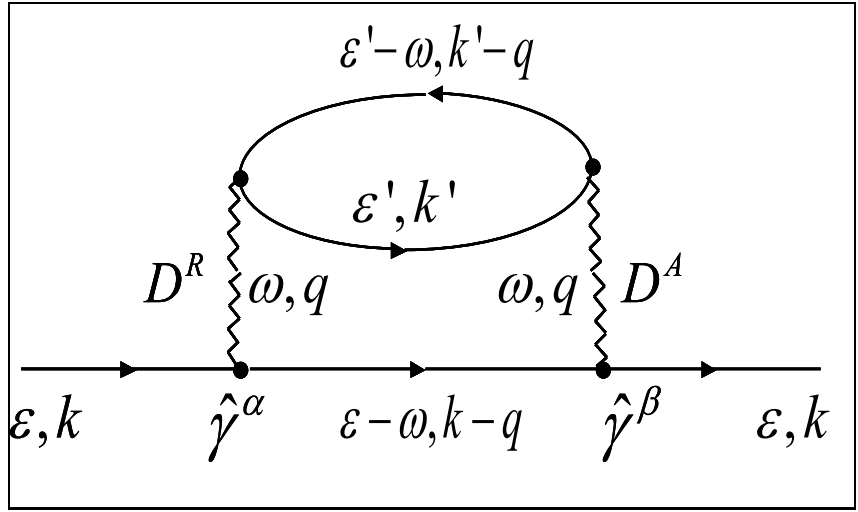}
 \hspace{0.1\hsize} \vss }}
\label{fig_collision}
 \caption{Structure of the four--fermion collision integral. The
 full lines are fermionic Green functions; the wavy lines are
 the RPA screened interaction potential. The fermionic loop represents the
 polarization matrix, $\hat \Pi(\omega,q)$.
     }
\end{figure}

One may substitute now Eqs.~(\ref{KuboWT}) into
Eq.~(\ref{collisionferm}) to write down the closed kinetic
equation for the fermionic distribution function. Most
conveniently it is done in terms of the occupation numbers,
defined as ${\bf f_F}\equiv 1-2{\bf n}$
\footnote{To derive this expression one should add and subtract
${\bf n}_{\epsilon}{\bf n}_{\epsilon'-\omega}{\bf
n}_{\epsilon'}{\bf n}_{\epsilon-\omega}$ in the square brackets.}
:
\begin{equation}
 \hskip -.45cm
{\partial {\bf n}_\epsilon \over \partial\tau}\! =\!\!
\int\!\!\!\!\! \int\!\! d\omega d\epsilon' M \Big[ {\bf
n}_{\epsilon'}{\bf n}_{\epsilon\!-\!\omega}(1\!-\!{\bf
n}_{\epsilon})(1-{\bf n}_{\epsilon'\!-\!\omega}) -    {\bf
n}_{\epsilon}{\bf n}_{\epsilon'\!-\!\omega}(1\!-\!{\bf
n}_{\epsilon'})(1\!-\!{\bf n}_{\epsilon\!-\!\omega}) \Big] ,
                                    \label{kineticeqferm}
\end{equation}
where the transition probability is given by:
\begin{equation}
 \hskip -.45cm
M(\epsilon,\omega)= \!4\pi\! \sum\limits_{q,k'} |D^R(\omega,q)|^2
\,{\bf \Delta_{\sf g}}(\epsilon-\omega,k-q) {\bf \Delta_{\sf
g}}(\epsilon',k'){\bf \Delta_{\sf g}}(\epsilon'-\omega,k'-q).
                                                       \label{Mferm}
\end{equation}
Equation (\ref{kineticeqferm}) is a generic kinetic equation with
a ``four--fermion'' collisional relaxation. The first term in the
square brackets on its r.h.s. may be identified as ``in'', while
the second one as ``out''. Each of these terms consists of the
product of four occupation numbers, giving a probability of having
two initial states occupied and two final states  empty. For ${\bf
n}(\epsilon)$ given by the Fermi function the ``in'' and the
``out'' terms cancel each other. Therefore in  thermal equilibrium
the components of the {\em dressed} fermionic Green function must
satisfy the FDT:
\begin{equation}
{\bf {\cal G}^K}= \tanh {\epsilon \over 2\,T}\, \big({\bf {\cal
G}^R} - {\bf {\cal G}^A}\big)\,.
                                       \label{FDTferm}
\end{equation}

The structure of the transmission probability $M$ is illustrated
in Fig.~\ref{fig_collision}. The three factors of $\Delta_{\sf g}$
enforce that all three intermediate fermionic particles must
satisfy energy--momentum conservation (stand on mass--shell), up
to the quasi-particle life--time. The real factor $|D^R|^2$ is
associated with the square of the matrix element of the screened
interaction potential (in the RPA).

\vskip .5cm

\section{Disordered fermionic systems}
\label{sec_6}

\subsection{Disorder averaging}
\label{sec_61}

We consider fermions in the field of a static (quenched)
space--dependent scalar potential $U_{dis}(r)$. The potential is
meant to model the effect of random static impurities,
dislocations, etc. Since one does not know the exact form of the
potential, the best one can hope for is to evaluate the
statistical properties of various observables, assuming some
statistics for $U_{dis}(r)$. It is usually a reasonable guess to
prescribe a Gaussian distribution for the potential. Namely, one
assumes that the relative probability for a realization of the
potential to appear in nature is given by:
\begin{equation}
{\cal P}[U_{dis}] \sim e^{-\pi\nu\,\tau\!\! \int \!\!dr\,
U_{dis}^2(r)}\, ,
                                   \label{weightdisorder}
\end{equation}
where $\nu$ is the bare fermionic DOS  at the Fermi level and
$\tau$, called the {\em mean--free time},  measures the strength
of the random potential.

In this chapter we concentrate on  non--interacting fermions. We
would like to evaluate, say, the response function, $\Pi^R$, in
presence of the random potential  and average it over the
realizations of $U_{dis}$ with the weight given by
Eq.~(\ref{weightdisorder}). The crucial observation is that  the
response function, $\Pi^R$, may be defined as variation of the
generating function, Eq.~(\ref{linear}), and {\em not the
logarithm} of the generating function. More precisely, the two
definitions with, Eq.~(\ref{respmatrix}), and without,
Eq.~(\ref{linear}), the logarithm coincide due to the fundamental
normalization, Eq.~(\ref{knorm}). This is {\em not} the case in
the equilibrium formalism, where the presence of the logarithm
(leading to the factor $Z^{-1}$ after differentiation) is
unavoidable in order to have the correct normalization. Such a
factor $Z^{-1}=Z^{-1}[U_{dis}]$ formidably complicates the
averaging over $U_{dis}$. Two techniques were invented to perform
the averaging: the replica trick \cite{Edwards75} and the
super-symmetry (SUSY) \cite{Efetov}. The first one utilizes the
observation that $\ln Z=\lim_{n\to 0}(Z^n-1)/n$, to perform
calculations for an integer number, $n$, replicas of the same
system and take $n\to 0$ at the end of the calculations. The
second one is based on the fact that $Z^{-1}$ of the {\em
non--interacting} fermionic system equals to $Z$ of a bosonic
system in the same random potential. One thus introduces an
additional bosonic replica of the fermionic system at hand. Both
of these ideas have serious drawbacks: the replica technique
requires analytical continuation, while the SUSY is not applicable
to interacting systems.

The Keldysh formalism provides an alternative to these two methods
by insuring that $Z=1$ by construction. One may thus directly
perform the averaging of the generating function,
Eq.~(\ref{generating}), over realizations of $U_{dis}$. Since the
disorder potential possesses only the classical component (it is
exactly the same on both branches of the contour), it is coupled
only to $\hat\gamma^{cl}=\hat 1$. The disorder--dependent term in
the averaged generating function has the form:
\begin{eqnarray}
                              \label{averagedgenerating}
&&\int\!\! {\cal D}U_{dis}\,\,e^{ -\,\,\int
\!\!dr\left[\!\pi\nu\tau U_{dis}^2(r) - i U_{dis}(r)
\!\!\!\int\limits_{-\infty}^{\infty} \!\!\! dt\, \hat{
\bar\psi}_t\hat\gamma^{cl}\hat \psi_t\right]} \\
&&= e^{ -{1\over 4\pi\nu\tau} \int
\!\!dr\!\!\int\!\!\!\!\!\!\int\limits_{-\infty}^{\infty} \!\!\!
dtdt'\, \left( \bar\psi^a_t \psi^a_t\right)\left(
\bar\psi^b_{t'}\psi^b_{t'}\right)}\,,\nonumber
\end{eqnarray}
where $a,b=1,2$, and there is a summation over repeated indexes.
One can rearrange the expression in the exponent on the r.h.s. of
the last equation as $\left( \bar\psi^a_t \psi^a_t\right)\left(
\bar\psi^b_{t'}\psi^b_{t'}\right)=-\left( \bar\psi^a_t
\psi^b_{t'}\right)\left( \bar\psi^b_{t'}\psi^a_{t}\right)$
 \footnote{The minus sign originates from commuting the Grassmann
 numbers.}
and then use   the Hubbard--Stratonovich matrix field, $\hat
Q=Q^{ab}_{t,t'}(r)$:
\begin{equation}
 \hskip -.45cm
e^{ {1\over 4\pi\nu\tau}\! \int
\!\!dr\!\!\int\!\!\!\!\!\!\int\limits_{-\infty}^{\infty} \!\!\!
dtdt' \left(\! \bar\psi^a_t \psi^b_{t'}\!\right)\left(\!
\bar\psi^b_{t'}\psi^a_{t}\!\right)}\!\!\!\! = \!\!\! \int\!\!
{\cal D}\hat Q\,\,e^{-
 \int \!\!dr\!\left[\!
 {\pi\nu\over 4\tau} \Tr\{\hat Q\circ \hat Q\} - {i\over 2\tau}\!\int\!\!\!
\!\!\!\int\limits_{-\infty}^{\infty} \!\!\! dt dt' Q^{ab}_{t,t'}
\bar\psi_{t'}^b\psi^a_t\!\right]} ,
                                   \label{averagedHS}
\end{equation}
where the spatial coordinate, $r$, is suppressed in both $\hat Q$
and $\hat \psi$. At this stage the {\em average} action becomes
quadratic in the Grassmann variables and they may be integrated
out leading to the determinant of the corresponding quadratic
form: $\hat{\cal G}_0^{-1} + V_{\alpha}\hat\gamma^{\alpha}
+{i\over 2\tau}\hat Q$. All the matrices here should be understood
as having a $2\times 2$ Keldysh structure along with an $N\times
N$ structure in  discrete time. One thus finds for the {\em
disorder averaged} generating function:
\begin{equation}
Z[\hat V] = \int\!\! {\cal D}\hat Q\,\,e^{\,iS[\hat Q;\hat V]}\, ,
                                              \label{averagedgen}
\end{equation}
where
\begin{equation}
iS[\hat Q;\hat V]= -
 {\pi\nu\over 4\tau}\!\int \!\!dr\, \Tr\{\hat Q^{\,2}\} + \Tr\ln\left[
\hat{\cal G}_0^{-1}  +{i\over 2\tau}\hat Q +
V_{\alpha}\hat\gamma^{\alpha}\right]\, .
                                         \label{NLSMaction}
\end{equation}
As a result, one has traded the initial functional integral over
the static field $U_{dis}(r)$ for the functional integral over the
dynamic matrix field $\hat Q_{t,t'}(r)$. At first glance, it does
not strike as a terribly bright idea. Nevertheless, there is a
great simplification hidden in this procedure. The point is that
the disorder potential, being $\delta$--correlated, is a rapidly
oscillating function. On the other hand, as one will see  below,
the $\hat Q$--matrix field is a slow (both in space and time)
function. Thus it represents the true {\em macroscopic} (or
hydrodynamic) degrees of freedom of the system, that happen to be
the diffusively propagating modes.

\subsection{Non--linear $\sigma$--model}
\label{sec_62}

To execute this program, one first looks for the stationary
configurations of the action (\ref{NLSMaction}). Taking the
variation over $\hat Q_{t,t'}(r)$, one obtains:
\begin{equation}
\underline{\hat Q}_{t,t'}(r)= \frac{i}{\pi\nu} \left. \Big[\hat{
{\cal G}}_0^{-1}+ \frac{i}{2\tau} \underline{\hat Q} \Big]^{-1}
\right|_{t,t';r,r}\, ,
                                                              \label{NLSMsaddle}
\end{equation}
where $\underline{\hat Q}$ denotes a stationary configuration of
the fluctuating field $\hat Q$. For the purpose  of finding the
stationary configurations one has omitted the small source field,
$\hat V$. It is important to notice that the spatially non--local
operator $\big[\hat{ {\cal G}}_0^{-1}+ \frac{i}{2\tau}
\underline{\hat Q} \big]^{-1}(t,t';r,r')$ on the r.h.s. is taken
at  coinciding spatial points $r'=r$.

The strategy is to find first a spatially uniform and
time--translationally invariant solution of
Eq.~(\ref{NLSMsaddle}): $\underline{\hat Q}_{t-t'}$, and then
consider space and time--dependent  deviations from such a
solution. This strategy is adopted from  the theory of magnetic
systems, where one first finds  the uniform static magnetized
configurations and then treats spin--waves as smooth perturbations
on top of such static uniform solutions. From the structure of
Eq.~(\ref{NLSMsaddle}) one expects that the saddle--point
configuration $\underline{\hat Q}$ possesses the same structure as
the fermionic self--energy, Eq.~(\ref{Greeninverseferm}) (more
accurately, one expects that among the possible saddle points
there is a ``classical'' one, that satisfies the causality
structure, Eq.~(\ref{Greeninverseferm})). One looks, therefore,
for a solution of Eq.~(\ref{NLSMsaddle}) in the form of:
\begin{equation}
\underline{\hat Q}_{\epsilon}\equiv \hat \Lambda_\epsilon = \left(
\begin{array}{cc}
\Lambda_\epsilon^R & \Lambda_\epsilon^K\\
0                  & \Lambda_\epsilon^A
 \end{array} \right)\, .
                                  \label{lambda}
\end{equation}
Substituting this expression in Eq.~(\ref{NLSMsaddle}), one finds
\begin{equation}
 \hskip -.4cm
\Lambda^{R(A)}_\epsilon\! =\!\frac{i}{\pi\nu}\! \left.
\frac{1}{\left[{\cal G}^{R(A)}_0\right]^{-1}\!\!+\!
\frac{i}{2\tau} \Lambda^{R(A)}_\epsilon}\right|_{r,r}
\!\!=\frac{i}{\pi\nu}\!
\sum\limits_k\frac{1}{\epsilon\!-\!\xi_k\!+\!\frac{i}{2\tau}
\Lambda^{R(A)}_\epsilon} =\pm 1 ,
                                           \label{RAdis}
\end{equation}
where $\xi_k\equiv k^2/(2m)-\mu$ and one adopts
$\sum_k\ldots=\nu\int d\xi_k\ldots$, where $\nu$ is the DOS at the
Fermi surface. The summation over momentum appears because the
matrix on the r.h.s. is taken at  coinciding spatial points. The
signs are chosen so as to respect  causality: the retarded
(advanced) Green function is analytic in the entire upper (lower)
half-plane of complex energy $\epsilon$. One has also assumed that
$1/(2\tau)\ll \mu$.  The Keldysh component, as always, may be
parameterized through a Hermitian distribution function matrix:
$\Lambda^K = \Lambda^R\circ {\cal F} - {\cal F}\circ \Lambda^A=
2\,{\cal F}_\epsilon$, where the distribution function ${\cal
F}_\epsilon$ is not fixed by the saddle point equation
(\ref{NLSMsaddle}) and must be determined through the boundary
conditions. As a result one obtains:
\begin{equation}
\hat \Lambda_\epsilon = \left(
\begin{array}{cc}
1 & 2\,{\cal F}_\epsilon\\
0                  & -1
 \end{array} \right)\, .
                                  \label{lambda1}
\end{equation}
Transforming back to the time representation, one obtains
$\Lambda^{R(A)}_{t-t'}=\pm \delta(t-t'\mp 0)$, where $\mp 0$
indicates that the $\delta$--function is shifted below (above) the
main diagonal, $t=t'$. As a result, $\Tr \,
\hat\Lambda_\epsilon=0$ and $S[\hat\Lambda]=0$, as it should be,
of course, for any purely {\em classical} field configuration,
Eq.~(\ref{lambda}). There is, however, a wider class of
configurations, that leave the action (\ref{NLSMaction}) invariant
(zero). Indeed, any field configuration of the form:
\begin{equation}
\hat Q  = \hat{{\cal T}}\circ \hat \Lambda \circ \hat{{\cal
T}}^{-1}\, ,
                                                            \label{q7}
\end{equation}
where $\hat{{\cal T}}_{t,t'}(r)=\hat{{\cal T}}_{t-t'}$, and thus
commutes with $\hat{\cal G}_0$, obviously does not change the
action (\ref{NLSMaction}). This is the zero--mode Goldstone
manifold. The standard way to introduce the massless modes
(``spin--waves'') is to allow the deformation matrices $\hat{{\cal
T}}$ to be  slow functions of $t+t'$ and $r$. Thus the expression
(\ref{q7}) parameterizes the soft modes manifold of the field
$\hat Q$. One may thus restrict oneself  {\em only} to the field
configurations given by Eq.~(\ref{q7}) and disregard all others
(massive modes). An equivalent  way to characterize this manifold
is by the condition (cf.~Eq.~(\ref{lambda1})):
\begin{equation}
\hat Q^{\,2}  = \hat 1\, .
                                                            \label{q8}
\end{equation}

Our goal now is to derive an action for the soft--mode field
configurations given by Eqs.~(\ref{q7}) or (\ref{q8}). To this end
one substitutes $\hat Q  = \hat{{\cal T}}\circ \hat \Lambda \circ
\hat{{\cal T}}^{-1}$ into Eq.~(\ref{NLSMaction}) and cyclically
permutes the $\hat{{\cal T}}$ matrices under the trace sign. This
way one arrives at $\hat{{\cal T}}^{-1}\circ\hat{\cal
G}_0^{-1}\circ \hat{{\cal T}}= \hat{\cal G}_0^{-1} + \hat{{\cal
T}}^{-1}\circ[\hat{\cal G}_0^{-1},\hat{{\cal T}}]_- = \hat{\cal
G}_0^{-1} + i\hat{{\cal T}}^{-1}\circ[\partial_t +
v_F\nabla_r,\hat{{\cal T}}]_-$, where one has linearized the
dispersion relation near the Fermi surface $k^2/(2m)-\mu\approx
v_F k\to iv_F\nabla_r$. As a result, the desired action has the
form:
\begin{equation}
iS[\hat Q]=  \Tr\ln\left[1+ i\hat{\cal G} \hat{{\cal
T}}^{-1}[\partial_t,\hat{{\cal T}}]_- + i\hat{\cal G} \hat{{\cal
T}}^{-1}[v_F\nabla_r,\hat{{\cal T}}]_- \right] \, ,
                                         \label{NLSMactioninterm}
\end{equation}
where $\hat{\cal G}$ is the {\em impurity dressed} Green function,
defined as: $(\hat{\cal G}_0^{-1} +{i\over
2\tau}\hat\Lambda)\hat{\cal G}=\hat 1$. For practical calculations
it is convenient to write it as:
\begin{equation}
 \hskip -.35cm
\hat {\cal G}_\epsilon(k)\! =\! \left(\!\! \begin{array}{cc}
 {\cal G}_\epsilon^R(k) &   {\cal G}_\epsilon^K(k)
\\ 0&  {\cal G}_\epsilon^A(k) \end{array} \!\!\right)
                                                             \label{p2}
\!=\!{1\over 2}\left(  {\cal G}_\epsilon^R(k)\, [\hat 1
+\hat\Lambda_\epsilon] + {\cal G}_\epsilon^A(k)\, [\hat 1
-\hat\Lambda_\epsilon]\right) ,
\end{equation}
with
\begin{eqnarray}
  {\cal G}_\epsilon^{R(A)}(k)\!\!\!\! &=&
\left( \epsilon-\xi_k \pm i/(2\tau) \right)^{-1}\, ;
                                                           \label{p3}\\
{\cal G}_\epsilon^K(k) &=& {\cal G}_\epsilon^R(k)\, {\cal
F}_\epsilon - {\cal F}_\epsilon\, {\cal G}_\epsilon^A(k)\nonumber
\, .
\end{eqnarray}
Notice that $\sum_k \hat {\cal G}_\epsilon(k) = -i\pi\nu\,
\hat\Lambda_\epsilon$ and $\sum_k  {\cal G}_\epsilon^R(k) {\cal
G}_\epsilon^A(k)= 2\pi\nu \tau$, while the other combinations
vanish:  $\sum_k {\cal G}_\epsilon^R(k) {\cal
G}_\epsilon^R(k)=\sum_k {\cal G}_\epsilon^A(k) {\cal
G}_\epsilon^A(k)= 0$,  due to the complex $\xi_k$--plane
integration.

One can now expand  the logarithm in Eq.~(\ref{NLSMactioninterm})
to the first order in the $\partial_t$ term and to the second
order in the $\nabla_r$ term (the first order term in $\nabla_r$
vanishes due to the angular integration) and evaluate traces using
Eq.~(\ref{p3}). For the $\partial_t$ term one finds:
$\pi\nu\,\Tr\{\hat\Lambda \hat{{\cal
T}}^{-1}[\partial_t,\hat{{\cal T}}]_-\}=\pi\nu\,\Tr\{
\partial_t\hat Q\}$, where one used that $\Tr\{\partial_t \hat
\Lambda\}=0$.
For the  $\nabla_r$ term, one finds: $-{1\over 4}\pi\nu D
\Tr\{(\nabla_r \hat Q)^2\}$, where $D\equiv v_F^2\tau/d$ is the
diffusion constant and $d$ is the spatial dimensionality
\footnote{One uses that $v_F=k/m$ and  $\sum_k  {\cal
G}_\epsilon^R(k){k\over m} {\cal G}_\epsilon^A(k){k\over m}=
2\pi\nu \tau v_F^2/d=2\pi\nu D$, while the corresponding $R-R$ and
$A-A$ terms vanish. Employing Eq.~(\ref{p2}), one then arrives at
 ${1\over 4}\,\Tr \{[\hat 1 +\hat \Lambda_\epsilon](\hat{{\cal T}}^{-1}\nabla_r\hat{{\cal
 T}}) [\hat 1 -\hat \Lambda_\epsilon](\hat{{\cal T}}^{-1}\nabla_r\hat{{\cal
 T}}) \} =-{1\over 8}\,\Tr\{ (\nabla_r (\hat{{\cal T}}\hat\Lambda_\epsilon\hat{{\cal
 T}}^{-1}))^2\}= -{1\over 8}\,\Tr\{ (\nabla_r \hat Q)^2\}$.}.
Finally, one finds for the action of the soft--mode configurations
\cite{Kamenev99}:
\begin{equation}
 \hskip -.5cm
S[\hat Q]\!=\!i\pi\nu\! \int\!\! dr\, \Tr\!\left\{\!{1\over 4}\, D
\big(\nabla_r \hat Q(r)\big)^2\! -\! \, \partial_t\hat Q(r)
\!-\!iV_\alpha\hat\gamma^{\alpha}\hat Q(r)\!-\!{i\over\pi}\hat
V^T\hat\sigma_1 V\! \right\} ,
                                               \label{theNLSM}
\end{equation}
where the trace is performed over the $2\times 2$ Keldysh
structure as well as over the $N\times N$ time structure. In the
last expression we have restored the source term from
Eq.~(\ref{NLSMaction}). The last term, $\hat V^T\hat\sigma_1 V $
is the static compressibility of the electron gas. It originates
from the second order expansion of Eq.~(\ref{NLSMaction}) in $\hat
V$, while keeping the high energy part of the ${\cal G}^{R}{\cal
G}^{R}$ and ${\cal G}^{A}{\cal G}^{A}$ terms. Despite of the
simple appearance, the action (\ref{theNLSM}) is highly
non--linear due to the condition $\hat Q^{\,2}=1$. The theory
specified by Eqs.~(\ref{q8}) and (\ref{theNLSM}) is called the
{\em matrix non--linear $\sigma$--model} (NL$\sigma$M). The name
came from the theory of magnetism, where the unit--length {\em
vector}, $\sigma(r)$, represents a local (classical) spin, that
may rotate over the sphere $\sigma^2=1$.

\subsection{Usadel equation}
\label{sec__63}

Our goal is to investigate the physical consequences of the
NL$\sigma$M. As a first step, one wants to determine the most
probable (stationary) configuration, $\underline{\hat
Q}_{t,t'}(r)$, on the soft--modes manifold, Eq.~(\ref{q8}). To
this end one parameterizes deviations from $\underline{\hat
Q}_{t,t'}(r)$ as $\hat Q =\hat {\cal T}\circ \underline{\hat
Q}\circ \hat {\cal T}^{-1}$ and chooses $\hat {\cal T} = e^{\hat
{\cal W}}$, where $\hat {\cal W}_{t,t'}(r)$ is the generator of
rotations. Expanding to the first order in $\hat {\cal W}$, one
finds: $\hat Q = \underline{\hat Q} + [\hat {\cal
W}\circ,\underline{\hat Q}]_-$. One may now substitute  such a
$\hat Q$--matrix into the action  (\ref{theNLSM}) and require that
the term linear in $\hat{\cal W}$ vanishes. This leads to the
saddle--point equation for $\underline{\hat Q}$. For the first
term in the curly brackets on the r.h.s. of Eq.~(\ref{theNLSM})
one obtains: ${1\over 2}\,\Tr\{\hat{\cal W}\circ\nabla_r
D\big(\nabla_r \underline{\hat Q}\circ \underline{\hat Q} -
\underline{\hat Q}\circ\nabla_r \underline{\hat
Q}\big)\}=-\Tr\{\hat{\cal W}\circ\nabla_r D\big( \underline{\hat
Q}\circ\nabla_r \underline{\hat Q}\big)\}$, where one employed
that $\nabla_r \underline{\hat Q}\circ \underline{\hat Q} +
\underline{\hat Q}\circ\nabla_r \underline{\hat Q}=0$, since
$\underline{\hat Q}^{\, 2}=\hat 1$. For the second term one finds:
$\Tr\{\hat{\cal W}_{t,t'}\big(\partial_{t'}+\partial_t\big)
\underline{\hat Q}_{t',t}\}$. It is written more compactly in the
energy representation, where $\partial_t\to -i\epsilon$, and thus
the second term is: $-i\Tr\{\hat{\cal W}\circ [\epsilon,
\underline{\hat Q}]_-\}$. Demanding that the linear term in
$\hat{\cal W}$ vanish, one finds:
\begin{equation}
\nabla_r\big(D \underline{\hat Q}\circ\nabla_r \underline{\hat
Q}\big) + i[\epsilon, \underline{\hat Q}]_- =0\, .
                                          \label{Usadel}
\end{equation}
This is the Usadel equation  for the stationary $\underline{\hat
Q}$--matrix, that must also satisfy $\underline{\hat Q}^{\,2}=\hat
1$. In the time representation $i[\epsilon, \underline{\hat
Q}]_-\to -\{\partial_t, \underline{\hat Q}\}_+$.

If one looks for  a solution of the Usadel equation (\ref{Usadel})
in the subspace of ``{\em classical}'' (having the causality
structure) configurations, then the condition $\underline{\hat
Q}^{\,2}=\hat 1$ restricts the possible solutions to $\hat
\Lambda$, Eq.~(\ref{lambda1}) (with a yet unspecified distribution
matrix ${\cal F}_{t,t'}(r)$). Therefore, in the
non--superconducting  case the Usadel equation is reduced to a
single equation for the distribution matrix ${\cal F}_{t,t'}(r)$.
It contains much more information for the superconducting case
(i.e. it also determines the local energy spectrum and
superconducting phase). Substituting Eq.~(\ref{lambda1}) into the
Usadel equation (\ref{Usadel}), one finds:
\begin{equation}
\nabla_r\, \big({D\nabla_r \cal F}\big)  + i[\epsilon, {\cal F}]_-
=0\, .
                                          \label{Usadel1}
\end{equation}
Finally, performing the time Wigner transform, ${\cal
F}_{t,t'}(r)\to {\bf f_F}(\tau,\epsilon;r)$, as explained in
section \ref{sec_35}, one obtains:
\begin{equation}
\nabla_r\, \big({D\nabla_r\bf f_F}\big)  -\partial_{\tau} {\bf
f_F} =0\, .
                                          \label{Usadel2}
\end{equation}
This is the kinetic equation for the fermionic distribution
function  ${\bf f_F}(\tau,\epsilon;r)$ of the disordered system.
It happens to be the diffusion equation. Notice, that it is the
same equation for any energy $\epsilon$ and different energies do
not ``talk'' to each other (in the adiabatic case, where the WT
works). This is a feature of  non--interacting systems. In the
presence of interactions, the equation acquires the collision
integral on the r.h.s. that mixes different energies between
themselves. It is worth mentioning that  elastic scattering does
not show up in the collision integral. It was already fully taken
into account in the  derivation of the Usadel equation and went
into the diffusion term, $D\nabla_r^2$.

As an example, let us consider a disordered one--dimensional wire
of  length $L$ \cite{Devoret}, attached to  two leads, that are
kept at  different voltages. There is a stationary current passing
through the wire. We look for the space dependent distribution
function, ${\bf f_F}(\epsilon;r)$, that satisfies $D\nabla_r^2{\bf
f_F}=0$ in a stationary setup (for a space independent diffusion
constant, $D$). As a result,
\begin{equation}
{\bf f_F}(\epsilon;r) =f_L(\epsilon) +
\left(f_R(\epsilon)-f_L(\epsilon)\right){r\over L}\, ,
                                         \label{wire}
\end{equation}
where $f_{L(R)}(\epsilon)$ are the distribution functions of the
left and right leads. The distribution function inside the wire
interpolates the two distributions linearly. At low temperatures
it looks like a two--step function, where the energy separation
between the steps is the applied voltage, $eV$, while the height
depends on position. Such a distribution was measured in a
beautiful experiment \cite{Devoret}. Comparing equation
(\ref{Usadel2}) with the continuity equation, one notices that the
current (at a given energy $\epsilon$) is given by
$J(\epsilon)=D\nabla_r {\bf f_F} =
D\left(f_R(\epsilon)-f_L(\epsilon)\right)/L$. And thus the total
current is $J=e\sum_k J(\epsilon) = e{\nu D\over L}\! \int\! d
\epsilon \left(f_R(\epsilon)-f_L(\epsilon)\right)= e^2{\nu D\over
L} V$. This is the Drude conductivity: $\sigma_D = e^2\nu D$.

\subsection{Fluctuations}
\label{sec_64}

Our next goal is to consider  fluctuations near the stationary
solution, $\underline{\hat Q}_{t,t'}(r)$. We restrict ourselves to
the soft--mode fluctuations  that satisfy $\hat Q^2=1$ only, and
neglect all massive modes that stay  outside this manifold. As was
already stated above these fluctuations of the $\hat Q$--matrix
may be parameterized as
\begin{equation}
\hat Q=e^{-{\cal W} }\circ \underline{\hat Q}\circ e^{{\cal W}} \,
.
                                                            \label{l2}
\end{equation}
The part of  ${\cal W}$ that commutes   with $\underline{\hat Q}$
does not generate any fluctuations, therefore one  restricts $\cal
W$ to  satisfy: ${\cal W}\circ \underline{\hat Q} +
\underline{\hat Q}\circ {\cal W} = 0$. Since $\underline{\hat Q}$
may be diagonalized according to:
\begin{equation}
\underline{\hat Q} = \left( \begin{array}{cc}
 1 &   2{\cal F}
\\ 0&  -1 \end{array} \right)
                                                             \label{p22}
=\left( \begin{array}{rr} 1&{\cal F} \\ 0&-1
\end{array}
\right)\circ \left( \begin{array}{cc} 1 & 0\\
0& -1 \end{array} \right)\circ \left(
\begin{array}{rr} 1&{\cal F}
\\ 0&-1 \end{array} \right)\, ,
\end{equation}
any generator ${\cal W}$ that anticommutes  with $\underline{\hat
Q}$ may be parameterized as
\begin{equation}
 \hskip -.45cm
{\cal W}\! =\! \left(\!\! \begin{array}{cc}
1 &  {\cal F} \\
0 & \!\! -1
\end{array}\!\! \right)\circ
\left( \!\! \begin{array}{cc}
          0 & \!\! w \\
\overline w &  0
\end{array} \!\!\right)\circ
\left( \begin{array}{cc}
1 &  {\cal F} \\
0 & \!\!-1
\end{array} \!\!\right)
                                                          \label{l4}
\!=\! \left(\!\! \begin{array}{cc}
{\cal F}\circ \overline w   &  {\cal F}\circ\overline w \circ {\cal F}-w\\
-\overline w    &    -\overline  w\circ {\cal F}
\end{array} \!\!\right)\!,
\end{equation}
where $\overline w_{t,t'}(r)$ and $w_{t,t'}(r)$ are arbitrary
Hermitian matrices in  time space. One, thus, understands the
functional integration over $\hat Q$ as an integration over
Hermitian $\overline w$ and $w$. The physical meaning of $w$ is a
deviation of the fermionic distribution function, ${\cal F}$, from
its stationary value. At the same time, $\overline w$ has no
classical interpretation. To a large extent it plays the role of
the quantum counterpart of $w$, that appears only as the internal
line in the diagrams.

One may now expand  the action,  Eqs.~(\ref{theNLSM}), in powers
of $\overline w$ and $w$. Since $\underline{\hat Q}$ was chosen to
be a stationary point,  the expansion starts from the second
order. In a spatially uniform case one obtains:
\begin{equation}
 \hskip -.4cm
iS^{(2)}[{\cal W}]\!= \!2\pi\nu\!\int\!\! dr\!\!\int\!\!\!\!\int
\!\!\frac{d\epsilon_1 d\epsilon_2}{4\pi^2}\,\, \overline
w_{\epsilon_1\epsilon_2}(r) \left[ -D \nabla_r^2 +
i(\epsilon_1-\epsilon_2) \right] w_{\epsilon_2\epsilon_1}(r)  .
                                                              \label{l5}
\end{equation}
The quadratic form  is diagonalized by transforming to the
momentum representation. As a result, the  propagator of small
$\hat Q$--matrix fluctuations is given by:
\begin{equation}
 \hskip -.25cm
\left\langle w_{\epsilon_2\epsilon_1}(q)\overline
w_{\epsilon_3\epsilon_4}(-q) \right\rangle_{{\cal W}}=
-\frac{1}{2\pi\nu}\,
\frac{\delta_{\epsilon_1\epsilon_3}\delta_{\epsilon_2\epsilon_4} }
{Dq^2+ i\omega }
                                                            \label{l5a}
\equiv-\frac{\delta_{\epsilon_1\epsilon_3}\delta_{\epsilon_2\epsilon_4}}{2\pi\nu}
\,\,D(\omega,q)\,  ,
\end{equation}
where $\omega\equiv\epsilon_1-\epsilon_2$ and the object
$D(\omega,q)=D(\epsilon_1-\epsilon_2,q)= (Dq^2
+i(\epsilon_1-\epsilon_2))^{-1}$ is called a {\em diffuson}. It is
an advanced (retarded) function of its first (second) energy
argument, $\epsilon_{1(2)}$, (or correspondingly $t_{1(2)}$).  The
higher order terms of the action's  expansion  describe
non--linear interactions of the diffusons with  vertices called
{\em Hikami boxes}. These non--linear terms are responsible for
the localization corrections. If the distribution function ${\cal
F}$ is spatially non--uniform, there is an additional term in the
quadratic action $i\tilde S^{(2)}[{\cal W}]=-2\pi\nu D \Tr\{
\overline w \nabla_r {\cal F} \overline w \nabla_r {\cal F}\}$.
This term generates non--zero correlations of the type $\langle
\overline w \overline w\rangle$ and is actually necessary for the
convergence of the functional integral over $\overline w$ and $w$.
In the spatially uniform case, such a convergence term is pure
regularization (the situation that was already encountered
before).

One can now derive  the linear density response to the applied
scalar potential. According to the general expression,
Eq.~(\ref{respmatrix}), the retarded response is given by
\begin{eqnarray}
&&\Pi^{R}(t,t';r,r') = -{i\over 2}\, \left. {\delta^2 Z[\hat V]
\over \delta V_{cl}(t';r') \delta V_{q}(t;r)} \right|_{\hat
V=0}\nonumber \\
 &&= \nu\delta_{t,t'}\delta_{r,r'} +
{i\over 2}(\pi\nu)^2\left\langle
\Tr\{\hat\gamma^{q}Q_{t,t}(r)\}\Tr\{\hat\gamma^{cl}Q_{t',t'}(r')\}
\right\rangle ,
                                              \label{Qdensdens}
\end{eqnarray}
where the angular brackets stand for the averaging over the action
(\ref{theNLSM}). In the Fourier representation the last expression
takes the form:
\begin{eqnarray}
 \hskip -.4cm
\Pi^{R}(\omega;q)\!=\!\nu\! + \!{i\over
2}(\pi\nu)^2\!\!\!\int\!\!\!\!\!\int\!\!\frac{d\epsilon
d\epsilon'}{(2\pi)^2} \left\langle
\Tr\{\hat\gamma^{q}Q_{\epsilon,\epsilon+\omega}(q)\}
\Tr\{\hat\gamma^{cl}Q_{\epsilon'+\omega,\epsilon'}\!(-q)\}
\right\rangle\! .
                                              \label{Qdensdensmomentum}
\end{eqnarray}
Employing Eq.~(\ref{l4}), one finds the  linear in ${\cal W}$
terms:
\begin{eqnarray}
                                                        \label{Qgammamomentum}
&&\hskip -1cm
\Tr\{\hat\gamma^{q}Q_{\epsilon,\epsilon+\omega}(q)\}\!\sim
\!2\left( {\cal F}_\epsilon \overline
w_{\epsilon,\epsilon+\omega}(q)-\overline
w_{\epsilon,\epsilon+\omega}(q){\cal F}_{\epsilon+\omega}
\right)\,;
\\
&&\hskip -1cm
\Tr\{\hat\gamma^{cl}Q_{\epsilon'+\omega,\epsilon'}(q)\}\!\sim \!
 2\!\left(
{\cal F}_{\epsilon'+\omega} \overline
w_{\epsilon'+\omega,\epsilon'}\!(q){\cal F}_{\epsilon'} -\overline
w_{\epsilon'+\omega,\epsilon'}\!(q)+
w_{\epsilon'+\omega,\epsilon'}\!(q)\right). \nonumber
\end{eqnarray}
For a spatially uniform distribution  $\langle \overline w
\overline w\rangle =0$ and only the last term of the last
expression contributes to the correlator. The result is:
\begin{eqnarray}
 \Pi^{R}(\omega;q)&=&\nu + {i\over
2}(\pi\nu)^24\!\!\int\!\!\frac{d\epsilon}{2\pi}\, \left({\cal
F}_\epsilon - {\cal F}_{\epsilon+\omega}\right) \left\langle
w_{\epsilon'+\omega,\epsilon'}(-q) \overline
w_{\epsilon,\epsilon+\omega}(q) \right\rangle \nonumber\\
&=&\nu\left[1+\frac{i\omega}{Dq^2-i\omega}\right]
=\nu\,\frac{Dq^2}{Dq^2-i\omega} \,,
                                              \label{Qdensdensmomentum}
\end{eqnarray}
where we have used the fact that for any reasonable fermionic
distribution ${\cal F}_{\pm\infty}=\pm 1$ and therefore $\int
d\epsilon({\cal F}_\epsilon - {\cal F}_{\epsilon+\omega})
=-2\omega$. The fact that $\Pi(\omega,0)=0$ is a consequence of
the particle number conservation. One has obtained the diffusion
form of the density--density response function. Also notice  that
this function is indeed retarded (analytic in the upper
half--plane of complex $\omega$), as it should be. The
current--current response function, $K^R(\omega;q)$ may be
obtained using the continuity equation $qj+\omega\varrho=0$ and is
$K^R(\omega;q)=\omega^2\Pi^{R}(\omega;q)/q^2$. As a result the
conductivity is given by
\begin{equation}
\sigma(\omega;q)={e^2\over i\omega}\,K^R(\omega;q)=e^2\nu D\,
\frac {-i\omega}{Dq^2-i\omega}\,.
                                         \label{condomegaq}
\end{equation}
In the uniform limit $q\to 0$, one obtains the Drude result:
$\sigma(\omega;0)=e^2\nu D$.

\subsection{Spectral statistics}
\label{sec_65}

Consider a piece of  disordered metal of size $L$ such that $L\gg
l$, where $l\equiv v_F\tau$ is the elastic mean free path. The
spectrum of the Schr\"odinger equation consists of a discrete set
of levels, $\epsilon_n$, that may be characterized by the {\em
sample--specific} DOS, $\nu(\epsilon)\sim
\sum_n\delta(\epsilon-\epsilon_n)$. This quantity fluctuates
wildly and usually cannot (and need not) be calculated
analytically. One may average it over the realizations of disorder
to obtain a mean DOS: $\overline{\nu(\epsilon)}$. The latter is a
smooth function of energy on the scale of the Fermi energy and
thus at low temperature  may be taken as a constant
$\overline{\nu(\epsilon_F)} \equiv \nu$. This is exactly the DOS
that was used  in the previous sections.

One may wonder how to sense fluctuations of the sample--specific
DOS $\nu(\epsilon)$ and, in particular,  a given spectrum at one
energy $\epsilon$ is correlated with itself at another energy
$\epsilon'$. To answer this question one may calculate the
spectral correlation function:
\begin{equation}
R(\epsilon,\epsilon')\equiv \overline{
\nu(\epsilon)\nu(\epsilon')} -\nu^2\, .
                                      \label{R}
\end{equation}
This function was calculated in the seminal paper of Altshuler and
Shklovskii \cite{Altshuler86} in 1986. Here we derive it using the
Keldysh NL$\sigma$M.

The DOS is  $\nu(\epsilon) = i\sum_k({\cal G}_k^R(\epsilon)- {\cal
G}_k^A(\epsilon))/(2\pi)=(\langle \psi_1\bar\psi_1\rangle -\langle
\psi_2\bar\psi_2\rangle)/(2\pi) = -\hat{\bar\psi}\hat\sigma_3
\hat\psi/(2\pi)$, where the angular brackets denote quantum (as
opposed to disorder) averaging and the indexes are in  Keldysh
space. To generate the DOS at any given energy one adds a source
term $-\int d\epsilon/(2\pi) J_\epsilon\int dr
\hat{\bar\psi}_\epsilon(r) \hat\sigma_3
\hat\psi_\epsilon(r)=-\!\int\!\!\int dt dt'\int dr
\hat{\bar\psi}_t(r) J_{t-t'} \hat\sigma_3 \hat\psi_{t'}(r) $ to
the fermionic action. Then the DOS is obtained by $\nu(\epsilon) =
\delta Z[J]/\delta J_\epsilon$. After averaging over disorder and
changing to the $\hat Q$--matrix representation in exactly the
same manner as above, the source term is translated to $\pi
\nu\int d\epsilon/(2\pi) J_\epsilon\int dr\Tr\{\hat
Q_{\epsilon,\epsilon}(r)\hat\sigma_3\}$. The derivation is the
same as the derivation of Eq.~(\ref{theNLSM}). It is now clear
that $\overline{\nu(\epsilon)}={1\over 2}\nu \langle \Tr\{\hat
Q_{\epsilon,\epsilon}\hat\sigma_3\}\rangle_Q$. Substituting $\hat
Q_{\epsilon,\epsilon}=\hat \Lambda_\epsilon$ one finds
$\overline{\nu(\epsilon)}=\nu$, as it should be, of course. It is
also easy to check that the fluctuations around $\hat \Lambda$ do
not change the result (all the fluctuation diagrams cancel due to
the causality constraints). We are now in the position to
calculate the correlation function:
\begin{equation}
R(\epsilon,\epsilon')\equiv \frac{\delta^2 Z[J]}{\delta J_\epsilon
\delta J_{\epsilon'}} -\nu^2 = \nu^2\left[{1\over 4} \langle
\Tr\{\hat Q_{\epsilon,\epsilon}\hat\sigma_3\}\Tr\{\hat
Q_{\epsilon',\epsilon'}\hat\sigma_3\}\rangle_Q -1\right]   \, .
                                      \label{R1}
\end{equation}
Employing the parameterization of Eqs.~(\ref{l2})--(\ref{l4}), one
finds up to the second order in ${\cal W}$ :
\begin{equation}
\Tr\{\hat Q\hat\sigma_3\} = 2\left[ 1 + {\cal F} \circ\overline
w+\overline w \circ{\cal F} + w\circ\overline w+\overline w\circ w
\right]         \, .
                                            \label{TrQsigma3}
\end{equation}
Since $\langle \overline w \overline w\rangle=0$, the only
non--vanishing terms contributing to  Eq.~(\ref{R1}) are those
with no $w$ and $\overline w$ at all (they cancel $\nu^2$ term)
and those of the type $\langle w\overline w w\overline w\rangle$.
Collecting the latter terms one finds:
\begin{eqnarray}
                                   \label{R2}
&& \hskip -.8cm R(\epsilon,\epsilon')\\
&&\hskip -.8cm=\nu^2\!\!\int \!\! dr \!\! \int\!\!\!\!\int\!\!
\frac{d\epsilon_1 d\epsilon_2}{4\pi^2} \, \left\langle
 (w_{\epsilon,\epsilon_1}\overline w_{\epsilon_1,\epsilon}\!+\!
 \overline  w_{\epsilon,\epsilon_1} w_{\epsilon_1,\epsilon})
 (w_{\epsilon',\epsilon_2}\overline w_{\epsilon_2,\epsilon'}\!+\!
 \overline  w_{\epsilon',\epsilon_2} w_{\epsilon_2,\epsilon'})
 \right\rangle_Q   . \nonumber
                                      \label{R2}
\end{eqnarray}
Finally, performing the Wick  contractions according to
Eq.~(\ref{l5a}) and taking into account that $\int d\epsilon_1
D^2(\epsilon-\epsilon_1;q)=0$, due to the integration of a
function that is analytic in the entire upper half--plane of
$\epsilon_1$, one finds:
\begin{equation}
R(\epsilon,\epsilon')={1\over (2\pi)^2}  \sum_q \, \left[
D^2(\epsilon-\epsilon';q)+D^2(\epsilon'-\epsilon;q)\right]\,,
                                      \label{R2}
\end{equation}
where the $q$--summation stands for a summation over the discrete
modes of the {\em diffusion} operator $D\nabla^2_r$ with the zero
current (zero derivative) at the boundary of the metal.  This is
the result of Altshuler and Shklovskii for the unitary symmetry
class. Notice that the correlation function depends  on the energy
difference $\omega=\epsilon-\epsilon'$ only.

For a small energy difference $\omega < E_{Thouless}\equiv D/L^2$
only the lowest homogenous mode, $q=0$, of the diffusion operator
(the so called zero--mode) may be retained and thus: $R(\omega) =
-1/(2\pi^2\omega^2)$. This is the universal random matrix result.
The negative  correlations mean energy levels' repulsion. Notice
that the correlations decay very slowly -- as the  inverse square
of the energy distance. One may notice that the true random matrix
result
$R_{RMT}(\omega)=-(1-\cos(2\pi\omega/\delta))/(2\pi^2\omega^2)$,
where $\delta$ is the mean level spacing, contains also an
oscillatory function of the energy difference. These oscillations
reflect discreteness of the underlying energy spectrum. They {\em
cannot} be found by the perturbation theory in small fluctuations
near the $\hat \Lambda$ ``point''. However, they may be recovered
once additional stationary points (not possessing the  causality
structure) are taken into account \cite{Altland00}. The
saddle--point method and perturbation theory  work as long as
$\omega> \delta$. Currently it is not known  how to work with the
Keldysh NL$\sigma$M at $\omega<\delta$.

In the opposite limit, $\omega>E_{Thouless}$, the summation over
modes may be replaced by an integration and thus
$R(\omega)=-c_d/\omega^{2-d/2}$, where $c_d$ is a positive
dimensionality dependent constant. This algebraic decay of the
correlations is reflected by many experimentally observable
phenomena generally known as {\em mesoscopic fluctuations}.



The purpose of these notes is to give the reader a general
perspective of the Keldysh formalism, its structure, guiding
principles, its strength and its limitations.  Due to  space
limitations, I could not include many topics of contemporary
research  interests into this introductory course. I hope to
fulfill some of the gaps on future occasions.

I am indebted to V. Lebedev, Y. Gefen, A. Andreev, A. I. Larkin,
M. Feigelman, L. Glazman, I. Ussishkin, M. Rokni  and many others
for numerous discussions that shaped these notes. I am sincerely
grateful to the school organizers for their invitation.  The work
was supported by the A.P. Sloan foundation and the NSF grant
DMR--0405212.





\appendix
\section{Gaussian integration}
\label{app_Gaussian}

For {\em any} complex $N\times N$ matrix $A_{ij}$, where
$i,j=1,\ldots N$, such that all its eigenvalues, $\lambda_i$, have
a positive real part: $\Re \lambda_i>0$, the following statement
holds:
\begin{equation}
 \hskip -.4cm
Z[J]\!=\! \int
\!\!\!\!\int\limits_{\!\!\!-\infty}^{\infty}\!\!\prod\limits_{j=1}^N
\!\frac{d\Re z_j  d\Im z_j}{\pi}\, e^{\,-\sum\limits_{ij}^N \bar
z_i A_{ij} z_j  + \sum\limits_{j}^N \left[\bar z_j J_j + \bar J_j
z_j\right] }\!\! =\! \frac{ e^{\,\,\sum\limits_{ij}^N \bar J_i
(A^{-1})_{ij} J_j}}{\det A} ,
                                            \label{Gauss}
\end{equation}
where $J_j$ is an arbitrary complex vector. To prove it, one may
start from a Hermitian matrix, that is diagonalized by a unitary
transformation: $A=U^\dagger\Lambda U$, where
$\Lambda=\mbox{diag}\{\lambda_j\}$. The identity is then easily
proven by a  change of  variables (with unit Jacobian) to
$w_i=U_{ij}z_j$. Finally, one notices that the r.h.s. of
Eq.~(\ref{Gauss}) is an analytic function of both $\Re A_{ij}$ and
$\Im A_{ij}$. Therefore, one may  continue them analytically to
the complex plane to reach an arbitrary complex matrix $A_{ij}$.
The identity (\ref{Gauss}) is thus valid as long as the integral
on its l.h.s. is well defined, that is all the eigenvalues of
$A_{ij}$ have a positive real part.

The Wick theorem deals with the average value of a string  $
z_{a_1}\ldots z_{a_k}\bar z_{b_1}\ldots \bar z_{b_k}$ weighted
with the factor $\exp\left\{\,-\sum\limits_{ij}^N \bar z_i A_{ij}
z_j\right\}$. The theorem states  that this average is given by
the sum of all possible products of pair-wise averages. For
example,
\begin{equation}
 \hskip -.4cm
\left\langle z_{a} \bar z_{b}\right\rangle \equiv \left.
\frac{1}{Z[0]} \, \frac{\delta^2  Z[J]}{\delta \bar J_a\delta
J_b}\right|_{J=0}= \left(A^{-1}\right)_{ab}\, ;
                                           \label{Wick1}
\end{equation}

\begin{equation}
 \hskip -.4cm
\left\langle z_{a_1}z_{a_2} \bar z_{b_1}\bar z_{b_2} \right\rangle
\!\equiv\! \left. \frac{1}{Z[0]}\, \frac{\delta^4  Z[J]}{\delta
\bar J_{a_1}\bar J_{a_2} \delta J_{b_1} J_{b_2}
}\right|_{J=0}\!\!\!=\! A^{-1}_{a_1b_1}A^{-1}_{a_2b_2}
+A^{-1}_{a_1b_2}A^{-1}_{a_2b_1} ,
                                           \label{Wick2}
\end{equation}
etc.

The Gaussian identity for integration over real variables  has the
form:
\begin{equation}
 \hskip -.2cm
Z[J]=\int\limits_{-\infty}^{\infty}\prod\limits_{j=1}^N \frac{d
x_j}{\sqrt{\pi}}\,\,\, e^{\,-\sum\limits_{ij}^N  x_i A_{ij} x_j +
2\sum\limits_{j}^N   x_j J_j  } = \frac{ e^{\,\,\sum\limits_{ij}^N
J_i (A^{-1})_{ij} J_j}}{\sqrt{\det A}}\, ,
                                            \label{Gaussreal}
\end{equation}
where $A$ is a {\em symmetric} complex matrix with all its
eigenvalues having a positive real part. The proof is similar to
the proof in the case of complex variables: one starts from a real
symmetric matrix, that may be diagonalized by an orthogonal
transformation. The identity (\ref{Gaussreal}) is then easily
proved by the change of variables. Finally, one may analytically
continue the r.h.s. (as long as the integral is well defined) from
a real symmetric matrix $A_{ij}$, to a {\em complex symmetric}
one.

For an integration over two sets of {\em independent} Grassmann
variables, $\bar \xi_j$ and $\xi_j$, where $j=1,2,\ldots,N$, the
Gaussian identity is valid for {\em any invertible} complex matrix
$A$:
\begin{eqnarray}
&& \hskip -.8cm Z[\bar\chi,\chi]
                                       \label{Gaussgrassmann}\\
&&\hskip -.8cm = \int \!\!\!\!\int\!\! \prod\limits_{j=1}^N
d\bar\xi_j d\xi_j \,\,\, e^{\,-\sum\limits_{ij}^N \bar \xi_i
A_{ij} \xi_j + \sum\limits_{j}^N \left[\bar \xi_j \chi_j + \bar
\chi_j \xi_j\right] } = \det A\,\, e^{\,\,\sum\limits_{ij}^N \bar
\chi_i (A^{-1})_{ij} \chi_j}\, . \nonumber
\end{eqnarray}
Here $\bar \chi_j$ and $\chi_j$ are two additional mutually
independent (and independent from $\bar \xi_j$ and $\xi_j$) sets
of Grassmann numbers. The proof may be obtained by e.g. brute
force expansion of the exponential factors, while noticing that
only terms that are linear in {\em all} $2N$ variables $\bar
\xi_j$ and $\xi_j$ are non--zero. The Wick theorem is formulated
in the same manner  as for the bosonic case, with the exception
that every combination is multiplied by the parity of the
corresponding permutation. E.g. the first term on the r.h.s. of
Eq.~(\ref{Wick2}) comes with a minus sign.

\section{Single particle quantum mechanics}
\label{app_singleparticle}

The simplest many--body system of a single bosonic state
(considered in Chapter \ref{sec_1}) is of course equivalent to a
single--particle harmonic oscillator. To make this connection
explicit,  consider the Keldysh contour action Eq.~(\ref{e2}) with
the correlator Eq.~(\ref{Gcontinious}) written in terms of the
complex field $\phi(t)$. The latter may be parameterized by its
real and imaginary parts as:
\begin{eqnarray}
                                                     \label{qp}
\phi(t) &=& {1\over\sqrt{2\,\omega_0}}\,\big({p(t)} -i\,
{\omega_0}\, q(t)  \big)\, ;
\nonumber \\
\bar \phi(t) &=& {1\over\sqrt{2\,\omega_0}}\,\big( {p(t)}+i\,
{\omega_0}\, q(t)  \big)\, .
\end{eqnarray}
In terms of the real fields $p(t)$ and $q(t)$ the action,
Eq.~(\ref{e2}), takes the form:
\begin{equation}
S[p,q]=\int\limits_{{\cal C}}\!\! dt\left[p\,\dot q - {1\over
2}\left(p^2+\omega_0^2 q^2\right) \right]\, ,
                                       \label{Hamiltonoscil}
\end{equation}
where the full time derivatives of $p^2$, $q^2$ and $p\,q$ were
omitted, since they contribute only to the boundary terms, not
written  explicitly in the continuous notation (they have to be
kept  for  proper regularization). Equation (\ref{Hamiltonoscil})
is nothing but the action of the quantum harmonic oscillator in
the Hamiltonian form. One may perform the Gaussian integration
over the $p(t)$ field to obtain:
\begin{equation}
S[q]= \int\limits_{{\cal C}}\!\! dt\left[ {1\over 2}\, \dot q^2 -
{\omega_0^2\over 2}\, q^2 \right]\, .
                                       \label{Lagroscil}
\end{equation}
This is the Feynman Lagrangian action of the harmonic oscillator,
written on the Keldysh contour. It may be generalized for  an
arbitrary single particle potential $U(q)$:
\begin{equation}
S[q(t)]= \int\limits_{{\cal C}}\!\! dt\left[ {1\over 2}\,\big(\dot
q(t)\big)^2 - U\big(q(t)\big) \right]\, .
                                       \label{Lagrpotential}
\end{equation}

One may split the $q(t)$ field into two components: $q_+(t)$ and
$q_-(t)$, residing on the forward and backward branches of the
contour, and then perform the Keldysh rotation: $q_{\pm}=q_{cl}\pm
q_{q}$. In terms of these fields the action takes the form:
\begin{equation}
S[q_{cl},q_q] = \int\limits_{-\infty}^\infty\!\!
dt\left[-2\,q_q{d^{\,2} q_{cl}\over dt^2} -
U\left(q_{cl}+q_q\right) + U(q_{cl}- q_q) \right]\, ,
                                       \label{KeldyshFeynman}
\end{equation}
where  integration by parts was performed in the term $\dot
q_q\dot q_{cl}$. This is the Keldysh form of the Feynman path
integral. The omitted boundary terms provide a convergence factor
of the form $\sim i0q_q^2$.

If the fluctuations of the quantum component $q_q(t)$ are regarded
as small, one may expand  the potential to the first order and
find for the action:
\begin{equation}
S[q_{cl},q_q] = \int\limits_{-\infty}^\infty\!\!
dt\,\left[-2\,q_q\left({d^{\,2} q_{cl}\over dt^2} + {\partial
U\left(q_{cl}\right)\over \partial q_{cl} }\right) + i0 q_q^2
+O(q_q^3)\right]\, .
                                       \label{Feynmanclass}
\end{equation}
In this limit the integration over the quantum component, $q_q$,
may be explicitly performed, leading to a functional
$\delta$--function of the expression in the round brackets. This
$\delta$--function enforces the classical Newtonian dynamics of
$q_{cl}\,$:
\begin{equation}
{d^{\,2} q_{cl}\over dt^2} =- {\partial U\left(q_{cl}\right)\over
\partial q_{cl} }\,\, .
                                       \label{Newton}
\end{equation}
For this  reason  the symmetric (over forward and backward
branches) part of the Keldysh field is called  the classical
component. The quantum mechanical information is contained in the
higher order terms in $q_q$, omitted in Eq.~(\ref{Feynmanclass}).
Notice, that for the harmonic oscillator potential the terms
denoted as $O(q_q^3)$ are absent identically. The quantum
(semiclassical) information resides, thus, in the convergence
term, $i0q_q^2$, as well as in the retarded regularization of the
$d^{\,2}/(dt^2)$ operator in Eq.~(\ref{Feynmanclass}).

One may generalize the single particle quantum mechanics onto a
chain (or lattice) of harmonically coupled particles by assigning
an index $r$ to particle coordinates: $q_r(t)$, and adding the
spring potential energy: ${v_s^2\over 2}(q_{r+1}(t)-q_r(t))^2$.
Changing to  spatially continuous  notations: $\varphi(t;r)\equiv
q_r(t)$, one finds for the Keldysh action of the real (phonon)
field:
\begin{equation}
S[\varphi]= \int \!\! dr\! \int\limits_{{\cal C}}\!\! dt\left[
{1\over 2}\,\dot \varphi^{\,2} - {v_s^2\over 2}\,(\nabla_r \varphi
)^2 - U\big(\varphi\big) \right]\, ,
                                       \label{Keldyshphonons}
\end{equation}
where the constant $v_s$ has the meaning of the sound velocity.
Finally, splitting the field into $(\varphi_+,\varphi_-)$
components and performing the Keldysh transformation:
$\varphi_\pm=\varphi_{cl}\pm \varphi_q$, and integrating by parts,
one obtains:
\begin{equation}
\hskip -.4cm
 S[\varphi_+,\varphi_-]\!=\!\! \int \!\! dr\!\!\!\!
\int\limits_{-\infty}^\infty\!\!\! dt\left[ 2\varphi_q\big(
v_s^2\,\nabla_r^2 -\partial_t^2 \big) \varphi_{cl}\! -\!
U(\varphi_{cl}\!+\!\varphi_q) \!+\!U(\varphi_{cl}\!-\!\varphi_q)
\right] .
                                       \label{Keldyshphonons1}
\end{equation}
According to the general structure of the Keldysh theory the
differential operator on the r.h.s., $\big( -\partial_t^2 +
v_s^2\,\nabla_r^2\big)$, should be understood as the retarded one.
This means it is a lower triangular matrix in the time domain.
Actually, one may symmetrize the action by performing the
integration by parts, and write it as: $\varphi_q\big(
-\partial_t^2 + v_s^2\,\nabla_r^2\big)^R \varphi_{cl}+
\varphi_{cl}\big( -\partial_t^2 + v_s^2\,\nabla_r^2\big)^A
\varphi_{q}$.

\newpage

%
%
%

%

\begin{thebibliography}{99}




\bibitem{Keldysh65} L.~V.~Keldysh, Zh. Eksp. Teor. Fiz. {\bf 47}, 1515 (1964);
[Sov. Phys. JETP {\bf 20}, 1018  (1965)].

\bibitem{Schwinger61} J. Schwinger, J. Math. Phys. {\bf 2}, 407 (1961).

\bibitem{Feynman63} R. P. Feynman and F. L Vernon Jr.,
Ann. Phys. {\bf 24}, 118 (1963).


\bibitem{Wyld} H.~W.~Wyld, Ann. Phys., {\bf 14}, 143 (1961).


\bibitem{MSR} P.~C.~Martin, E.~D.~Siggia, and H.~A.~Rose,
Phys. Rev. {\bf A 8},  423  (1973); DeDominics, J. Physique
(Paris), {\bf 37}, C1 (1976).

\bibitem{Landau}E.~M.~Lifshitz, and L.~P.~Pitaevskii,
{\em Statistical Physics, part II}, Pergamon Press (1980).

\bibitem{Mahan} G.~D.~Mahan, {\em Many--particle physics}, Plenum Press, NY, 1990.

\bibitem{Rammer} J.~Rammer, and H.~Smith, Rev. Mod. Phys.
{\bf 58}, 323, (1986).


\bibitem{Levitov}L.S. Levitov, {\em The Statistical Theory of Mesoscopic
Noise}, in  {\em Quantum Noise}, edited by Yu. V. Nazarov and Ya.
M. Blanter, Kluwer 2003.

\bibitem{Nazarov} Yu.~V. Nazarov, Ann. Phys. (Leipzig), {\bf 8}, 507 (1999);
M.~Kindermann and Yu.~V.~Nazarov, {\em Full counting statistics in
electric circuits} in {\em Quantum Noise}, edited by Yu. V.
Nazarov and Ya. M. Blanter, Kluwer 2003.

\bibitem{Sompolinsky}H.~Sompolinsky, Phys. Rev. Lett {\bf 47},  935  (1981);
H.~Sompolinsky, and A.~Zippelius, Phys. Rev. {\bf B 25},  6860
(1982).

\bibitem{Feigelman} V.~S.~Dotsenko, M.~V.~Feigelman, and L.~B.~Ioffe,
{\em Spin glasses and related problems}, v. 15, pt. 1 {\em  Soviet
scientific reviews} {\bf 15} (Harwood Academic, New--York 1990).


\bibitem{Cugliandolo} L. F. Cugliandolo,
Lecture notes in {\it Slow Relaxation and non equilibrium dynamics
in condensed matter}, Les Houches Session 77 July 2002, eds. J.-L.
Barrat, J. Dalibard, J. Kurchan, M. V. Feigel'man,
cond-mat/0210312.



\bibitem{Kamenev99}A. Kamenev, and A. Andreev, Phys. Rev. {\bf B 60}, 2218, (1999).

\bibitem{Shamon99}C.~Chamon, A.~W. W. Ludwig, and C.~Nayak,
Phys. Rev. {\bf B 60}, 2239, (1999).




\bibitem{AGD} A.~A. Abrikosov, L.~P.~Gorkov, I.~E.~Dzyaloshinski,
{\em Methods of quantum field theory in statistical physics},
Dover, NY, 1963.

\bibitem{Negele}
J.W. Negele and H. Orland, {\it Quantum Many-Particle Systems},
Adelison-Wesley, 1988.


\bibitem{Leggett01}A.~J.~Leggett, Rev. Mod. Phys. {\bf 73}, 307 (2001).

\bibitem{CL}A.~O.~Caldeira, A.~J.~Leggett, Ann. Phys. (NY) {\bf 149},
374 (1983).

\bibitem{LO} A.~I.~Larkin, and Yu.~N.~Ovchinnikov, {\em Vortex
motion in Superconductors}, in {\em Nonequilibrium
Superconductivity}, eds. D.~N.~Langenberg, A.~I.~Larkin, Elsevier
1986.

\bibitem{Edwards75} S.~F.~Edwards and P.~W.~Andreson,
J. Phys. F. {\bf 5}, 89 (1975).


\bibitem{Efetov}K. B. Efetov, Adv. Phys. {\bf 32}, 53 (1983);
K. B. Efetov, {\em Supersymmetry in Disorder and Chaos}, Cambridge
University Press, 1997.

\bibitem{Devoret} F. Pierre, H. Pothier, D. Esteve, and
M.H.~Devoret, J. Low Temp. Phys. {\bf 118}, 437 (2000).

\bibitem{Altshuler86}B. L. Altshuler, and B. I. Shklovskii, Zh. Eksp. Teor.
Fiz. {\bf 91}, 220 (1986) [Sov. Phys. JETP {\bf 64}, 127 (1986)].

\bibitem{Altland00} A. Altland and A. Kamenev,
    Phys. Rev. Lett., {\bf 85} 5615 (2000).


\end{thebibliography}
\end{document}